\documentclass[twocolumn,tighten]{aastex61}

\input epsf.tex

\begin{document}
\title{The Gemini/HST galaxy cluster project: Redshift 0.2-1.0 cluster sample, X-ray data and optical photometry catalog}
\author[0000-0003-3002-1446]{Inger J{\o}rgensen}
\affil{Gemini Observatory, 670 N.\ A`ohoku Pl., Hilo, HI 96720, USA}
\author{Kristin Chiboucas}
\affil{Gemini Observatory, 670 N.\ A`ohoku Pl., Hilo, HI 96720, USA}
\author{Pascale Hibon}
\affil{European Southern Observatory, 3107 Alonso de C\'{o}rdova, Vitacura, Santiago, Chile}
\author{Louise D. Nielsen}
\affil{Geneva Observatory, Switzerland}
\author{Marianne Takamiya}
\affil{University of Hawaii, Hilo, Hawaii, USA}

\correspondingauthor{Inger J{\o}rgensen}

\email{ijorgensen@gemini.edu, kchiboucas@gemini.edu,
phibon@eso.org, Louise.Nielsen@unige.ch, takamiya@hawaii.edu} 

\accepted{February 21, 2018}
\submitjournal{Astrophysical Journal Supplement}
%\date{Received ; accepted }
%\date{\today}

\begin{abstract}
The  Gemini/HST Galaxy Cluster Project (GCP) covers 14 $z=0.2-1.0$ clusters with 
X-ray luminosity of $L_{500} \ge 10^{44}\,{\rm ergs\,s^{-1}}$ in the 0.1-2.4 keV band.
In this paper we provide homogeneously calibrated X-ray luminosities, masses and radii,
and we present the complete catalog of the ground-based photometry for the GCP clusters. 
The clusters were observed with Gemini North or South in three or four of the optical passbands
$g'$, $r'$, $i'$ and $z'$.
The photometric catalog includes consistently calibrated total magnitudes, colors, and geometrical
parameters. The photometry reaches $\approx 25$ mag in the passband closest to rest frame $B$-band.
We summarize comparisons of our photometry with data from the Sloan Digital Sky Survey.
We describe the sample selection for our spectroscopic observations, 
and establish the calibrations to obtain rest frame magnitudes and colors.
Finally, we derive the color-magnitude relations for the clusters and briefly discuss these in the context
of evolution with redshift. Consistent with our results based on spectroscopic data, 
the color-magnitude relations support passive evolution of the red-sequence galaxies.
The absence of change in the slope with redshift, constrains the allowable age variation
along the red sequence to $<0.05$ dex between the brightest cluster galaxies and those four magnitudes fainter.  
The paper serves as the main reference for the GCP cluster and galaxy selection, X-ray data and ground-based photometry.
\end{abstract}


\section{Introduction}

Galaxy evolution can be studied through observations of galaxies at different redshifts.
Systematic surveys of clusters published in the mid- to late-1990s investigated the evolution of the 
galaxy population out to $z\approx 1$ using photometric measurements, in 
some cases combined with low resolution spectroscopic data.
Examples include the Canadian Network for Observational Cosmology (CNOC) surveys (Yee et al.\ 1996, 2000) and
the ``MORPHS'' project led by Smail and Dressler (Smail et al.\ 1997; Dressler et al.\ 1999).

The goal of the CNOC cluster survey was to establish the mass distribution within the clusters. 
However, the data, combined with CNOC2 field galaxy data, were also used for investigations of the evolution of galaxies
from $z\approx 0.6$ to the present. 
Schade et al.\ (1996ab) studied the evolution of luminosities as a function of redshift and sizes, and 
tested for environmental effects. The results supported passive evolution for bulge-dominated galaxies, 
and show no environmental dependencies in the evolution of disk- nor bulge-dominated galaxies.
Balogh et al.\ (1997, 1998) focused on the star formation rates (SFR) as measured from the $[$\ion{O}{2}$]$ emission lines
and demonstrated the significantly lower SFR present in cluster disk galaxies compared to similar
galaxies in the field. 

The MORPHS project provided imaging with {\it Hubble Space Telescope} ({\it HST}) 
(Smail et al.\ 1997) and low resolution spectroscopy (Dressler et al.\ 1999) of 10 clusters at $z=0.37-0.56$.
The data have been used for studies of morphological evolution (Dressler et al.\ 1997),
evolution of $(U-V)$ colors with redshift (Ellis et al.\ 1997), as well as studies of star formation history.
In particular,  Dressler et al.\ (2004) used stacked MORPHS spectra, combined with similar data for
higher redshift clusters, to establish that younger stellar populations were present in the higher redshift clusters.

With increased access to 8-meter class telescopes, a number of surveys were carried out
focused on more detailed studies of the spectral properties of the cluster galaxies.
The European Southern Observatory (ESO) large project ESO Distant Cluster Survey (EDisCS) 
targeted clusters at $z=0.4-0.9$ (White et al.\ 2005).
Based on these data, S\'{a}nchez-Bl\'{a}zquez et al.\ (2009) studied the stellar populations from stacked spectra,
while Sagila et al.\ (2010) investigated size evolution and established the Fundamental Plane 
(Dressler et al.\ 1987; Djorgovski \& Davis 1987) for the clusters. The results support passive evolution
of the bulge-dominated galaxies, but also indicate that a large fraction of the now passive galaxies 
entered the red sequence between $z\approx 0.8$ and $\approx 0.4$.

The Gemini Cluster Astrophysics Survey (GCLASS) consists of spectroscopic follow up of ten of the 
richest $z\approx 1.1$ clusters from the Spitzer Adaptation of the Red Sequence Survey (SpARCS) survey 
(Wilson et al.\ 2009; Muzzin et al.\ 2009).
One of the key results from GCLASS concerns the relative roles of environment or galaxy
mass as the driver of the evolution of the galaxies, Muzzin et al.\ (2012). These authors
conclude that the environment primarily affects the fraction of star-forming galaxies,
while the galaxy mass determines the stellar populations.

Beyond $z\approx 1$, deep spectroscopic observations become very challenging.
The survey GOGREEN ({\it Gemini Observations of Galaxies in Rich Early ENvironments}) 
aims to study the stellar populations of both red sequence and star forming galaxies, and 
to cover a large range in galaxy masses (Balogh et al.\ 2017).
The project includes 12 clusters and 9 groups at $z=0.8-1.5$.
The spectroscopy is of sufficient spectral resolution to study absorption lines, but cannot
be used to determine velocity dispersions of the galaxies.

Another approach is to use primarily imaging data at these redshifts.
For example, the HAWK-I Cluster Survey (PI: Lidman) covers nine clusters at $z=0.8-1.5$ with near-IR 
imaging obtained with VLT, see project summary in Cerulo et al.\ (2016).
The project aims to study galaxy populations of $z>0.8$ clusters, primarily from multi-band
photometry. The sample includes some of the most massive known clusters at these redshifts.

The above brief summary of large projects is by no means a complete list of the past and ongoing effort, 
but serves to show examples of the different approaches taken in this field.
Ultimately all the projects aim to establish aspects of the galaxy evolution from high 
redshift to the present by quantifying the galaxy properties at different redshifts. 

Our project, the Gemini/HST Galaxy Cluster Project (GCP) shares this aim.  
The clusters in the GCP are significantly more massive than the bulk of the 
clusters in EDisCS and GOGREEN. 
The GCP data include multi-band optical photometry obtained with Gemini,
high-resolution imaging primarily from {\it HST}, and deep ground-based optical spectroscopy.
Our spectroscopic observations have higher signal-to-noise (S/N) than 
reached by other projects covering similar redshifts, and
have sufficient spectral resolution for reliable measurements of velocity
dispersions and absorption line indices for individual galaxies. 
The original GCP, which is the topic of this paper, covers $z=0.2-1$ (J\o rgensen \& Chiboucas 2013;
J\o rgensen et al.\ 2017). 
Our high-redshift extension of the project, xGCP, is aimed at $z=1.2-2.0$. The first results
for $z>1$ galaxies include measurements
of galaxy velocity dispersions and line strengths, and we establish for the first time the
Fundamental Plane for a significant cluster sample at $z=1.3$ (J\o rgensen et al.\ 2014). 
Future papers will provide more detail and results for the xGCP.

In this paper we present the X-ray data and catalog of the ground-based photometry for the $z=0.2-1.0$ GCP clusters.
We start by describing the main science goals, methods and observing 
strategy of the GCP, Section \ref{SEC-GCP}. The section 
also details the cluster selection, 
contains an overview of previously published papers originating from the project, 
describes the calibration of the cluster X-ray data and summarizes
the properties of the GCP clusters.
The processing of the ground-based imaging and the determination and calibration
of the photometry are covered in Sections \ref{SEC-DATA}-\ref{SEC-PHOT}.
Comparisons with photometry from Sloan Digital Sky Survey (SDSS) are used
to ensure consistently calibrated magnitudes and colors.
Section \ref{SEC-FINALPHOT} presents the fully calibrated photometry. 
The catalog is available as a machine readable table.
In our analysis of the GCP data we make use of photometry calibrated to the 
rest frame $B$-band, as well as rest frame colors $(U-B)$ and $(B-V)$. 
These calibrations are established in Section \ref{SEC-BREST}.
In Sections \ref{SEC-CM} and \ref{SEC-CMREDSHIFT}, we 
describe the sample selection for our spectroscopic observations,
establish the  color magnitude relations in the observed bands, 
and finally discuss the evolution of the color magnitude relations as a function of redshift.
Section \ref{SEC-SUMMARY} summarizes the paper.

Throughout this paper we adopt a $\Lambda$CDM cosmology with 
$\rm H_0 = 70\,km\,s^{-1}\,Mpc^{-1}$, $\Omega_{\rm M}=0.3$, and $\Omega_{\rm \Lambda}=0.7$.

% Updated 2017jun08 with recalibrated Xray data
\begin{deluxetable*}{llrrrrrr}
\tablecaption{Cluster Properties \label{tab-redshifts} }
\tabletypesize{\scriptsize}
\tablewidth{0pt}
%\tablenum{8}
\tablehead{
\colhead{Cluster} & \colhead{Redshift} & \colhead{$\sigma _{\rm cluster}$} &
\colhead{$L_{500}$} & \colhead{$M_{500}$} & \colhead{$R_{500}$} & \colhead{N$_{\rm member}$} & \colhead{Ref.}\\
 & & $\rm km~s^{-1}$ & $10^{44} \rm{erg\,s^{-1}}$ & $10^{14}M_{\sun}$ & Mpc & \\
\colhead{(1)} & \colhead{(2)} & \colhead{(3)} & \colhead{(4)} & \colhead{(5)} & \colhead{(6)} & \colhead{(7)} & \colhead{(8)} }
\startdata
Abell 1689/RXJ1311.4--0120 & $0.1865\pm 0.0010$ & $2182_{-163}^{+150}$ & 12.524 & 8.392 & 1.350 &  72 & J2017 \\ % updated 2016mar16, from biweight method
RXJ0056.2+2622/Abell 115  & $0.1922\pm 0.0008$ & $1444_{-140}^{+119}$ &  7.485 & 6.068 & 1.206 &  58 & J2017 \\ % updated 2014jul24, from biweight method
RXJ0056.2+2622N\tablenotemark{a} & $0.1932\pm 0.0010$ & $1328_{-334}^{+213}$ &  3.935 & 4.100 & 1.058 &  12 & J2017 \\ % updated 2017apr25, from biweight method, sample limited
RXJ0056.2+2622S\tablenotemark{a} & $0.1929\pm 0.0010$ & $1218_{-206}^{+164}$ &  4.094 & 4.200 & 1.067 &  22 & J2017 \\ % updated 2017apr25, from biweight method, sample limited
RXJ0142.0+2131              & $0.2794\pm 0.0009$ & $1283_{-138}^{+166}$ &  5.587 & 4.761 & 1.079 &  30 & B2005 \\ % was 0.2796+-0.0008, 1278+-134, updated 2015dec13 from the spectro-table
RXJ0027.6+2616              & $0.3650\pm 0.0009$ & $1232_{-165}^{+122}$ &  8.376 & 5.684 & 1.108 &  34 & J2017\\  % updated 2014may13, from biweight method
RXJ0027.6+2616 group        & $0.3404\pm 0.0003$ & $172_{-47}^{+29} $ & \nodata & \nodata & \nodata & 9 & J2017\\  % updated 2014may13, from biweight method
Abell 851                   & $0.4050\pm 0.0008$ & $1391_{-112}^{+102} $ &  4.907 & 3.970 & 0.980 & 50 & H2017 \\  % updated 2017jun13 based on updated table from Pascale
RXJ1347.5-1145\tablenotemark{b} & $0.4506\pm 0.0008$ & $1259_{-250}^{+210} $ & 8.278 & 5.264 & 1.046 &  43 & J2017 \\ % updated 2017apr25, from biweight method, incl 3 more galaxies
RXJ2146.0+0423              & $0.532$  & \nodata  &  1.912 & 2.015 & 0.736 &  \nodata & In prep. \\ %
MS0451.6--0305              & $0.5398\pm 0.0010$ & $1450_{-159}^{+105}$  & 15.352 & 7.134 & 1.118 &  47 & J2013 \\
RXJ0216.5--1747             & $0.578$  & \nodata  &  2.267 & 2.147 & 0.738 & 36 & In prep. \\
RXJ1334.3+5030              & $0.620$  & \nodata  &  3.406 & 2.656 & 0.779 & \nodata & In prep. \\
RXJ1716.6+6708\tablenotemark{c} & $0.809$  & \nodata   & 4.368 & 2.623 & 0.718 & \nodata & In prep. \\
MS1610.4+6616 field\tablenotemark{d}    & $0.8300 \pm 0.0011$ & $681_{-195}^{+132}$ & \nodata & \nodata & \nodata & 12 & In prep. \\
RXJ0152.7--1357             & $0.8350\pm 0.0012$ & $1110_{-174}^{+147}$ &  6.291 & 3.222 & 0.763 &  29 & J2005 \\
RXJ0152.7--1357N\tablenotemark{b} & $0.8372\pm 0.0014$ & $681 \pm 232$ &  1.933 & 1.567 & 0.599 &  7 & J2005 \\
RXJ0152.7--1357S\tablenotemark{b} & $0.8349\pm 0.0020$ & $866 \pm 266$ &  2.961 & 2.043 & 0.657 &  6 & J2005 \\
RXJ1226.9+3332              & $0.8908\pm 0.0011$ & $1298_{-137}^{+122}$  & 11.253 & 4.386 & 0.827 &  55 & J2013 \\
RXJ1415.1+3612\tablenotemark{e} & $1.0269\pm 0.0010$ &  $676_{-77}^{+69}$    &  7.773 & 3.109 & 0.698 &  18 & In prep. \\ % updated 2015dec13 based on original process
\enddata
\tablecomments{Column 1: Galaxy cluster. Column 2: Cluster redshift. Column 3: Cluster velocity dispersion.
Column 4: X-ray luminosity in the 0.1--2.4 keV band within the radius $R_{500}$. X-ray data are from Piffaretti et al.\ (2011) except as noted.
Column 5: Cluster mass derived from X-ray data within the radius $R_{500}$.
Column 6: Radius within which the mean over-density of the cluster is 500 times the critical density at the cluster redshift.
Column 7: Number of member galaxies for which spectroscopy has been obtained.
Column 8: References for redshifts, velocity dispersions and spectroscopic data. 
B2005 -- Barr et al.\ (2005), updated to use consistent method for determination the velocity dispersion.
J2005 -- J\o rgensen et al.\ (2005);
J2013 -- J\o rgensen \& Chiboucas (2013);
J2017 -- J\o rgensen et al.\ (2017); 
H2017 -- Hibon et al.\ (In prep.).
In prep.: Papers in preparation. Except for MS1610.4+6616 and RXJ1415.1+3612, our spectroscopic data for these clusters are not fully processed. 
Thus, we do not list the cluster velocity dispersions.
}
\tablenotetext{a}{Re-calibrated X-ray data from Mahdavi et al.\ (2013, 2014), see Section \ref{SEC-xraycalib}.} % RXJ0056 subclusters
\tablenotetext{b}{Re-calibrated X-ray data from Ettori et al.\ (2004), see Section \ref{SEC-xraycalib}.} % RXJ0152 subclusters, RXJ1347
\tablenotetext{c}{Average of re-calibrated X-ray data from Ettori et al.\ (2004, 2009), see Section \ref{SEC-xraycalib}.} % RXJ1716
\tablenotetext{d}{MS1610.4+6616 is not a rich galaxy cluster. There are 27 galaxies in the redshift interval 0.60--0.86, 12 of which
are clustered at $z=0.83$.}
\tablenotetext{e}{Average of re-calibrated X-ray data from Ettori et al.\ (2009), Stott et al.\ (2010), and Pascut \& Ponman (2015), see Section \ref{SEC-xraycalib}.} % RXJ1415
\end{deluxetable*}

\section{The Gemini/HST galaxy cluster project \label{SEC-GCP} }

\subsection{Science Goals, Cluster Sample, Observing Strategy, and Methods}

The Gemini/HST Galaxy Cluster Project (GCP) 
was designed to study the evolution of the bulge-dominated passive galaxies in
very massive clusters. 
The main scientific goals of the project is to investigate to what extent 
these galaxies share a common evolutionary path, and map such a path. In the process,
we can quantify dependencies on galaxy properties and possibly cluster properties.
The present paper serves as the main reference for GCP cluster selection, project description,
X-ray data and the catalog of the ground-based photometry.

The original cluster selection was based on X-ray luminosities and spectroscopic
redshifts as available in the literature in the period 2000--2004.
Fifteen clusters were selected for the project, with the aim to have 
3-4 clusters for each 0.2 interval in redshift from $z=0.2$ to $z=1.0$. 
MS1610.4+6616 selected as a cluster at $z=0.65$ turned out not to be a 
massive cluster.  The apparently extended X-ray emission
detected by the {\it Einstein} satellite likely originates from several point sources.
Our spectroscopic data of galaxies in this field show no well-defined concentration in redshift space
consistent with a massive cluster.
This leaves us with fourteen clusters, and also the effect that 
the redshift interval $z=0.6-0.8$ is rather sparsely covered by our sample.

Using the X-ray data from Piffaretti et al.\ (2011),
the luminosity limit for the sample is $L_{500} = 10^{44}\,{\rm erg\,s^{-1}}$ in the 0.1-2.4 keV band
and within the radius $R_{500}$.
The radius $R_{500}$ is the radius within which the average cluster over-density is 500 times
the critical density of the Universe at the redshift of the cluster.
The cluster properties are summarized in Table \ref{tab-redshifts},
including information on $L_{500}$, $R_{500}$ and the corresponding masses $M_{500}$.

For each cluster we have obtained ground-based imaging in three or four passbands of $g'$, $r'$, $i'$ and $z'$.
The photometry typically reaches a limiting magnitude of 25 mag in the passband closest to the $B$-band
in the rest frame of the clusters.
The photometry is used for (1) sample selection for spectroscopic observations and
(2) calibration of both the ground-based photometry and photometry from higher spatial resolution imaging
to rest frame magnitudes and colors.

The spectroscopic samples contain 30-60 candidate members in each cluster.
This usually results in spectroscopic data for 20 or more passive bulge-dominated members in each cluster.
The S/N and resolution of the spectra are sufficient to reliably
measure velocity dispersions and absorption line strength for individual galaxies.
Our samples reach from the brightest cluster galaxies with typical dynamical masses
of ${\rm Mass} \approx 10^{12.6}\, M_{\sun}$ to galaxies with
dynamical masses of ${\rm Mass} \approx 10^{10.3} M_{\sun}$, equivalent
to a velocity dispersion of about $\rm 100\, km\,s^{-1}$.
All collection of ground-based imaging and spectroscopy was done in the period 2001--2005 with
Gemini North and South, using the Gemini Multi-Object Spectrographs GMOS-N and GMOS-S. 
See Hook et al.\ (2004) for a description of the instruments.

The GCP makes use of high spatial resolution imaging of the clusters primarily from 
the Advanced Camera for Surveys (ACS) or the Wide Field and Planetary Camera 2
(WFPC2) on board {\it Hubble Space Telescope} ({\it HST}). 
The data are in part archive data obtained for other programs and in part a result of 
our approved programs. % in Cycles 14 and 15.
RXJ0056.2+2622 was covered by high-resolution ground-based imaging in the $r'$-band obtained with Gemini North. 
The high spatial resolution imaging is used to measure half light radii, mean surface brightnesses
and total magnitudes from fits with S\'{e}rsic profiles (S\'{e}rsic 1968) and $r^{1/4}$ profiles. 
The S\'{e}rsic indices are used to ensure that our final samples for the analysis are indeed bulge-dominated galaxies.
Details on our methods for determining these parameters can be found in Chiboucas et al.\ (2009).
Table \ref{tab-hst} gives an overview of the relevant {\it HST} data. Additional two-dimensional photometry
derived from these data will be included in future papers. For some of the clusters, data 
are available for shorter wavelength filters 
than listed in the table, but these are not used in the GCP.

Our main methods for analysis so far have been to (1) study how the scaling relations 
like the Fundamental Plane and velocity dispersion--line strength relations evolve with redshift,
and (2) investigate the distributions of ages, metallicities and abundance ratios as well as 
establish how these parameters depend on galaxy velocity dispersion, redshift, and possibly the cluster environment.
Our previous papers detail the results of this analysis of eight of the clusters, see
J\o rgensen et al.\ (2005, 2006, 2007), Barr et al.\ (2005), J\o rgensen \& Chiboucas (2013) and 
J\o rgensen et al.\ (2017).
The next section provides a brief overview of these and other papers relevant for the project.

\begin{deluxetable*}{lrrrrrr}
\tablecaption{Relevant {\it Hubble Space Telescope} Imaging data \label{tab-hst} }
\tabletypesize{\scriptsize}
\tablewidth{0pt}
\tablehead{
\colhead{Cluster} & \colhead{Prg.\ ID} & \colhead{Instrument} & \colhead{Filters} & \colhead{Exptime} & \colhead{Ref.} \\
\colhead{(1)} & \colhead{(2)} & \colhead{(3)} & \colhead{(4)} & \colhead{(5)} & \colhead{(6)} }
\startdata
Abell 1689      &  9289 & ACS & F625W, F775W, F850LP & 9500, 11800, 14200 & J2018 \\
Abell 1689      & 11710 & ACS & F814W & 75180 & J2018 \\
Abell 1689      &  5993 & WFPC2 & F602W, F814W & 1800, 2300 & J2018 \\
RXJ0142.0+2131  &  9770 & ACS & F775W & 4420 & C2009 \\
RXJ0027.6+2616  &  9770 & ACS & F775W & 7185 & J2018 \\
Abell 851       &  10418 & ACS & F814W & 5464 & H2018 \\
Abell 851       &  5190, 6480 & WFPC2 & F702W & 4400-18900 & H2018 \\
Abell 851       &  5378 & WFPC2 & F814W & 12600 & H2018 \\
RXJ1347.5-1145  & 12104 & ACS & F606W, F625W, F775W & 1939, 1924, 2048 & J2018 \\
RXJ1347.5-1145  & 10492, 11591 & ACS & F814W, F850LP & 7340, 5280 & J2018 \\
RXJ2146.0+0423  & 10152 & ACS & F814W & 2150 & In prep. \\
MS0451.6--0305  & 9836 & ACS & F814W & 4072 & JC2013 \\
RXJ0216.5--1747 & 9770 & ACS & F775W & 10120 & In prep. \\
RXJ1334.3+5030  & 9770 & ACS & F775W & 10120 & In prep. \\
RXJ1716.6+6708  & 7293 & WFPC2 & F555W, F814W & 5600, 2700 & In prep. \\
MS1610.4+6616   & 10826 & WFPC2 & F702W & 24000 & In prep. \\
RXJ0152.7--1357 & 9290 & ACS & F625W, F775W, F850LP & 4750, 4800, 4750 & C2009 \\
RXJ1226.9+3332  & 9033 & ACS & F606W, F814W & 4000, 4000 & C2009 \\
RXJ1415.1+3612  & 10496 & ACS & F850LP & 9920 & In prep. \\
\enddata
\tablecomments{Column 1: Cluster name.
Column 2: {\it HST} program ID.
Column 3: {\it HST} instrument.
Column 4: Filters for imaging.
Column 5: Total exposure times in seconds for each of the filters.
Column 6: Reference for our 2-dimensional photometry. 
C2009 -- Chiboucas et al.\ (2009),
JC2013 -- J\o rgensen \& Chiboucas (2013),
J2018 -- J\o rgensen et al.\ (2018, in prep.),
H2018 -- Hibon et al.\ (2018, in prep.).
}
\end{deluxetable*}

\subsection{Previous Papers from the GCP}

In our first paper from the GCP, J\o rgensen et al.\ (2005), we presented results for RXJ0152.7--1157 ($z=0.84$) based on
the ground-based photometry and spectroscopy. The data support passive evolution, but also
highlighted that the cluster appears to contain galaxies with unusually high abundance ratios, [$\alpha$/Fe].
The paper contains all spectroscopic measurements for the cluster members, as well as a grey-scale
image showing the sample and the X-ray data.

Barr et al.\ (2005, 2006) studied the $z=0.28$ cluster RXJ0142.0+2131. 
The cluster has scaling relations with unusually high scatter, and may be a merging cluster.
At the time of publication, no {\it XMM-Newton} or {\it Chandra} X-ray data existed
of the cluster, making it difficult to evaluate the presence of a cluster merger.
Barr et al.\ (2005) present all spectroscopic measurements for the cluster members.

In J\o rgensen et al.\ (2006, 2007), we establized the Fundamental Plane (FP) for the two
clusters  RXJ0152.7--1157 ($z=0.84$) and RXJ1226.9+3332 ($z=0.89$). 
Our results showed for the first time
that the FP, when viewed as a relation between the dynamical mass-to-light ratios and the dynamical masses,
is steeper at higher redshift than found for our local reference sample.
We interpreted this to be due to the presence of younger stellar population in the lower mass
galaxies, than in the higher mass galaxies.

In order to provide a homogeneous photometric calibration to apply to all the ground-based
photometry used in the GCP, we processed all standard star observations obtained with GMOS-N
in the period August 2001 to December 2003. The magnitude zero points and color terms 
established from these data were presented in J\o rgensen (2009) and are used in the present paper.

The methods used for deriving 2-dimensional photometry from the {\it HST} data are described
in Chiboucas et al.\ (2009). The paper contains measurements of effective radii, total magnitudes and
S\'{e}rsic (1968) indices for our sample galaxies in RXJ0142.0+2131, RXJ0152.7--1157 and RXJ1226.9+3332.

In J\o rgensen \& Chiboucas (2013), we presented the joint analysis of the spectroscopic and 
photometric data of the three clusters MS0451.6--0305 ($z=0.54$), RXJ0152.7--1157 and RXJ1226.9+3332.
We do not detect any size evolution of the galaxies from $z\approx 0.9$ to the present. 
Our results based on the FP indicated a lower formation redshift than we found from the 
Balmer absorption lines. We speculated that the difference may be due to 
evolution in the dark matter content affecting the FP result.
The paper contains all spectroscopic measurements of cluster members in MS0451.6--0305 and
RXJ1226.9+3332, as well as photometric parameters for galaxies in MS0451.6--0305 based on the 
available {\it HST} imaging.
We also provide grey-scale images of MS0451.6--0305 and RXJ1226.9+3332 showing the samples and the X-ray imaging
of the clusters.

Woodrum et al.\ (2017) analyzed the stellar populations of the non-member galaxies in the
fields of MS0451.6--0305, RXJ0152.7--1157 and RXJ1226.9+3332. The data show an absence of
size evolution also for the field galaxies, a FP in agreement with our results for the
cluster galaxies, and formation redshifts also consistent with our results for cluster galaxies.
The paper contains the spectroscopic measurements for all non-member galaxies in the three fields.

Our analysis in J\o rgensen et al.\ (2017) is focused on the seven most massive clusters in
the GCP $z=0.2-1.0$ sample. We analyzed the joint spectroscopic data for the clusters
Abell 1689 ($z=0.19$), RXJ0056.2+2622 ($z=0.19$), RXJ0027.6+2616 ($z=0.37$), RXJ1347.5--1145 ($z=0.45$), MS0451.6--0305, RXJ0152.7--1157,
and RXJ1226.9+3332.
In addition to revisiting the formation redshift of the passive galaxies, we also established
the age-velocity dispersion, [M/H]-velocity dispersion, and [$\alpha$/Fe]-velocity dispersion relations.
We found a flat age-velocity dispersion in apparent disagreement with results for local galaxies.
The two other relations are steep and tight, in agreement with results for local galaxies.
The paper contains all spectroscopic measurements for the cluster members in 
Abell 1689, RXJ0056.2+2622, RXJ0027.6+2616, and RXJ1347.5--1145, as well as grey scale images 
of these clusters showing the samples and the X-ray imaging.

\begin{deluxetable*}{llrrrrr}
\tablecaption{Calibration of Cluster X-ray Measurements \label{tab-xraycomp} }
\tabletypesize{\scriptsize}
\tablewidth{0pt}
\tablehead{
\colhead{Catalog} & \colhead{Primary measure} & \colhead{N} &
\colhead{Mean} & \colhead{Median} & \colhead{rms} & \colhead{$\Delta$} \\
\colhead{(1)} & \colhead{(2)} & \colhead{(3)} & \colhead{(4)} & \colhead{(5)} & \colhead{(6)} & \colhead{(7)} }
\startdata
Ettori et al.\ (2004) & $R_{500}$, $M_{500}$  & 14 & 0.23 &  0.23 & 0.21 &  0.23 \\
Ettori et al.\ (2009) & $R_{500}$, $M_{500}$  & 31 & 0.24 &  0.29 & 0.23 &  0.29 \\
Stott et al.\ (2010) & $T_{\rm X}$, $M_{500}$ &  7 & 0.18 &  0.27 & 0.19 &  0.18 \\
Mahdavi et al.\ (2013, 2014) & $M_{\rm Hydro}$& 29 & 0.02 &  0.00 & 0.14 &  0.00 \\
Pascut \& Ponman (2015) & $R_{500}$, $M_{500}$& 20 & 0.08 &  0.08 & 0.13 &  0.08 \\
\enddata
\tablecomments{Column 1: Reference for X-ray data. 
Column 2: Primary parameter from catalog, see text.
Column 3: Number of clusters in common with Piffaretti et al.\ (2011). 
Column 4: Mean of the differences in $\log M_{\rm 500}$, differences are calculated as {\it Catalog} -- Piffaretti.
Column 5: Median of the differences.
Column 6: rms of the differences.
Column 7: Adopted offset in $\log M_{\rm 500}$ to reach consistency with Piffaretti et al.
}
\end{deluxetable*}

\begin{figure}
\epsfxsize 8.5cm
\epsfbox{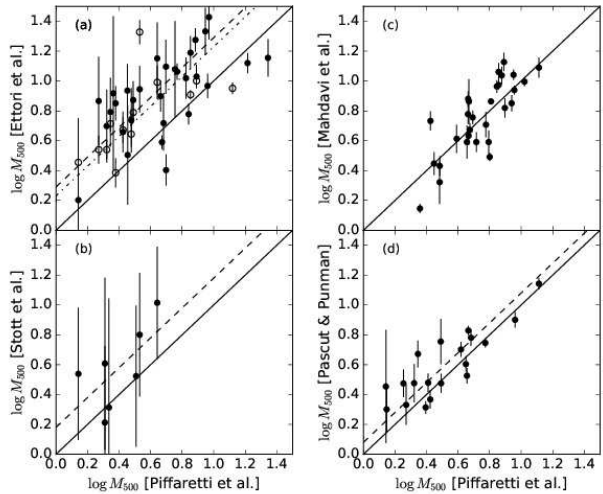}
\caption{
Cluster masses $M_{\rm 500}$ based on data from Ettori et al.\ (2004, 2009), Stott et al.\ (2010),
Mahdavi et al.\ (2013, 2014) and Pascut \& Ponman (2015) versus $M_{\rm 500}$ from
Piffaretti et al.\ (2011). Solid lines -- one-to-one relations. 
Table \ref{tab-xraycomp} summarizes the comparisons.
The offsets are shown as dashed lines on the panels.
In panel (a) data from Ettori et al.\ (2004) are shown as open symbols, with the short-dashed
line showing the offset. Data from Ettori et al.\ (2009) are shown as filled symbols, with the
long-dashed line showing the offset.
Data from Mahdavi et al.\ (2013, 2014) are consistent with Piffaretti et al.
\label{fig-XrayM500_calib} }
\end{figure}

\subsection{Calibration of X-ray Data \label{SEC-xraycalib} }

The comprehensive X-ray cluster catalog by Piffaretti et al.\ (2011) provides consistently calibrated X-ray data for the majority
of the GCP clusters. However, RXJ1716.6+6708 and RXJ1415.1+3612 are not included in this catalog, 
and it treats the binary clusters RXJ0056.2+2622 and RXJ0152.7--1357 as single clusters.
To cover these clusters, 
we calibrate X-ray data from Ettori et al.\ (2004, 2009), Stott et al.\ (2010), Mahdavi et al.\ (2013, 2014), 
and Pascut \& Ponman (2015) to consistency with Piffaretti et al.
In addition, we use updated (and calibrated) values for RXJ1347.5--1145 from Ettori et al.\ (2004), who
correct the X-ray measurements for diffuse emission from an infalling sub-cluster to the south-east of the main cluster,
see also J\o rgensen et al.\ (2017) for discussion of this cluster.

In the calibration, we use conversions between radii $R_{500}$, masses $M_{500}$, and luminosities $L_{500}$ 
as given by Piffaretti et al.\ in their equations (2) and (3). We reproduce these here for clarity.

\begin{equation}
h(z)^{-7/3} \left( \frac{L_{500}} {10^{44}\,{\rm erg\,s^{-1}}} \right) =
C \left( \frac{M_{500}} {3\cdot 10^{14} M_{\sun}} \right)^{\alpha}
\end{equation}
where $h(z)$ is the Hubble factor at redshift $z$, $\log C = 0.274$, and $\alpha = 1.64$.

\begin{equation}
\label{eq-m500}
M_{500} = \frac{4\pi}{3} \, R_{500}^3 500 \rho _c (z)
\end{equation}
where $\rho _c (z) = 3 H(z)^2 / (8\pi G)$ is the critical density of the Universe
at redshift $z$.

We convert the X-ray data from literature to $M_{500}$.
As needed we also adopt the following conversions from Piffaretti et al.\
$L_{500} = 0.91 L_{\rm total}$, $R_{200} = 1.52 R_{500}$, $L_{500}=0.96 L_{200}$.
The relation between $R_{200}$ and $R_{500}$ is equivalent to 
$M_{200} = 1.40 M_{500}$, cf.\ equation (\ref{eq-m500}).
When other conversions are used in the literature, we remove those and apply 
the above conversions before comparing with data from Piffaretti et al.

We determine the offsets in $\log M_{500}$ between the other catalogs and Piffaretti et al.\
to establish the best offset for each set of data.
Table \ref{tab-xraycomp} and Figure \ref{fig-XrayM500_calib} summarize the comparisons
and the adopted offsets.

\begin{figure}
\epsfxsize 8.2cm
\epsfbox{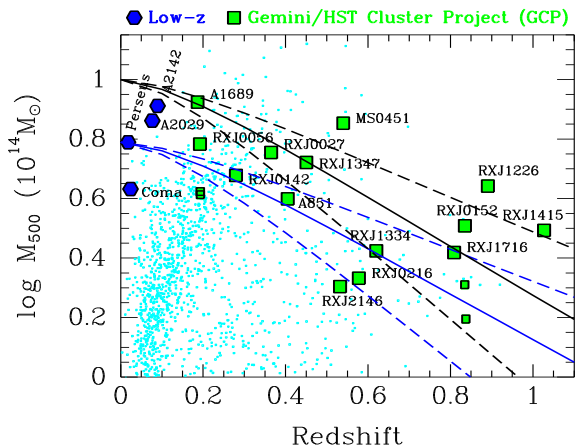}
\caption{The cluster masses, $M_{\rm 500}$, based on X-ray data versus the redshifts of the clusters,
adopted from J\o rgensen et al.\ (2017).
Blue -- our local reference sample.
Green - our $z=0.2-1$ cluster sample.
The pairs of slightly smaller points at the same redshifts as RXJ0056.6+2622
and RXJ0152.7--1357 show the values for the sub-clusters of these binary clusters.
$M_{\rm 500}$ is from Piffaretti et al.\ (2011), except  as described in the text and in the 
notes of Table \ref{tab-redshifts}.
All data have been calibrated to consistency with Piffaretti et al.\ (2011), see text.
Small cyan points -- all clusters from Piffaretti et al.\ shown for reference.
Blue and black lines -- mass development of clusters based on numerical simulations
by van den Bosch (2002).
The black lines terminate at Mass=$10^{15} M_{\sun}$ at $z=0$ roughly
matching the highest mass clusters at $z=0.1-0.2$.
The blue lines terminate at Mass=$10^{14.8} M_{\sun}$ at $z=0$
matching the mass of the Perseus cluster.
The dashed lines represent the typical uncertainty in the
mass development represented by the numerical simulations.
\label{fig-M500redshift} }
\end{figure}

\subsection{Cluster Properties}

Table \ref{tab-redshifts} summarizes the properties for all GCP clusters.
In Figure \ref{fig-M500redshift} we show the cluster masses versus redshifts for these clusters.
For reference the figure also shows our local reference sample and the clusters
from Piffaretti et al.\ (2011).
We show sample models for the growth of cluster masses with time, based on simulations
from van den Bosch (2002). 
These models are in general agreement with newer and more detailed analysis of the results from
the Millennium simulations (Fakhouri et al.\ 2010).

Grey scale optical images with X-ray data overlaid of clusters for which we previously have published results
are available in those papers as follows: 
RXJ0152.7--1357 is published in J\o rgensen et al.\ (2005), 
MS0451.6--0305 and RXJ1226.2+3332  are published J\o rgensen \& Chiboucas (2013), and
Abell 1689, RXJ0056.2+2622, RXJ0027.6+2616, and RXJ1347.5--1147 are published in J\o rgensen et al.\ (2017).
The remaining clusters are shown in Appendix \ref{SEC-GREYSCALE} of the current paper, Figures \ref{fig-RXJ0142grey}-\ref{fig-RXJ1415grey}. 
The grey scale images show the spectroscopic samples, and when available at this time, information 
about cluster membership and galaxy properties.
In Appendix \ref{SEC-GREYSCALE} we also describe each of the clusters, including providing the
original references for their discovery and main references for substantial results on the 
cluster properties and, when available, the star formation history of their members.

\section{Ground-Based Imaging \label{SEC-DATA} }

Ground-based imaging of the clusters was obtained with GMOS-N and GMOS-S in
the period from 2001 to 2005. Each cluster was observed in three or four filters
of $g'$, $r'$, $i'$ and $z'$. 
Table \ref{tab-inst} summarizes the instrument information, while Table \ref{tab-imdata} 
in Appendix \ref{SEC-OBSLOG} gives detailed information on the available observations, including the Gemini program IDs.
That table also lists the adopted Galactic extinction for each of the fields and filters.
Abell 1689 and RXJ0056.2+2622,
were observed with two pointings, while all other clusters have data for one pointing.

\begin{deluxetable*}{lrr}
\tablecaption{Instrumentation\label{tab-inst} }
\tabletypesize{\scriptsize}
\tablehead{Parameter & GMOS-N & GMOS-S }
\startdata
CCDs            & 3 $\times$ E2V 2048$\times$4608 & 3 $\times$ E2V 2048$\times$4608 \\
r.o.n.\tablenotemark{a}          & (3.5,3.3,3.0) e$^-$  & (4.0,3.7,3.3) e$^-$      \\
gain\tablenotemark{a}            & (2.04,2.3,2.19) e$^-$/ADU  &  (2.33,2.07,2.07) e$^-$/ADU \\
Unbinned pixel scale     & 0.0727arcsec/pixel  & 0.073arcsec/pixel \\
Field of view   & $5\farcm5\times5\farcm5$ & $5\farcm5\times5\farcm5$ \\
Imaging filters  & $g', r', i', z'$  & $g', r', i', z'$\\
\enddata
\tablenotetext{a}{Values for the three detectors in the array.}
\end{deluxetable*}

\begin{deluxetable*}{lcrr rrrr}
\tablecaption{Photometry Overview \label{tab-photoverview} }
\tabletypesize{\scriptsize}
\tablewidth{0pt}
%\tablenum{8}
\tablehead{
\colhead{Cluster} & \colhead{Detection band} & \colhead{Thresholds} &
\colhead{Apertures} & \colhead{5-$\sigma$ limit} & \colhead{$N_{\rm star}$} & \colhead{$N_{\rm galaxy}$} & \colhead{Area} \\
\colhead{(1)} & \colhead{(2)} & \colhead{(3)} & \colhead{(4)} & \colhead{(5)} & \colhead{(6)} & \colhead{(7)} & \colhead{(8)} }
\startdata
Abell 1689           &  $g'$ & \multicolumn{1}{l}{25.5, 24.5, 24.0, ...}  & 4.35  & 24.8 &  90 & 1981 & 44.9\tablenotemark{a} \\
RXJ0056.2+2622       &  $g'$ & \multicolumn{1}{l}{25.5, 24.5, 24.0, ...}  & 4.18  & 25.1 & 113 &  991 & 53.3\tablenotemark{a} \\
RXJ0142.0+2131       &  $r'$ & \multicolumn{1}{l}{27.0, 25.5, 24.9, ...}  & 3.08  & 24.8 &  55 & 1546 & 28.6 \\ 
RXJ0027.6+2616       &  $r'$ & \multicolumn{1}{l}{27.0, 25.5, 24.9, ...}  & 2.66  & 25.2 &  75 & 1291 & 28.0\tablenotemark{a} \\  
Abell 851            &  $r'$ & \multicolumn{1}{l}{27.0, 25.5, 24.9, ...}  & 2.61  & 25.2 &  60 & 1684 & 28.5\\  
RXJ1347.5-1145       &  $r'$ & \multicolumn{1}{l}{27.0, 25.5, 24.8, ...}  & 3.09, 2.51 & 25.0 & 100 &  952 & 27.1\tablenotemark{a} \\ 
RXJ2146.0+0423       &  $r'$ & \multicolumn{1}{l}{27.0, 25.5, 24.5, ...}  & 2.17  & 25.0 & 156 & 1223 & 28.8\\
MS0451.6--0305       &  $r'$ & 27.0, 25.5, 24.5, 24.1 & 2.44 & 25.4 &  95 & 1453 & 27.5\tablenotemark{a} \\
RXJ0216.5--1747      &  $i'$ & 27.0, 25.5, 24.5, 24.1 & 2.24 & 24.8 &  49 &  710 & 26.8\tablenotemark{a} \\
RXJ1334.3+5030       &  $i'$ & ... , 26.3, 25.3, 24.9 & 2.74, 2.04 & 25.1 &  50 &  967 & 27.1\tablenotemark{a} \\
RXJ1716.6+6708       &  $i'$ & ... , 26.5, 25.3, 24.6 & 2.14 & 24.9 & 104 & 1126 & 27.2\tablenotemark{a}\\
MS1610.4+6616 field  &  $i'$ & ... , 26.3, 25.3, 24.9 & 2.25 & 25.1 &  98 & 1180 & 27.8 \\
RXJ0152.7--1357      &  $i'$ & ... , 26.5, 25.3, 24.6 & 2.06 & 24.7 &  51 & 1732 & 28.6 \\
RXJ1226.9+3332       &  $i'$ & ... , 26.5, 25.3, 24.6 & 2.17 & 25.0 &  61 & 1056 & 27.8 \\
RXJ1415.1+3612       &  $z'$ & ... , 26.5, 25.4, 24.7 & 2.04 & 24.2 &  76 & 1132 & 26.8\tablenotemark{b}\\ 
\enddata
\tablecomments{Column 1: Galaxy cluster.  
Column 2: Filter used for detections. 
Column 3: Thresholds in $\rm mag\, arcsec^2$ in the order ($g'$, $r'$, $i'$, $z'$).
Column 4: Diameter of apertures in arcsec. For RXJ1347.5-1145 3.09 arcsec was use for $g'$ and 2.52 arcsec for $r'$ and $i'$, 
while for RXJ1334.3+5030 2.74 arcsec was used for $r'$ and $i'$ and 2.04 arcsec was used for $z'$, see text. 
Column 5: 5-sigma detection limit in magnitudes in the detection band, see Section \ref{SEC-NOISE}.
Column 6: Number of stars in catalog. 
Column 7: Number of galaxies in catalog. 
Column 8: Area observed in $\rm arcmin^2$. Abell 1689 and RXJ0056.2+2622 were both observed with two
slightly overlapping GMOS-N fields. The other clusters were covered with one field. 
Small variations final area are due to differences in dither patterns and vignetting from
the OIWFS as noted. 
}
\tablenotetext{a}{Areas affected by vignetting by the OIWFS are excluded from the total area.}
\tablenotetext{b}{Areas around bright foreground galaxies are excluded from the total area.}
\end{deluxetable*}

\subsection{Processing of Imaging Data}

The basic processing of the data was done in a standard fashion using the Gemini IRAF package.
We followed procedures similar to those described for RXJ0152.7--1357 in 
J\o rgensen et al.\ (2005), involving the following steps:
\begin{enumerate}
\item
Bias subtraction with master bias frame for the month of the observations.
\item
Flat fielding with normalized twilight flat created from 10-20 individual twilight flats.
\item
For $i'$- and $z'$-band, fringe correction with scaled fringe frames established
from the science data.
\item
For $g'$- and $r'$-band, as needed, scattered light correction with scaled 
scattered light images established from the science data.
\item
Mosaicing of the images from the three GMOS detectors into one image, using the
transformations available in the Gemini IRAF package task {\tt gmosaic}.
\item
Stacking of images taken in the same filter to obtain a co-added cosmic-ray cleaned image,
normalized to one of the exposures taken in photometric conditions.  This was done
using the Gemini IRAF package task {\tt imcoadd}. The stack made as the average 
of good pixels was used for all photometry.
\item 
Observations taken unbinned were rebinned to $2 \times 2$.
This applies to the observations taken during 2001 and to the RXJ0216.5--1747 $z'$-band observations.
The resulting pixel scale for all GMOS-N observations is 0.1454 ${\rm arcsec\,pixel^{-1}}$,
while the GMOS-S observations have a pixel scale of 0.146 ${\rm arcsec\,pixel^{-1}}$.
\item
The images were calibrated to astrometric consistency with the USNO catalog (Monet et al.\ 1998) by
means of simple offsets. Only linear calibrations were used. The rms scatter of the calibrations
is $\approx 0.7$ arcsec.
\end{enumerate}

We refer to the final stacked images as the ``co-added images''.
The original processing of the data as described in J\o rgensen et al.\ (2005) used
prototypes of the later released tasks for fringe correction of GMOS data.
Because the released tasks provide better object cleaning of the fringe 
correction frames than the prototypes, and
because the fringes in the $z'$-band are quite strong (5\% peak-to-peak for GMOS-N), 
we have reprocessed $z'$-band imaging from the raw data available in the 
Gemini Observatory Archive, using currently released tasks for the processing.
The $r'$-band observations of RXJ1415.1+3612 were also reprocessed in order to 
achive a better correction for the scattered light in these observations.

\subsection{Derived Photometric Parameters}

The co-added images were processed with SExtractor version 2.8.6 (Bertin \& Arnouts 1996).
We used SExtractor in dual-image mode, with the images pre-registered to
each other. The image in the filter closest to the rest frame $B$-band was used for detections, while
the images in the other filters were used only for photometry.
For consistency between clusters, the threshold for detection was defined as a surface brightness. 
The detection thresholds combined with the requirement of meeting the threshold
over a minimum of 9 pixels correspond to a signal-to-noise (S/N) of 8-10.
The analysis threshold in the other bands were then defined using the approximate 
expected colors of the cluster members on the red sequence.  
In all cases, we maintain analysis thresholds corresponding to S/N of 5-6 or better 
over the minimum detection area of 9 pixels.
Thus, for objects on
the red sequence roughly the same aperture size in each band is used to derive 
the geometrical parameters. The geometrical parameters are used by SExtractor
to derive the {\tt class\_star} parameter, which we use to separate galaxies and stars.
Table \ref{tab-photoverview} lists the filter used for detection and the adopted thresholds.

The SExtractor background mesh size was adjusted to avoid systematic effects from the
galaxies with the largest angular size. We typically use a background mesh size of 
256 pixels, with a filter size of 5 pixels.
We used 64 sub-thresholds and a minimum contrast for the deblending of only 0.0005 
(the default is 0.005). This enables deblending of objects in these fairly crowded fields.
For Abell 1689 and RXJ1334.3+5030 even lower minimum contrast for deblending of 0.00002
was needed to deblend fainter objects in the vicinity of either the brightest galaxies
in the cluster center (Abell 1689) or at close angular distance from bright stars 
(RXJ1334.3+5030). For Abell 1689 the lower deblending contrast is used only within
30 arcsec of the cluster core (for this purpose defined as the position of the galaxy with ID 626).
For RXJ1334.3+5030 the detections were done in
the $i'$-band, which has image quality of FWHM=0.87 arcsec (measured as the full-width-half maximum from a 
Gaussian fit to stars in the image). We used the better 
seeing $z'$-band image (FWHM=0.54 arcsec) to check that the deblending was correct.
In all cases, we use the SExtractor convolution file {\tt gauss\_2.0\_3x3.conv}. 
We visually inspected all fields to ensure that galaxies in our spectroscopic samples were correctly deblended.

SExtractor was run without a weight image. However, the catalogs were cleaned of
spurious detections along the edges of the field and along the edges of any
vignetting from the GMOS on-instrument-wave-front-sensor (OIWFS), when this is inside the field of view.
Table \ref{tab-photoverview} lists the effective area for object detection after 
such cleaning.

We adopt {\tt mag\_auto} from SExtractor as the total 
magnitudes of the objects, as these magnitudes are consistently derived based 
on apertures 2.5 times the Kron radii, $r_{\rm Kron}$ (Kron 1980). 
See Graham \& Driver (2005) for a discussion of the implementation of the Kron radius
in SExtractor.
In some cases of close neighbors, the magnitudes may be affected by these.
We also provide the isophotal magnitudes, {\tt mag\_iso}, in the photometry table.
In Section \ref{SEC-APERTURE} we discuss to what extent {\tt mag\_auto} is 
different from true total magnitudes and we derive aperture corrections for point sources.

Differences in image quality between the observations in the different passbands can complicate the 
determination of colors of the galaxies.
Various techniques to address this issue have been used in the past. One approach
is to used fixed size apertures, but to convolve all images of a given field to the a common (worst) 
resolution, or a less drastic approach of convolving the images only in pairs as described by
Meyers et al.\ (2012) and used by, e.g., Cerulo et al.\ (2016).
Alternatively, one may obtain ``global'' colors of the galaxies, using {\tt mag\_auto} as the  basis for the colors. 
We note that {\tt mag\_auto} are also the only choice of the SExtractor ``total'' magnitudes that use the same 
size aperture for all frames.

Instead of convolving the images, we take the approach of measuring aperture colors, using 
aperture sizes chosen to minimize the effect of image quality differences on the 
measured colors. 
To decide on the aperture sizes, we first 
estimate aperture diameters in arcsec using both the image quality, FWHM,
and the half light radii of typical small galaxies in the clusters. Specifically the
aperture diameter in arcsec is chosen as
\begin{equation}
\label{eq-daper}
D_{\rm app} = 2 \cdot 2.355 \left ( ({\rm FWHM}/2.355)^2 + r_{\rm galaxy}^2 \right )^{0.5}
\end{equation}
where $r_{\rm galaxy}$ is the half light radius in arcsec at the redshift of the cluster, 
corresponding to a physical size of 2.5 kpc. 
For those clusters with differences between $D_{\rm app}$ in the different
passbands of less than 10 percent, we then used the largest of those diameters as the aperture 
size for all the passbands. 
For RXJ1347.5--1145 and RXJ1334.3+5030 the differences 
between $D_{\rm app}$ for the different passbands were larger than 10 percent. 
Thus, for these clusters we used two different aperture sizes.
Aperture sizes are listed in Table \ref{tab-photoverview}.
In the catalog table (see Section \ref{SEC-FINALPHOT}) we give the aperture magnitudes.
For reference, we also provide aperture magnitudes within an aperture with diameter 2.5 arcsec.

Other adjustments to the SExtractor parameters were trivial adjustments
for the magnitude zero points and the image quality.

\begin{figure}
\epsfxsize 8.0cm
\epsfbox{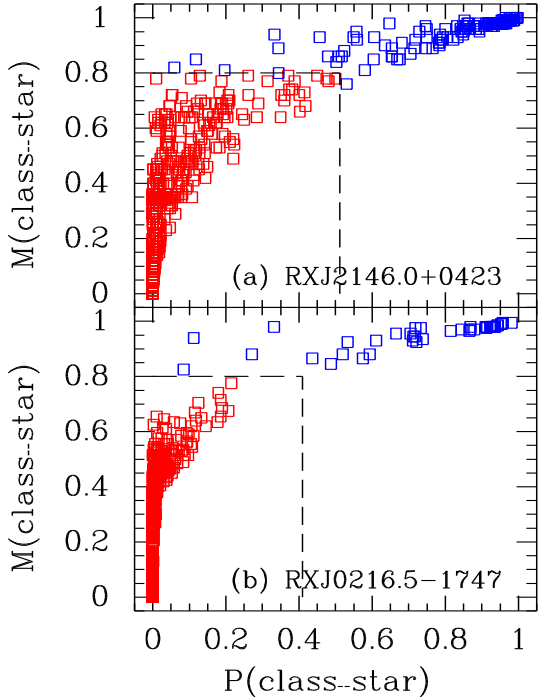}
\caption{
P(class\_star) versus M(class\_star) for two of the clusters. Red -- objects classified
as galaxies; blue - objects classified as stars. The dashed lines mark the cut off
in P(class\_star) and M(class\_star)  for the two clusters. 
RXJ2146.0+0423 and RXJ0216.5--1747 have photometry in three and four filters, respectively,
leading to cut off values in P(class\_star) of $0.8^3=0.51$ and $0.8^4=0.41$, respectively.
\label{fig-class_star} }
\end{figure}

\begin{figure*}
\epsfxsize 17.0cm
\epsfbox{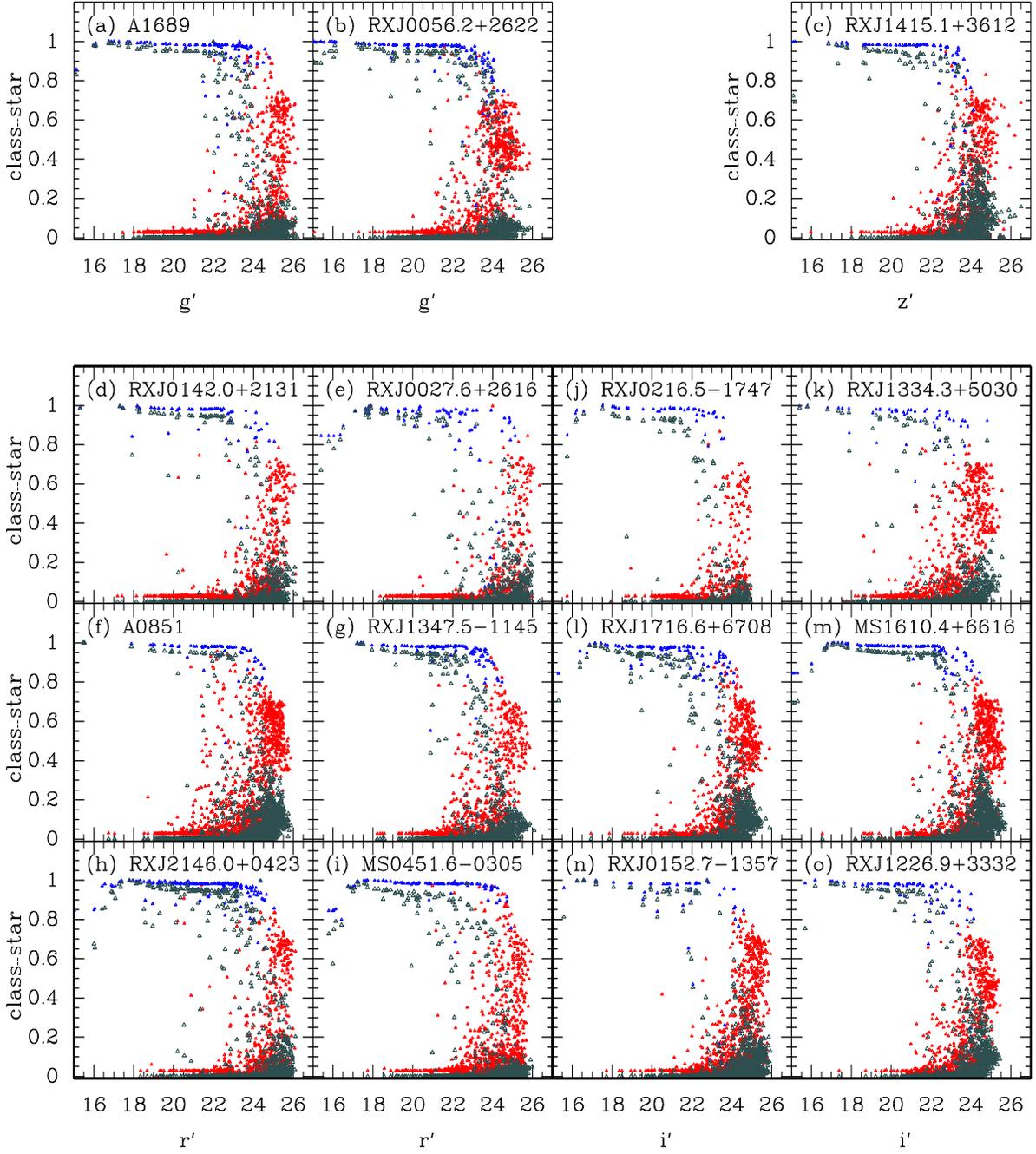}
\caption{
Total magnitude {\tt mag\_auto} in the detection band versus {\tt class\_star}.  
Grey -- product $P({\tt class\_star})$ of {\tt class\_star} from all available filters, see equation (\ref{eq-pstar}).
Red -- {\tt class\_star}  in the detection bands for those objects classified as galaxies; 
blue -- {\tt class\_star}  in the detection bands for those objects classified as stars.
\label{fig-mag_class} }
\end{figure*}

Stars and galaxies were separated based on the SExtractor classification
parameters {\tt class\_star} for all available bands. 
We derive the product and the median of those available, and define
\begin{equation}
\label{eq-pstar}
P({\tt class\_star}) \equiv \prod {\tt class\_star}
\end{equation}
\begin{equation}
M({\tt class\_star}) \equiv {\rm median}({\tt class\_star})
\end{equation}
Objects are classified as stars if they meet the criterium
\begin{equation}
P({\tt class\_star}) \ge 0.8^N \, || \, M({\tt class\_star}) \ge 0.8
\end{equation}
while all other objects were considered galaxies. 
$N$ is the number of bands available for a given field. 
The classification of saturated stars was set manually.
Figure \ref{fig-class_star} shows $M$({\tt class\_star}) versus $P$({\tt class\_star}) for two
of the clusters, RXJ2146.0+0423 with photometry in three bands
and RXJ0216.5--1747 with photometry in four bands.
The image quality for the observations of these two clusters is comparable, 0.50-0.65 arcsec.
The parameters $P({\tt class\_star})$ and $M({\tt class\_star})$ are included in the table of the photometric data 
(see Section \ref{SEC-FINALPHOT}), allowing users to reclassify the objects based on the available data.
Figure \ref{fig-mag_class} shows $P$({\tt class\_star}) and {\tt class\_star} in the detection
band versus magnitudes,
illustrating how the use of $P$({\tt class\_star}) aids in the 
classification of especially faint objects within $\approx 2$ mag of the detection limit.
In our original sample selection of targets for spectroscopic observations we 
required {\tt class\_star}$<$0.80 in the detection band. In a few cases, our refined
classification would have excluded a spectroscopic sample target from the sample. 
In all such cases, except the Seyfert galaxy RXJ1415.1+3612 ID 983, 
the spectra confirm that the objects are indeed stars.
In Table \ref{tab-photoverview}, we list the number of objects classified as galaxies 
and as stars in each cluster field.

While our ground-based imaging in general is not of sufficient spatial resolution to
warrant detailed 2-dimensional photometry, we do provide measures of sizes as an
isophotal radius, $r_{\rm iso}$, as well as position angles and ellipticities, which may be useful for 
sample selections for other follow up studies of the clusters. 
We use the isophotal area {\tt iso\_area\_image} determined by SExtractor in pixels to 
derive a circularized isophotal radius in arcseconds as
\begin{equation}
\label{eq-riso}
r_{\rm iso} = \left ( {\tt iso\_area\_image} / \pi \right ) ^{0.5} pixelscale
\end{equation}
where $pixelscale$ is the pixel scale for the image in $\rm arcsec\,pixel^{-1}$.
The surface brightnesses used for determinations are listed in Table \ref{tab-photoverview}.

\begin{deluxetable*}{l rrrr rrrr rrrr rrrr}
\tablecaption{Noise Parameters \label{tab-noise} }
\tabletypesize{\scriptsize}
\tablewidth{0pt}
%\tablenum{8}
\tablehead{
\colhead{Cluster} & \multicolumn{4}{c}{$g'$-band} & \multicolumn{4}{c}{$r'$-band} & \multicolumn{4}{c}{$i'$-band} & \multicolumn{4}{c}{$z'$-band}  \\
 & \colhead{$a$} & \colhead{$b$} & \colhead{$\sigma _{\rm sky}$} & \colhead{Factor} 
 & \colhead{$a$} & \colhead{$b$} & \colhead{$\sigma _{\rm sky}$} & \colhead{Factor} 
 & \colhead{$a$} & \colhead{$b$} & \colhead{$\sigma _{\rm sky}$} & \colhead{Factor} 
 & \colhead{$a$} & \colhead{$b$} & \colhead{$\sigma _{\rm sky}$} & \colhead{Factor} \\
\colhead{(1)} & \colhead{(2)} & \colhead{(3)} & \colhead{(4)} & \colhead{(5)} & \colhead{(6)} & \colhead{(7)} & \colhead{(8)} & \colhead{(9)} & 
                \colhead{(10)} & \colhead{(11)} & \colhead{(12)} & \colhead{(13)} & \colhead{(14)} & \colhead{(15)} & \colhead{(16)} &  \colhead{(17)}
}
\startdata
A1689  F1         & 1.42 & 0.196 & 0.048 & 4.41 & 0.85 & 0.280 & 0.071 & 5.13 & 0.78 & 0.163 & 0.210 & 3.26 & \nodata & \nodata & \nodata & \nodata \\
A1689  F2         & 1.25 & 0.223 & 0.047 & 4.65 & 0.62 & 0.259 & 0.091 & 4.56 & 0.87 & 0.111 & 0.213 & 2.56 & \nodata & \nodata & \nodata & \nodata \\
RXJ0056p2p2622 F1 & 0.96 & 0.075 & 0.084 & 2.09 & 1.18 & 0.148 & 0.100 & 3.44 & 1.61 & 0.057 & 0.136 & 2.48 & \nodata & \nodata & \nodata & \nodata \\
RXJ0056p2p2622 F2 & 1.15 & 0.045 & 0.080 & 1.83 & 1.21 & 0.140 & 0.095 & 3.34 & 1.39 & 0.162 & 0.115 & 3.85 & \nodata & \nodata & \nodata & \nodata \\
RXJ0142p0p2131    & 1.23 & 0.162 & 0.028 & 3.70 & 0.86 & 0.169 & 0.087 & 3.44 & 1.37 & 0.081 & 0.083 & 2.61 & \nodata & \nodata & \nodata & \nodata \\
RXJ0027p6p2616    & 1.24 & 0.168 & 0.029 & 3.80 & 0.79 & 0.121 & 0.068 & 2.64 & 1.12 & 0.341 & 0.074 & 6.32 & \nodata & \nodata & \nodata & \nodata \\
A0851             & 1.16 & 0.146 & 0.026 & 3.39 & 1.03 & 0.094 & 0.085 & 2.47 & 1.57 & 0.088 & 0.077 & 2.90 & \nodata & \nodata & \nodata & \nodata \\
RXJ1347p5m1145    & 1.24 & 0.249 & 0.038 & 5.04 & 0.97 & 0.130 & 0.117 & 2.96 & 1.27 & 0.193 & 0.098 & 4.21 & \nodata & \nodata & \nodata & \nodata \\
RXJ2146p0p0423    & 1.38 & 0.166 & 0.020 & 3.91 & 0.70 & 0.190 & 0.070 & 3.59 & 1.33 & 0.099 & 0.063 & 2.84 & \nodata & \nodata & \nodata & \nodata \\
MS0451p6m0305     & 1.30 & 0.261 & 0.017 & 5.28 & 0.84 & 0.155 & 0.047 & 3.20 & 1.50 & 0.213 & 0.061 & 4.75 & 1.52 & 0.131 & 0.033 & 3.52 \\
RXJ0216p5m1747    & 1.78 & 0.087 & 0.030 & 3.11 & 1.65 & 0.041 & 0.065 & 2.28 & 1.26 & 0.048 & 0.115 & 2.00 & 1.70 & 0.042 & 0.059 & 2.34 \\ 
RXJ1334p3p5030    & \nodata & \nodata & \nodata & \nodata & 1.28 & 0.201 & 0.030 & 4.33 & 1.14 & 0.063 & 0.072 & 2.09 & 1.50 & 0.053 & 0.045 & 2.30 \\
RXJ1716p6p6708    & \nodata & \nodata & \nodata & \nodata & 0.94 & 0.259 & 0.044 & 4.89 & 0.97 & 0.097 & 0.094 & 2.44 & 1.47 & 0.089 & 0.068 & 2.82 \\
MS1610p4p6616     & \nodata & \nodata & \nodata & \nodata & 1.88 & 0.064 & 0.041 & 2.86 & 1.16 & 0.046 & 0.101 & 1.87 & 1.97 & 0.032 & 0.055 & 2.45 \\
RXJ0152p7m1357    & \nodata & \nodata & \nodata & \nodata & 1.71 & 0.081 & 0.035 & 2.94 & 2.12 & 0.090 & 0.065 & 3.49 & 1.61 & 0.071 & 0.026 & 2.69 \\
RXJ1226p9p3332    & \nodata & \nodata & \nodata & \nodata & 1.69 & 0.094 & 0.026 & 3.12 & 0.78 & 0.198 & 0.054 & 3.80 & 1.44 & 0.066 & 0.126 & 2.44 \\
RXJ1415p1p3612    & \nodata & \nodata & \nodata & \nodata & 1.09 & 0.282 & 0.033 & 5.40 & 1.08 & 0.367 & 0.056 & 6.67 & 1.51 & 0.077 & 0.051 & 2.69 \\
\enddata
\tablecomments{Column 1: Galaxy cluster and field.  
Column 2: $g'$-band noise correction coefficient $a$. Typical uncertainties are 0.11.
Column 3: $g'$-band noise correction coefficient $b$. Typical uncertainties are 0.009.
Column 4: $g'$-band sky noise per pixel, $\sigma _{\rm sky}$, normalized to 1 second.
Column 5: $g'$-band noise correction factor $a_i + b_i  A^{1/2}$ for an aperture with a diameter of 2.5 arcsec.
Columns 6--9: Same information for the $r'$-band.
Columns 10--13: Same information for the $i'$-band.
Columns 14--17: Same information for the $z'$-band.
}
\end{deluxetable*}

\subsection{Uncertainties on Magnitudes \label{SEC-NOISE} }

The uncertainties on the magnitudes estimated by SExtractor are based on the sky noise per pixel,
$\sigma _{\rm sky}$, combined with the Poisson noise of the signal from the objects.
It is assumed that the noise from the sky scales with the area, $A$, of the aperture in pixels such that 
the total uncertainty on the flux, $F$, measured from an object can be expressed as
\begin{equation}
\sigma _{\rm object} = \left ( A \sigma _{\rm sky}^2 + F gain^{-1} \right ) ^{1/2}
\end{equation}
where $F$ is in counts and $gain$ is the gain of the image in $\rm e^-$/counts.
Several studies have shown that these uncertainties in general are underestimated. 
In particular, Labb\'{e} et al.\ (2003) find that even for {\it HST} imaging and small apertures 
the effect can be a factor two. 
Due to imperfect corrections for scattered light and/or fringing ground-based imaging 
often has stronger large scale variations of the background, than typically found in {\it HST} imaging.
We adopted a combination of the method used by Labb\'{e} et al.\ and that by Guo et al.\ (2013)
to determine the correction factor for the noise estimates, given the sizes of the apertures.

For each field and filter combination, we mask out pixels containing signal from objects. 
We then place empty apertures with areas from 16 to 400 pixels, equivalent to 
aperture diameters of 0.65--3.3 arcsec. For the largest size apertures, we typically place 
200-300 empty apertures across the masked images, while for the smaller sizes 600-800 empty apertures
were used.
The background was subtracted using the sky image produced by SExtractor. 
We then measure the flux in each of the empty apertures.
For each size aperture we determine the scatter of the fluxes from a Gaussian fit to their distribution.
As also found by Labb\'{e} et al., the scatter for given field and filter can be parameterized as
\begin{equation}
\sigma_i (A) = A^{1/2} \sigma _{\rm sky} ( a_i + b_i  A^{1/2} )
\end{equation}
We determine the coefficients $a_i$ and $b_i$ from least squares fits.
Table \ref{tab-noise} summarizes the determinations, the sky noise per pixel $\sigma _{\rm sky}$ 
normalized to one second, and the resulting correction factor $a_i + b_i  A^{1/2}$ for an aperture with a diameter of 2.5 arcsec.
The correction factors are between 1.8 and 6.7, with a median value of 3.2.
The median of the coefficients $a_i$ is 1.25, reflecting the typical correlation of noise between pixels due to
the stacking of multiple frames, cf.\ Labb\'{e} et al. The coefficients $b_i$, reflecting the typical
large-scale variations in the background, have a median value of 0.13.

\begin{figure}
\epsfxsize 7.0cm
\epsfbox{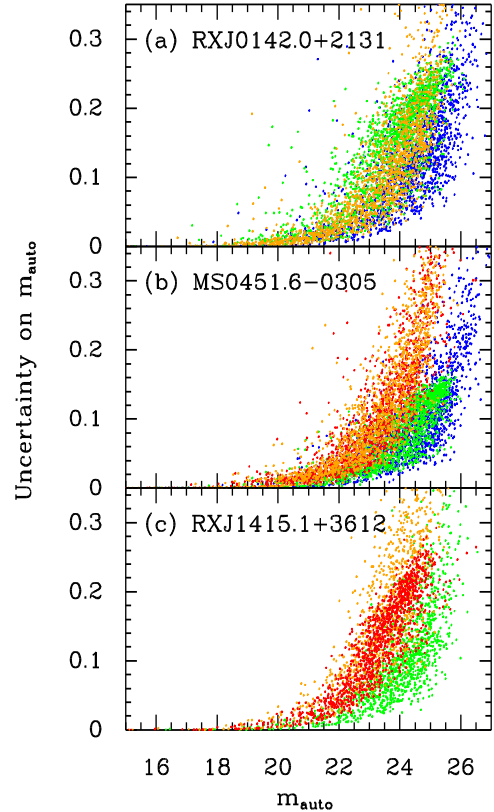}
\caption{
The uncertainties on the magnitudes  {\tt mag\_auto} versus the magnitude {\tt mag\_auto} 
in the detection band for three of the clusters, RXJ0142.0+2131 at $z=0.28$,
MS0451.6--0305 at $z=0.54$ and RXJ1415.1+3612 at $z=1.03$. 
The figure illustrates the depth of the data at different redshifts.
All objects detected in the fields are included. 
The points are color coded by filter: Blue -- $g'$, green -- $r'$, orange -- $i'$, and
red -- $z'$. 
\label{fig-mag_uncertainty} }
\end{figure}

\begin{figure*}
\epsfxsize 17.0cm
\epsfbox{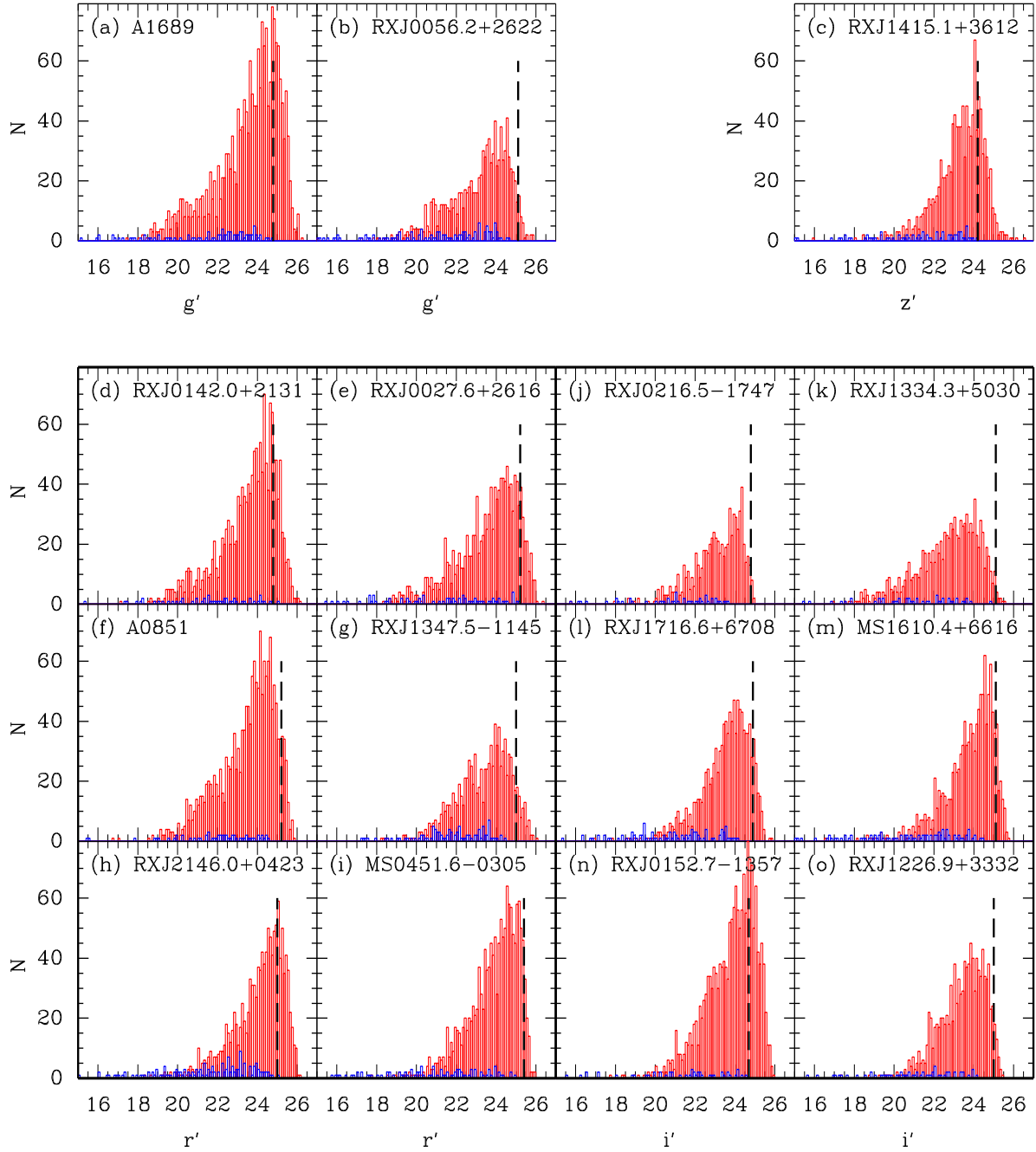}
\caption{
Distribution of {\tt mag\_auto} for stars (blue) and galaxies (red) in the fields. 
The vertical dashed lines mark the 5-sigma detection limit, see Table \ref{tab-photoverview}. 
\label{fig-mag_hist} }
\end{figure*}

The uncertainties of all magnitudes were then derived using the coefficients and the relevant sizes of the apertures.
Figure \ref{fig-mag_uncertainty} shows uncertainties on {\tt mag\_auto} as a function of {\tt mag\_auto}
for three of the clusters spanning the redshift range of the sample and
illustrating the typical depth of the data as a function of redshift and passband.
In Figure \ref{fig-mag_hist}, we show the magnitude distribution of the 
objects for each field in the detection band. 
The 5-sigma detection limits are listed in Table \ref{tab-photoverview} and marked on the figure. 

\begin{deluxetable*}{l rr rr rr rr}
\tablecaption{Aperture Corrections for Point Sources \label{tab-aperture} }
\tabletypesize{\scriptsize}
\tablewidth{0pt}
%\tablenum{8}
\tablehead{
\colhead{Cluster} & \multicolumn{2}{c}{$g'$-band} & \multicolumn{2}{c}{$r'$-band} & \multicolumn{2}{c}{$i'$-band} & \multicolumn{2}{c}{$z'$-band}  \\
 & \colhead{$\Delta m_{\rm aper}$} & \colhead{$\Delta m_{2.5}$} 
 & \colhead{$\Delta m_{\rm aper}$} & \colhead{$\Delta m_{2.5}$} 
 & \colhead{$\Delta m_{\rm aper}$} & \colhead{$\Delta m_{2.5}$} 
 & \colhead{$\Delta m_{\rm aper}$} & \colhead{$\Delta m_{2.5}$} \\
\colhead{(1)} & \colhead{(2)} & \colhead{(3)} & \colhead{(4)} & \colhead{(5)} & \colhead{(6)} & \colhead{(7)} & \colhead{(8)} & \colhead{(9)} 
}
\startdata
Abell 1689 F1     & -0.024 & -0.060 & -0.039 & -0.166 & -0.044 & -0.163 & \nodata & \nodata \\
Abell 1689 F2     & -0.028 & -0.060 & -0.042 & -0.153 & -0.031 & -0.106 & \nodata & \nodata \\
RXJ0056.2+2622 F1 & -0.043 & -0.131 & -0.040 & -0.111 & -0.036 & -0.096 & \nodata & \nodata \\
RXJ0056.2+2622 F2 & -0.061 & -0.207 & -0.050 & -0.135 & -0.101 & -0.296 & \nodata & \nodata \\
RXJ0142.0+2131    & -0.055 & -0.064 & -0.041 & -0.052 & -0.038 & -0.051 & \nodata & \nodata \\
RXJ0027.6+2616    & -0.091 & -0.099 & -0.074 & -0.079 & -0.051 & -0.052 & \nodata & \nodata \\
Abell 851         & -0.088 & -0.097 & -0.064 & -0.067 & -0.064 & -0.071 & \nodata & \nodata \\
RXJ1347.5--1145   & -0.178 & -0.280 & -0.117 & -0.121 & -0.123 & -0.124 & \nodata & \nodata \\
RXJ2146.0+0423    & -0.094 & -0.076 & -0.075 & -0.057 & -0.071 & -0.054 & \nodata & \nodata \\
MS0451.6--0305    & -0.142 & -0.135 & -0.067 & -0.064 & -0.138 & -0.131 & -0.134 & -0.129 \\
RXJ0216.5--1747   & -0.114 & -0.096 & -0.094 & -0.079 & -0.076 & -0.062 & -0.085 & -0.075 \\
RXJ1334.3+5030    & \nodata & \nodata & -0.215 & -0.258 & -0.133 & -0.174 & -0.098 & -0.079 \\
RXJ1716.6+6708    & \nodata & \nodata & -0.158 & -0.116 & -0.173 & -0.130 & -0.174 & -0.132 \\
MS1610.4+6616     & \nodata & \nodata & -0.112 & -0.094 & -0.113 & -0.094 & -0.139 & -0.118 \\
RXJ0152.7--1357   & \nodata & \nodata & -0.093 & -0.071 & -0.112 & -0.089 & -0.108 & -0.090 \\
RXJ1226.9+3332    & \nodata & \nodata & -0.148 & -0.116 & -0.179 & -0.132 & -0.131 & -0.104 \\
RXJ1415.1+3612    & \nodata & \nodata & -0.107 & -0.080 & -0.200 & -0.139 & -0.120 & -0.089 \\
\enddata
\tablecomments{Column 1: Galaxy cluster and field.  
Column 2: $g'$-band aperture correction for adaptive aperture size. Typical uncertainties are 0.015 mag.
Column 3: $g'$-band aperture correction for aperture diameter of 2.5 arcsec. Typical uncertainties are 0.015 mag.
Columns 4--5: Same information for the $r'$-band.
Columns 6--7: Same information for the $i'$-band.
Columns 8--9: Same information for the $z'$-band.
}
\end{deluxetable*}

\subsection{SExtractor Magnitudes versus Total Magnitudes \label{SEC-APERTURE} }

The SExtractor magnitudes {\tt mag\_auto} are known to miss a small fraction of the total flux from
objects. Bertin \& Arnouts (1996) determined from simulations the loss to be 0.03-0.06 magnitudes.
Theoretical work by Graham \& Driver (2005) shows that the fraction lost for galaxies
depends on their luminosity profile. For galaxies with S\'{e}rsic (1968) profiles, the fraction lost is
$\approx 4$\% for an exponential profile increasing to $\approx 10$\% for an $r^{1/4}$ profile.
However, Graham \& Driver also point out that if the Kron radius (used as the basis for {\tt mag\_auto})
is derived from integration over only 1-2 effective radii of the galaxies, the lost flux can
be substantially larger.

\begin{figure}
\epsfxsize 8.0cm
\epsfbox{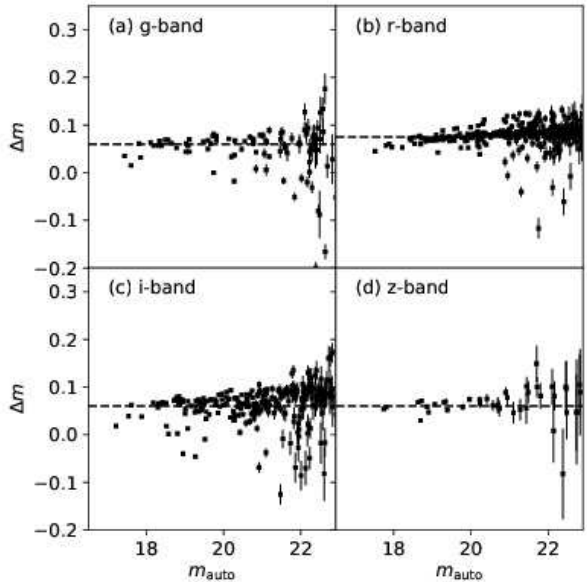}
\caption{
The flux missed by {\tt mag\_auto}, $\Delta m$, as a function of  {\tt mag\_auto} for stars observed in
the four filters.
The dashed lines mark the median missed flux, 0.06 mag in $g'$, $i'$, and $z'$ and 
0.075 mag in $r'$.
\label{fig-magmissed} }
\end{figure}

We first derived aperture corrections for the point sources in a standard fashion, using magnitudes
derived through large apertures for bright isolated stars. The resulting aperture corrections,
$\Delta m_{\rm aper}$ for the adaptive aperture sizes defined in Equation \ref{eq-daper} as well as for the fixed
aperture size of 2.5 arcsec diameter are listed in Table \ref{tab-aperture}.
As expected, the aperture correction
for the fixed aperture size is strongly correlated with the image quality, increasing from $\approx 0.05$ mag for the
best seeing images to $\approx 0.15$ mag at a seeing of 0.8 arcsec. 
The flux missed from {\tt mag\_auto} can then be derived as
\begin{equation}
\Delta m = {\tt mag\_auto} - ( {\tt mag\_aper} + \Delta m_{\rm aper} )
\end{equation}
Figure \ref{fig-magmissed} shows $\Delta m$ versus {\tt mag\_auto} for the unsaturated stars in the detection bands.
In median the missed flux is 0.06 mag in $g'$, $i'$, and $z'$. The missing flux for the $r'$-band
is slightly higher at 0.075 mag, presumably due to differences in the point-spread-functions.

To assess the fraction of flux lost from {\tt mag\_auto} of the galaxies in the observed fields, 
we performed detailed simulations matching five of the 15 clusters, spanning the relevant parameter space 
in redshifts, image quality, and filters. The simulations were created using python software by Peterson et al.\ (2018), 
produced during Peterson's internship with the GCP. The simulation software calls the galaxy fitting program 
GALFIT (Peng et al.\ 2002) to make the model galaxies.

For each cluster we made simulated images matching the 3 (or 4) available filters.
Each simulated image contains 500-1000 model galaxies. 
The model galaxies have distributions in total magnitudes matching the {\tt mag\_auto} distributions of
the real data, and 2-dimensional distributions in effective radii and total magnitudes, which once convolved 
with the point-spread-function (PSF) match the 2-dimensional distributions in ({\tt mag\_auto}, {\tt flux\_radius}) of the real data.
The galaxies were assumed to have S\'{e}rsic profiles. 
For each galaxy, values of the S\'{e}rsic indices, $n_{\rm ser}$, ellipticities and position angles
were chosen randomly from uniform distributions. 
We assumed $n_{\rm ser}$ between 0.5 and 5, ellipticities between zero and 0.7, and position angles between 
--90 and +90 degrees.
We then used GALFIT to create noiseless model images of each galaxy.
Each model galaxy was convolved with the empirical PSF of the real data. 
The PSFs were established from 10-15 isolated stars in the fields.
The model galaxies were randomly placed into empty images of the same size as our GMOS images
and with a background level matching the real data. 
Thus, these images have similar crowding of the objects as the real observations, 
except for the very center of the clusters.
Finally, noise was added taking into account read-out-noise and Poisson noise. We did not attempt to model
the correlated noise due to the image stacking, or the contribution from the non-flat sky background in the 
real data. We do not expect these effects to contribute significantly to the fraction of lost flux.

For each cluster we used the same seed for generation of the random samples for each of the 3 (or 4) filters. 
Thus, a given model galaxy will have identical $n_{\rm ser}$, ellipticity 
and position angle in the 3 (or 4) filters. The color of the model galaxy will represent the average color of
galaxies in the field at the given magnitude.
SExtractor was run in dual-image mode on the simulations, with all parameters set identically to
those used for the real data.
The simulations use the adaptive aperture sizes, Equation (\ref{eq-daper}), for aperture magnitudes and colors.

\begin{figure*}
\epsfxsize 17.0cm
\epsfbox{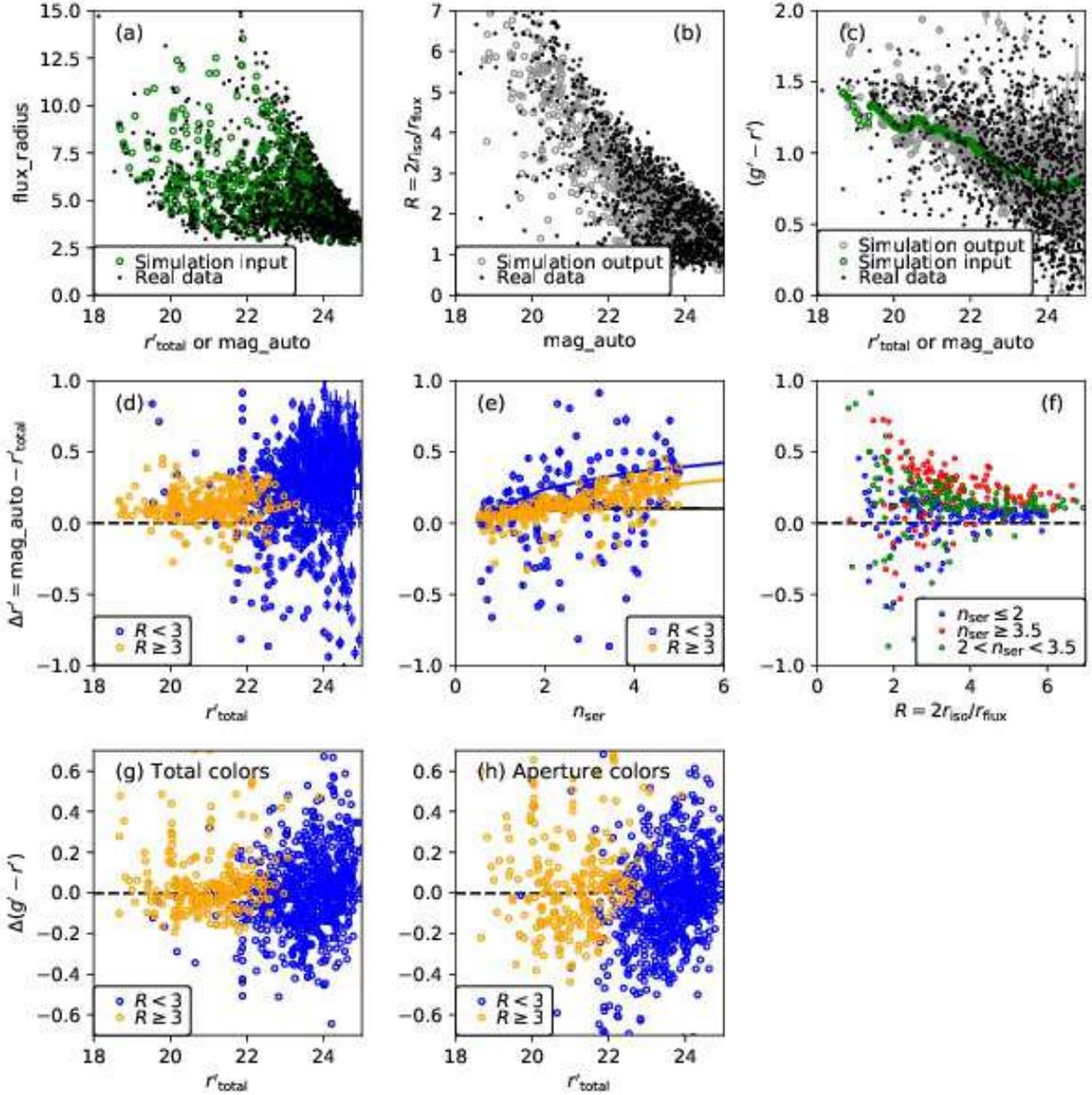}
\caption{
Galaxy simulations matching the observations of RXJ0142.0+2131 ($z=0.28$).
Panels (a)--(c) show the simulated data compared to the real data to illustrate that the simulations
span the relevant parameter space. Green open circles -- simulation input parameters; grey open cirles --
simulation output parameters; black points -- real data.
Panels (d)--(f) show the lost flux in magnitudes, $\Delta m = {\tt mag\_auto} - r'_{\rm total}$, as a 
function of the input magnitudes $r'_{\rm total}$, the input S\'{e}rsic indices $n_{\rm ser}$ and the 
output ratio $R= 2 r_{\rm iso} r_{\rm flux}^{-1}$ , see text for discussion.
In panels (d) and (e), yellow points are galaxies with $R\ge 3$ and blue points are galaxies with 
$R<3$. Panel (e) shows the expected values of $\Delta m$ from Graham \& Driver (2005) 
for $R=2$ (blue), $R=4$ (yellow), and $R=\infty$ (black).
In panel (f) the points are color coded for $n_{\rm ser}$. Blue points -- $n_{\rm ser}\le 2$;
green points -- $2 < n_{\rm ser} < 3.5$; and red points $n_{\rm ser}\ge 3.5$.
Panels (e) and (f) are limited to model galaxies with $r'_{\rm total}\le 23$.
Panels (g) and (h) show the effect on the colors $(g'-r')$ based on {\tt mag\_auto} and 
aperture magnitudes (using aperture sizes as defined in Eq. (\ref{eq-daper})). Symbols as in panels (d) and (e).
\label{fig-rxj0142sim} }
\end{figure*}

\begin{figure*}
\epsfxsize 17.0cm
\epsfbox{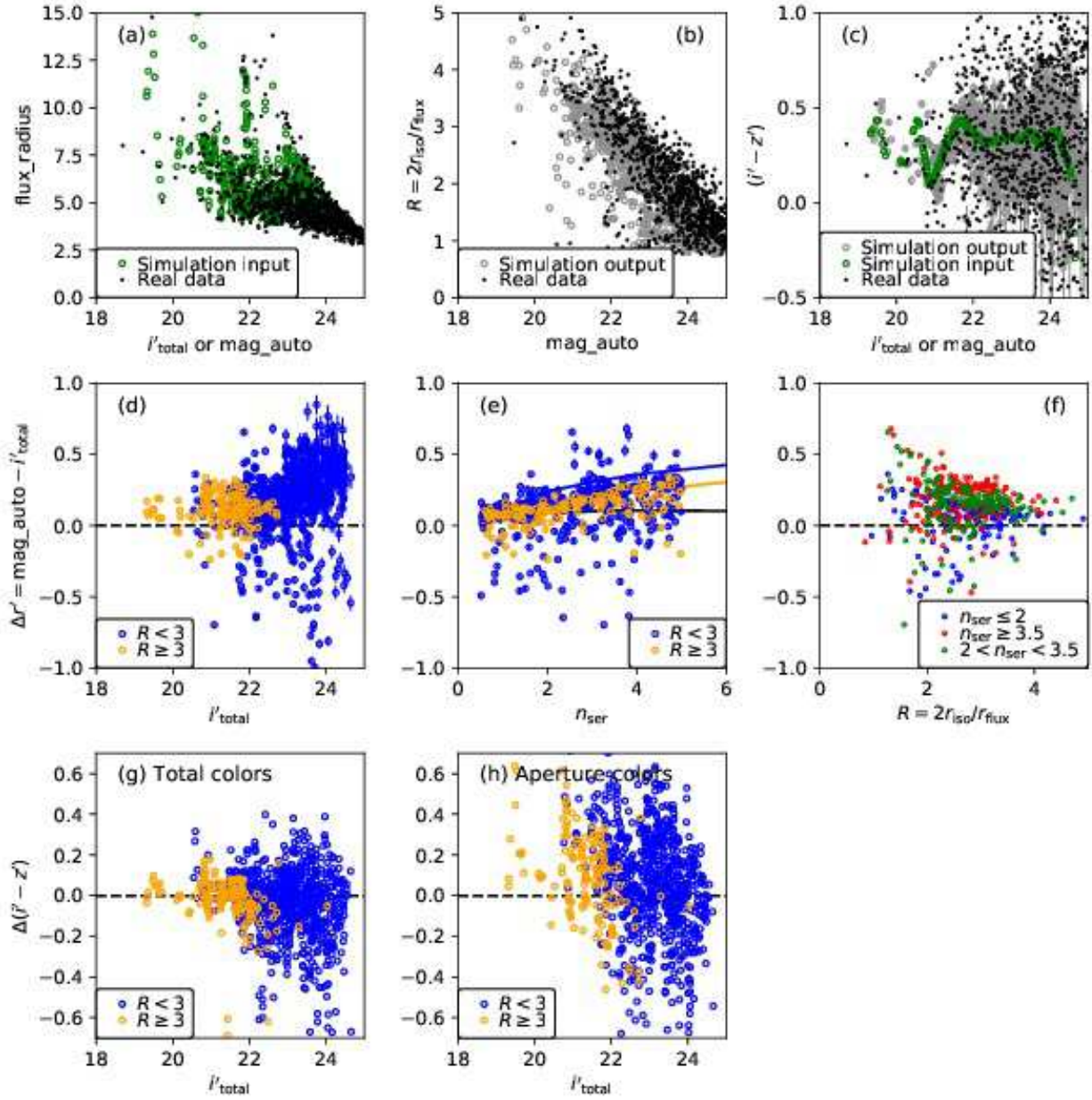}
\caption{
Galaxy simulations matching the observations of RXJ1226.9+3332 ($z=0.89$).
The layout and symbols are as in Figure \ref{fig-rxj0142sim}. For this cluster we show the $(i'-z')$ colors.
\label{fig-rxj1226sim} }
\end{figure*}

Figures \ref{fig-rxj0142sim} and \ref{fig-rxj1226sim} summarize the results from
the simulations matching RXJ0142.0+2131 ($z=0.28$)
and RXJ1226.9+3332 ($z=0.89$), serving as representative for the relevant parameter space.
Panels (a)--(c) on the figures show how the simulated data match the real data in magnitudes, sizes and 
colors. In particular, panels (b) show the ratio $R= 2 r_{\rm iso} r_{\rm flux}^{-1}$ between the 
aperture size within which the Kron radius is determined by SExtractor ($2 r_{\rm iso}$) and the 
effective radius here approximated with $r_{\rm flux}$ from SExtactor.

Panels (d)--(f) on the figures show the lost flux, $\Delta m$, as difference between the SExtractor 
{\tt mag\_auto} and the input total magnitude. 
Fainter galaxies have larger $\Delta m$. However, the main drivers for the difference are
the ratio $R$, which depends on the magnitudes of the galaxies (panels b), and the assumed S\'{e}rsic index. 
In panels (e) we show for galaxies brighter than 23 mag $\Delta m$ as a function of $n_{\rm ser}$. The points
are color-coded for $R$ larger (orange) or smaller (blue) than 3. 
The simulations follow the expected dependency established by Graham \& Driver, and shown on the figure. 
To further illustrate the dependency on both $R$ and $n_{\rm ser}$, panels (f) show the effect 
as $\Delta m$ versus the ratio $R$, color-coded for $n_{\rm ser}$. 
Based on our simulations for five clusters, we conclude that there is no significant
differences in $\Delta m$ due to differences in filters, image quality, or redshift of the clusters. 
In summary, the median $\Delta m$ for galaxies brighter than 23 mag and with $R\ge 3$ is 0.06, 0.13, and 0.21
for $n_{\rm ser} \le 2$, $2-3.5$ and $\ge 3.5$, respectively.
The galaxies included in our spectroscopic samples and our investigation of the red sequence (Section \ref{SEC-CM})
typically have $R\ge 3$.
In practice, we do not know  $n_{\rm ser}$ from our ground-based data. However, we expect that galaxies
on the red sequence have  $n_{\rm ser}\ge 2$ and therefore will have $\Delta m \approx 0.1-0.2$ mag.

In panels (g)--(h) we show the simulation results for the main color for the two clusters. In both cases colors
based on {\tt mag\_auto} reproduce the input colors better than colors based on aperture magnitudes, even when 
aperture sizes are chosen to match the seeing, cf.\ Equation (\ref{eq-daper}). 
However, when evaluating which to use for investigation of colors of the real galaxies, it should also be 
kept in mind that the aperture magnitudes usually have lower uncertainties due background noise. 
The simulations assumed no internal color gradients in the galaxies. Color gradients in the real galaxies
will of course cause differences between aperture colors and colors using {\tt mag\_auto}.
In our discussion of the color-magnitude relation for the clusters, Section \ref{SEC-CM}, we show both colors.

\begin{deluxetable*}{llrrrrr}
\tablecaption{GMOS-N Color Terms: Linear Relations from J\o rgensen (2009)\label{tab-color} }
\tabletypesize{\scriptsize}
\tablewidth{0pc}
%\tablenum{8}
\tablehead{
\colhead{No.} & \colhead{$\Delta {\rm m}$} & \colhead{rms} & \colhead{Color term fit} & \colhead{rms(fit)} & \colhead{N} & Color interval \\
\colhead{(1)} & \colhead{(2)} & \colhead{(3)} & \colhead{(4)} & \colhead{(5)} & \colhead{(6)} & \colhead{(7)}
}
\startdata
2  & $\Delta g_{\rm zero}$ & 0.044\tablenotemark{a} & $(0.066\pm 0.002) (g'-r') - (0.037\pm 0.002)$ & 0.034 & 794 & $-0.55\le (g'-r')\le 2.05$ \\
13 & $\Delta r_{\rm zero}$ & 0.045\tablenotemark{b} & $(0.027\pm 0.003) (g'-r') - (0.016\pm 0.002)$ & 0.043 & 1084 & $-0.55\le (g'-r')\le 2.05$\\
16 &                       &                        & $(0.042\pm 0.004) (r'-i') - (0.011\pm 0.002)$ & 0.043 & 1084 & $-0.35\le (r'-i')\le 2.4$ \\
22 & $\Delta i_{\rm zero}$ & 0.054\tablenotemark{b} & $(0.063\pm 0.005) (r'-i') - (0.013\pm 0.002)$ & 0.050 & 1081 & $-0.35\le (r'-i')\le 2.4$ \\
24 &                       &                        & $(0.113\pm 0.008) (i'-z') - (0.010\pm 0.002)$ & 0.049 & 1081 & $-0.3\le (i'-z')\le 0.85$ \\
34 & $\Delta z_{\rm zero}$ & 0.055\tablenotemark{c} & $(0.125\pm 0.012) (i'-z') - (0.014\pm 0.003)$ & 0.050 & 492 & $-0.3\le (i'-z')\le 0.68$ \\
35 &                       &                        & $(1.929\pm 0.177) (i'-z') - (1.188\pm 0.127)$ & 0.043 & 12 & $0.6\le (i'-z')\le 0.85$ \\
\enddata
\tablecomments{Column 1: Relation number from J\o rgensen 2009. 
Column 2: Residual zero point. 
Column 3: rms of $\Delta m$, equivalent to the expected uncertainty on the standard calibration if the color terms are ignored. 
Column 4: Linear fits to the color terms.
Column 5: rms of the linear fits. 
Column 6: Number of individual measurements included in the fits. 
Column 7: Color interval within which the linear fit applies.
}
\tablenotetext{a}{$-1.1\le (g'-i')\le 3.05$}
\tablenotetext{b}{$-0.7\le (r'-i')\le 2.5$}
\tablenotetext{c}{$-0.3\le (i'-z')\le 0.85$}  % $-0.7\le (r'-z')\le 2.25$ This should be in i-z ...}
\end{deluxetable*}

\begin{deluxetable}{llrr}
\tablecaption{GMOS-S Photometric Calibration \label{tab-photgmoss} }
\tabletypesize{\scriptsize}
\tablewidth{0pc}
\tablehead{
\colhead{Filter} & \colhead{$m_{\rm zp}$} & \colhead{Color term\tablenotemark{a}} \\
\colhead{(1)} & \colhead{(2)} & \colhead{(3)} 
}
\startdata
$g'$ & 28.638 & $-0.06 (g'-r')$ \\
$r'$ & 28.587 & $-0.02 (g'-r')$ \\
$i'$ & 28.088 & $-0.02 (i'-z')$ \\
$z'$ & 27.011 & 0.0  \\
\enddata
\tablenotetext{a}{From the Gemini web site.}
\end{deluxetable}

\section{Photometric Calibration \label{SEC-PHOT} }

\subsection{Initial Calibration \label{SEC-INITCAL} }

The photometry from GMOS-N observations has been calibrated using
magnitude zero points and color terms established in J\o rgensen (2009).
As described in that paper, the expected absolute accuracy of the
calibrations is $\approx 0.05$ mag.
For convenience, the specific relations used for the color terms are reproduced in 
Table \ref{tab-color} with the original calibration numbers from J\o rgensen (2009) noted.
The $i'$-band observations were calibrated using the $(r'-i')$ color terms,
except for the three highest redshift clusters for which the $i'$-band was 
calibrated using the $(i'-z')$ color terms.
This is done to avoid color terms spanning the 4000 {\AA} break at the redshifts
of the clusters.
Ideally, the calibration based on $(i'-z')$ color term should also have been used
for RXJ1716.6+6708.
However, the $z'$-band observation of this cluster is too shallow for the $(i'-z')$ color term
to provide a good calibration of the $i'$-band magnitudes.

RXJ1347.5--1147 was observed with GMOS-S.
We determined the calibration from standard stars observed the same night (UT 2005 Jan 11).
Color terms were adopted from the Gemini web site.
The calibrations are summarized in Table \ref{tab-photgmoss}.
For completeness we also list the magnitude zero point for the $z'$-band, though not used 
for our photometry.

All observed magnitudes are calibrated to AB magnitudes.
We adopted the mean atmospheric extinction for Mauna Kea as listed in J\o rgensen (2009),
$k_g =0.14$, $k_r =0.11$, $k_i =0.10$, and $k_z =0.05$.
For Cerro Pach\'{o}n we adopted extinction as listed on the Gemini web site:
$k_g =0.18$, $k_r =0.10$, $k_i =0.08$, and $k_z =0.05$.

\begin{deluxetable*}{lrrrrl}
\tablecaption{Offsets Added to Photometric Zero Points \label{tab-photoffsets} }
\tabletypesize{\scriptsize}
\tablewidth{0pt}
\tablehead{
\colhead{Cluster} & \colhead{$g'$} & \colhead{$r'$} & \colhead{$i'$} & \colhead{$z'$} & \colhead{Comments}  }
\startdata
Abell 1689           &  0.000 & --0.147 & --0.117 & \nodata & SDSS comparison \\
RXJ0056.2+2622  F1   &  0.000 &   0.000 &   0.095 & \nodata & Internal comparison \\
RXJ1334.3+5030       &  \nodata &   0.000 &   0.000 & --0.065 & SDSS comparison \\
RXJ0152.7--1357      &  \nodata &   0.000 &   0.000 & --0.100 & SDSS star colors\\
RXJ1226.9+3332       &  \nodata & --0.139 & --0.049 & --0.060 & SDSS comparison \\
RXJ1415.1+3612       &  \nodata & --0.092 & --0.159 & --0.095 & SDSS comparison \\ 
\enddata
\tablecomments{Clusters not listed in the table were standard calibrated with
the nominal zero points.}
\end{deluxetable*}

\begin{deluxetable*}{lrrrrrr rrrrrr}
\tablecaption{Comparison of GMOS Photometry with SDSS Photometry \label{tab-sdss} }
\tabletypesize{\scriptsize}
\tablewidth{0pc}
\tablehead{
\colhead{Field \& objects} & \multicolumn{3}{c}{$g'$-band} & \multicolumn{3}{c}{$r'$-band} & \multicolumn{3}{c}{$i'$-band} & \multicolumn{3}{c}{$z'$-band} \\
\colhead{} & \colhead{N} & \colhead{$\Delta$} & \colhead{rms} & \colhead{N} & \colhead{$\Delta$} & \colhead{rms} & 
\colhead{N} & \colhead{$\Delta$} & \colhead{rms} & \colhead{N} & \colhead{$\Delta$} & \colhead{rms} 
 }
\startdata
Abell 1689 galaxies\tablenotemark{a}  & 360 & --0.013 & 0.28 & 405 & --0.091 & 0.26 & 393 & --0.051 & 0.28 & \nodata & \nodata & \nodata \\  % updated 2016feb12 FINAL
Abell 1689 stars\tablenotemark{a}     &  62 &   0.052 & 0.21 &  67 & 0.097 & 0.19 &  63 & 0.051 & 0.20 & \nodata & \nodata & \nodata \\  % 
RXJ0056.2+2622 galaxies & 332 & --0.049 & 0.27 & 400 &   0.002 & 0.22 & 390 & 0.022 & 0.22 & \nodata & \nodata & \nodata \\  % updated 2016feb09 FINAL
RXJ0056.2+2622 stars    &  87 & --0.021 & 0.16 &  98 &   0.014 & 0.18 & 102 & 0.070 & 0.14 & \nodata & \nodata & \nodata \\  %  
RXJ0142.0+2131 galaxies &  99 & --0.103 & 0.24 & 109 & --0.059 & 0.28 & 180 & 0.026 & 0.21 & \nodata & \nodata & \nodata \\  % updated 2016feb08 FINAL maybe 
RXJ0142.0+2131 stars    &  21 &   0.018 & 0.12 &  20 & --0.220 & 0.22 &  29 & 0.057 & 0.13 & \nodata & \nodata & \nodata \\  % 
RXJ0027.6+2616 galaxies & 91 & 0.084 & 0.31 & 153 & 0.003 & 0.24 & 156 & --0.072 & 0.24 & \nodata & \nodata & \nodata \\  % updated 2016feb08 FINAL
RXJ0027.6+2616 stars    & 34 & 0.132 & 0.13 &  40 & 0.142 & 0.15 &  44 & 0.020 & 0.12 & \nodata & \nodata & \nodata \\  %   
Abell 851 galaxies  & 202 & 0.042 & 0.27 & 302 & 0.027 & 0.22 & 323 & --0.003 & 0.22 & \nodata & \nodata & \nodata \\  % updated 2016feb08 FINAL
Abell 851 stars     &  27 & 0.107 & 0.16 &  35 & 0.055 & 0.17 &  36 &   0.043 & 0.12 & \nodata & \nodata & \nodata \\  % 
RXJ2146.0+0423 galaxies & 80 & 0.061 & 0.28 & 112 & --0.017 & 0.26 & 104 &   0.026 & 0.23 & \nodata & \nodata & \nodata \\ % updated 2016feb05 FINAL 
RXJ2146.0+0423 stars    & 89 & 0.033 & 0.17 &  98 &   0.015 & 0.16 &  98 & --0.004 & 0.14 & \nodata & \nodata & \nodata \\ % 
RXJ1334.3+5030 galaxies\tablenotemark{a} & \nodata & \nodata & \nodata & 144 & 0.047 & 0.30 & 178 & 0.012 & 0.29 & 99 & 0.124 & 0.29 \\ % updated 2016feb04 FINAL 
RXJ1334.3+5030 stars\tablenotemark{a}    & \nodata & \nodata & \nodata &  32 & --0.030 & 0.21 & 33 & --0.006 & 0.19 & 25 & 0.008 & 0.15 \\ % 
RXJ1716.6+6708 galaxies &  \nodata & \nodata & \nodata & 37 & 0.087 & 0.27 & 48 & --0.026 & 0.27 & 24 & 0.178 & 0.32 \\  % updated 2016feb04 FINAL 
RXJ1716.6+6708 stars    &  \nodata & \nodata & \nodata & 43 & 0.059 & 0.15 & 45 & --0.010 & 0.10 & 43 & 0.053 & 0.13 \\  
RXJ1226.9+3332 galaxies\tablenotemark{a} & \nodata & \nodata & \nodata & 100 &   0.021 & 0.31 & 111 &   0.016 & 0.28 & 30 & 0.308 & 0.27 \\ % updated 2016feb04 FINAL
RXJ1226.9+3332 stars\tablenotemark{a}    & \nodata & \nodata & \nodata &  39 & --0.012 & 0.15 &  39 & --0.010 & 0.13 & 35 & 0.001 & 0.12 \\
RXJ1415.1+3612 galaxies\tablenotemark{a} & \nodata & \nodata & \nodata & 104 &   0.042 & 0.24 & 101 & --0.018 & 0.26 & 32 & 0.128 & 0.26 \\ % updated 2016feb02 FINAL
RXJ1415.1+3612 stars\tablenotemark{a}    & \nodata & \nodata & \nodata &  39 &   0.007 & 0.15 &  44 &   0.019 & 0.19 & 33 & 0.065 & 0.13 \\
\enddata
\tablecomments{Differences are ``GMOS''--``SDSS''.}
\tablenotetext{a}{Zero point offsets were applied to one of more passbands before final comparison,
see Table \ref{tab-photoffsets}.}
\end{deluxetable*}

\subsection{Comparison with SDSS Photometry \label{SEC-SDSS} }

We compared our photometry to that of the SDSS data release 12 (DR12).
For objects that from our observations are classified as stars we use SDSS {\tt psfMag}, 
while for objects classified as galaxies
we use the SDSS magnitude {\tt cmodelMag}, which is the magnitude from a linear combination
of the best fit exponential and $r^{1/4}$ profiles. 
In all cases, we compare to our standard calibrated magnitudes {\tt mag\_auto}.

We used two methods in the comparison:
(1) A direct comparison of magnitudes of objects in the ten clusters with available SDSS photometry,
and (2) a comparison of star colors to the Northern SDSS standard stars (Smith et al.\ 2002).
The second method enables us to evaluate the accuracy of the photometry of all the clusters.

\begin{figure*}
\begin{center}
\epsfxsize 16.0cm
\epsfbox{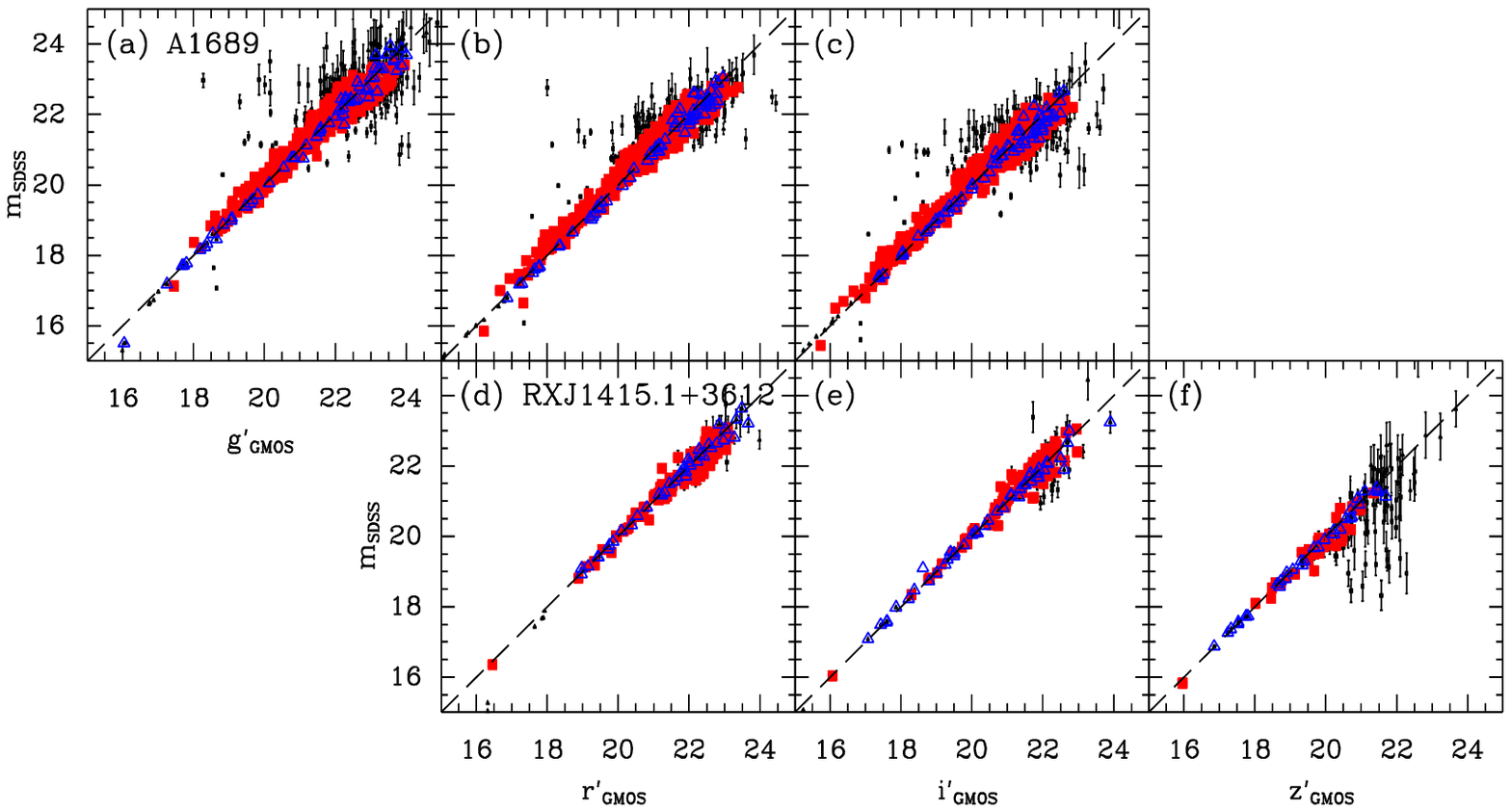}
\end{center}
\caption{
Comparison of our GMOS photometry  with available SDSS photometry for Abell 1689, 
and RXJ1415.1+3612.
This figures serves as an example of the typical comparisons with the SDSS photometry.
Red boxes -- galaxies included in the comparison; 
blue open triangles -- stars included in the comparison; 
black triangles -- stars excluded from the comparison, typically saturated stars; 
black squares -- galaxies excluded from the comparison, see text.
Dashed lines -- one-to-one relations.
\label{fig-sdssphotsampleplot} }
\end{figure*}

\begin{figure*}
\epsfxsize 17.0cm
\epsfbox{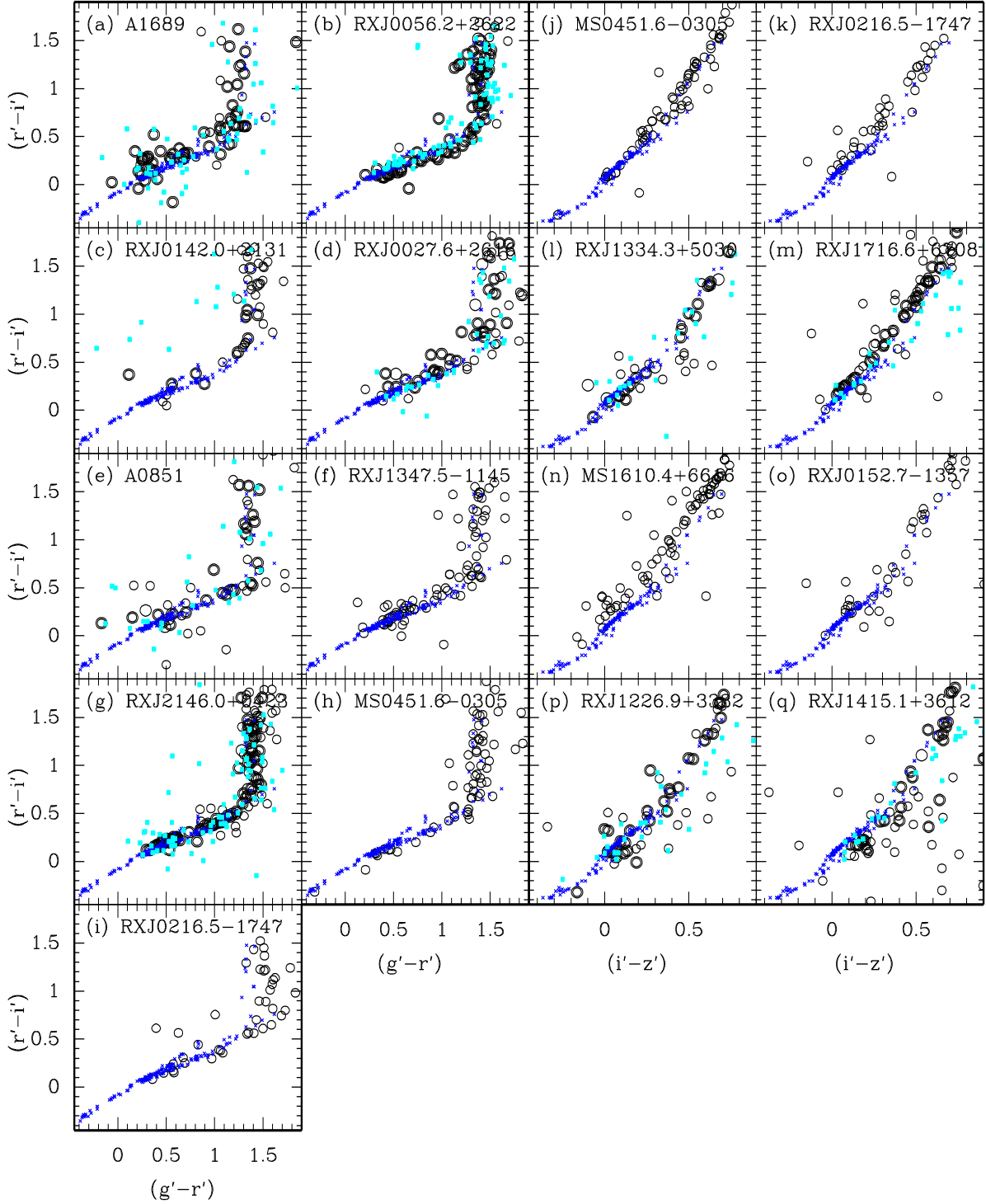}
\caption{
Star colors in the GCP fields, showing our GMOS photometry compared to the SDSS photometry.
Small blue crosses -- SDSS standard star data, these data are shown on all the panels and provide 
a reference for the location of stars in the color-color spaces; 
black circles -- photometry for stars in the cluster fields, colors based on total
magnitudes {\tt mag\_auto}, bolder circles are those stars also included in SDSS;
cyan -- photometry from SDSS for stars in each of the ten fields with SDSS photometry available.
The photometry shown on the figure has not been corrected for Galactic extinction. 
The correction is $<0.04$ for $(g'-r')$, and $<0.02$ in $(i'-z'$ for all clusters,
except for RXJ2145.0+0423 for which the correction of $(g'-r')$ is 0.06.
Offsets to obtain the best calibration have been applied to our data as described in the text.
\label{fig-sdssstar} }
\end{figure*}

Based on the comparison of SDSS photometry with our photometry calibrated using the initial calibrations (Section \ref{SEC-INITCAL}),
magnitude zero point offsets were applied as detailed in Table \ref{tab-photoffsets}.
Figures \ref{fig-sdssphotsampleplot}--\ref{fig-sdssstar} and Table \ref{tab-sdss} summarize 
the comparisons after these offsets were applied.
Figure \ref{fig-sdssphotsampleplot} shows comparisons for the lowest and the highest redshift cluster, only,
as all other comparisons look similar.
The resulting offsets and scatter of the comparisons listed in Table \ref{tab-sdss}
were derived from objects with SDSS magnitude uncertainties less than 0.2 mag, and excluding
saturated objects and objects for which our photometry deviates from the SDSS photometry with more than 0.7 mag.
The resulting scatter in the comparisons is typically 0.15-0.30 mag, lower for the stars than the galaxies.

The adopted magnitude zero point offsets (Table \ref{tab-photoffsets}) represent a compromise between offsets derived 
from the direct comparisons and achieving colors of the stars in the fields consistent 
with the locus of the SDSS standard stars in the color-color diagrams shown in Figure \ref{fig-sdssstar}.
In addition the direct comparisons show systematic differences between
the photometry for stars and that for galaxies in the fields. These differences likely
originate from differences between methods used in the SDSS to determine total magnitudes and 
those used in this paper. 
Based on our simulations presented in Section \ref{SEC-APERTURE},
we expect the differences to depend on the distribution of $n_{\rm ser}$ for the galaxies
included in the comparisons. If the comparisons are dominated by disk galaxies then the differences
should be close to zero, while for a mix of disk and bulge-dominated galaxies the differences
are expected to be $\approx 0.07$ mag, increasing to 0.15 mag for a sample of only bulge-dominated galaxies. 
The median value of the differences $g'$, $r'$ and $i'$ is --0.03 mag, 
with 73\% of the fields and filters within $\pm 0.07$ mag. 
The median of the differences for four $z'$-band comparisons is 0.12 mag.  
For three of the four fields with available SDSS $z'$-band photometry we adjusted the magnitude 
zero points based on primarily the photometry of the stars in the fields.
Thus, it is unlikely that this magnitude offset for the galaxies is related to problem with our zero points. 
Finally, the $z'$-band comparisons for the galaxies show no dependencies on magnitudes, 
colors or sizes of the galaxies.
For our purpose, we conclude that absolute calibrations of $g'$, $r'$ and $i'$
are consistent with the expected calibration consistency of $\approx 0.05$ mag obtainable with
GMOS when using standard methods for calibration, cf.\ J\o rgensen (2009). 
The $z'$-band magnitudes may only be consistent to $\approx 0.12$ mag. We discuss this further
when establishing the color-magnitude relations for the clusters, see Section \ref{SEC-CM}.

\subsection{Calibration Notes for Individual Clusters}

This section contains information on the calibration corrections made for the individual clusters
as well as any differences with 
previously published photometry in J\o rgensen et al.\ (2005), Barr et al.\ (2005)
and J\o rgensen \& Chiboucas (2013).
The photometry in J\o rgensen et al.\ (2017) originates from the consistently 
calibrated photometric catalog included in the present paper.
Only the ten fields with special considerations for the calibration and/or previously
published photometry are listed.

{\bf Abell 1689:} This cluster was covered with two GMOS-N pointings.
Magnitudes from the two fields are internally consistent.
The observations in $r'$-band were obtained on UT 2001 Dec 24 during which no standard stars
were observed. We adopted the zero point from UT 2001 Dec 25. 
However, comparison with SDSS photometry shows a significant offset for both the $r'$-band and the 
$i'$-band. The photometry was corrected for these offsets, cf.\ Table \ref{tab-photoffsets}. 
Photometric parameters for objects covered by both fields observed of this cluster were averaged.

\begin{splitdeluxetable*}{lrrr rrrr rrrr rrrr rr B rr rrrr rrrr rr r rrrr rr}
\tablecaption{GCP Photometric Catalog \label{tab-finalphot} }
\tabletypesize{\scriptsize}
\tablewidth{0pc}
\tablehead{
\colhead{Cluster} & \colhead{ID} & \colhead{R.A.\ (J2000)} & \colhead{Decl.\ (J2000)} & 
\colhead{$g'_{\rm total}$} &  \colhead{$\sigma _{g}$} &
\colhead{$r'_{\rm total}$} &  \colhead{$\sigma _{r}$} &
\colhead{$i'_{\rm total}$} &  \colhead{$\sigma _{i}$} &
\colhead{$z'_{\rm total}$} &  \colhead{$\sigma _{z}$} &
\colhead{$g'_{\rm aper}$} &  \colhead{$\sigma _{g}$} &
\colhead{$r'_{\rm aper}$} &  \colhead{$\sigma _{r}$} &
\colhead{$i'_{\rm aper}$} &  \colhead{$\sigma _{i}$} &
\colhead{$z'_{\rm aper}$} &  \colhead{$\sigma _{z}$} &
\colhead{$g'_{\rm 2.5}$} &  \colhead{$\sigma _{g}$} &
\colhead{$r'_{\rm 2.5}$} &  \colhead{$\sigma _{r}$} &
\colhead{$i'_{\rm 2.5}$} &  \colhead{$\sigma _{i}$} &
\colhead{$z'_{\rm 2.5}$} &  \colhead{$\sigma _{z}$} &
\colhead{$P$} &  \colhead{$M$} & \colhead{$\log r_{\rm iso}$} & 
\colhead{$g'_{\rm iso}$} & \colhead{$r'_{\rm iso}$} & \colhead{$i'_{\rm iso}$} & \colhead{$z'_{\rm iso}$} &
\colhead{$\epsilon$} & \colhead{PA}  \\
\colhead{(1)} & \colhead{(2)} & \colhead{(3)} & \colhead{(4)} & \colhead{(5)} & \colhead{(6)} & \colhead{(7)} & \colhead{(8)} & \colhead{(9)} & \colhead{(10)} &
\colhead{(11)} & \colhead{(12)} & \colhead{(13)} & \colhead{(14)} & \colhead{(15)} & \colhead{(16)} & \colhead{(17)} & \colhead{(18)} & \colhead{(19)} & \colhead{(20)} &
\colhead{(21)} & \colhead{(22)} & \colhead{(23)} & \colhead{(24)} & \colhead{(25)} & \colhead{(26)} & \colhead{(27)} & \colhead{(28)} & \colhead{(29)} & \colhead{(30)} &
\colhead{(31)} & \colhead{(32)} & \colhead{(33)} & \colhead{(34)} & \colhead{(35)} & \colhead{(36)} & \colhead{(37)} 
 }
\startdata
A1689 & 1 & 13:11:12.96 & -1:20:31.8 & 25.234 & 0.251 & 24.274 & 0.191 & 23.725 & 0.191 & \nodata & \nodata & \nodata & \nodata & 24.342 & 0.738 & 23.666 & 0.576 & \nodata & \nodata & 25.315 & 0.335 & 24.170 & 0.222 & 23.651 & 0.221 & \nodata & \nodata & 0.00 & 0.35 & -0.498 & 26.462 & 25.689 & 25.366 & \nodata & 0.594 & 128 \\
A1689 & 2 & 13:11:12.99 & -1:20:38.7 & 20.614 & 0.027 & 20.106 & 0.035 & 19.828 & 0.038 & \nodata & \nodata & 20.770 & 0.014 & 20.277 & 0.018 & 19.994 & 0.020 & \nodata & \nodata & 21.325 & 0.009 & 20.840 & 0.010 & 20.531 & 0.013 & \nodata & \nodata & 0.00 & 0.02 & 0.369 & 20.694 & 20.205 & 19.933 & \nodata & 0.349 & 171 \\
A1689 & 3 & 13:11:13.03 & -1:19:27.3 & 23.832 & 0.162 & 24.255 & 0.419 & 23.021 & 0.201 & \nodata & \nodata & 23.387 & 0.181 & \nodata & \nodata & 22.822 & 0.240 & \nodata & \nodata & 24.202 & 0.127 & 24.007 & 0.188 & 23.151 & 0.137 & \nodata & \nodata & 0.00 & 0.36 & -0.314 & 25.139 & 24.691 & 24.115 & \nodata & 0.120 & 164 \\
A1689 & 4 & 13:11:13.10 & -1:21:45.2 & 23.356 & 0.101 & 22.362 & 0.079 & 21.741 & 0.067 & \nodata & \nodata & 23.189 & 0.128 & 22.177 & 0.100 & 21.510 & 0.079 & \nodata & \nodata & 23.632 & 0.072 & 22.600 & 0.052 & 22.015 & 0.049 & \nodata & \nodata & 0.08 & 0.55 & -0.193 & 24.028 & 23.218 & 22.603 & \nodata & 0.050 & 24 \\
A1689 & 6 & 13:11:13.16 & -1:19:44.1 & 23.814 & 0.182 & 23.800 & 0.329 & 22.840 & 0.202 & \nodata & \nodata & 23.624 & 0.206 & 23.905 & 0.484 & 22.701 & 0.230 & \nodata & \nodata & 24.164 & 0.122 & 23.857 & 0.165 & 23.130 & 0.135 & \nodata & \nodata & 0.00 & 0.24 & -0.250 & 24.886 & 24.542 & 23.980 & \nodata & 0.096 & 163 \\
A1689 & 9 & 13:11:13.36 & -1:21:02.3 & 23.358 & 0.053 & 23.103 & 0.075 & 22.494 & 0.069 & \nodata & \nodata & 23.318 & 0.151 & 23.116 & 0.236 & 22.177 & 0.144 & \nodata & \nodata & 23.349 & 0.058 & 23.091 & 0.082 & 22.468 & 0.074 & \nodata & \nodata & 0.00 & 0.12 & -0.121 & 23.503 & 23.420 & 22.812 & \nodata & 0.074 & 19 \\
A1689 & 13 & 13:11:13.47 & -1:22:05.6 & 23.260 & 0.085 & 22.049 & 0.057 & 22.159 & 0.096 & \nodata & \nodata & 23.118 & 0.117 & 21.744 & 0.070 & 21.964 & 0.127 & \nodata & \nodata & 23.382 & 0.057 & 22.356 & 0.043 & 22.330 & 0.067 & \nodata & \nodata & 0.00 & 0.08 & -0.124 & 23.593 & 22.871 & 22.752 & \nodata & 0.127 & 30 \\
A1689 & 14 & 13:11:13.54 & -1:22:23.9 & 20.523 & 0.016 & 19.151 & 0.010 & 18.764 & 0.010 & \nodata & \nodata & 20.584 & 0.011 & 19.270 & 0.007 & 18.860 & 0.007 & \nodata & \nodata & 20.900 & 0.006 & 19.718 & 0.004 & 19.263 & 0.004 & \nodata & \nodata & 0.00 & 0.03 & 0.347 & 20.571 & 19.250 & 18.842 & \nodata & 0.106 & 24 \\
A1689 & 15 & 13:11:13.54 & -1:19:35.1 & 18.772 & 0.028 & 17.715 & 0.023 & 17.174 & 0.018 & \nodata & \nodata & 19.531 & 0.004 & 18.353 & 0.003 & 17.860 & 0.003 & \nodata & \nodata & 19.949 & 0.003 & 18.834 & 0.002 & 18.311 & 0.002 & \nodata & \nodata & 0.00 & 0.03 & 0.735 & 18.950 & 17.830 & 17.329 & \nodata & 0.068 & 108 \\
A1689 & 16 & 13:11:13.64 & -1:19:41.9 & 21.918 & 0.081 & 21.395 & 0.095 & 20.776 & 0.073 & \nodata & \nodata & 22.226 & 0.054 & 21.598 & 0.059 & 21.101 & 0.055 & \nodata & \nodata & 22.605 & 0.029 & 22.057 & 0.032 & 21.640 & 0.035 & \nodata & \nodata & 0.00 & 0.02 & 0.091 & 22.562 & 22.045 & 21.636 & \nodata & 0.204 & 134 \\
\enddata
\tablecomments{Table 12 is published in its entirety in the machine-readable format.  
A portion is shown here for guidance regarding its form and content.
Columns are explained in Section \ref{SEC-FINALPHOT}. }
\end{splitdeluxetable*}

{\bf RXJ0056.2+2622:} The cluster was covered with two GMOS-N pointings.
Since no standard stars were observed the night of the observations of 
RXJ0056.2+2622 F1, we first adopted the average of the zero points for the preceding
and following night. Comparison of the photometry of the 31 objects brighter than 
$i'\approx 22.5$ mag and included in both 
fields show offsets of $< 0.01$ mag for the $g'$- and $r'$-band.
However, we find a significant offset for the $i'$-band, and
offset the zero point for the RXJ0056.2+2622 F1 $i'$-band to reach 
consistency with the $i'$-band photometry of RXJ0056.2+2622 F2, cf.\ Table \ref{tab-photoffsets}.
Photometric parameters for objects covered by both fields observed
of this cluster were averaged.

{\bf RXJ0142.0+2131:} We use magnitude zero points corresponding to the night
of the observations, UT 2001 Oct 22.
These are 0.02--0.04 mag different from those adopted by Barr et al.\ (2005). 
This has no significant effect on our results in that paper. 
We note that the direct comparison to the SDSS photometry (Table \ref{tab-sdss}) show 
offsets of the galaxy magnitudes of 0.03-0.10 mag. However, the colors of the stars
follow the sequence of the SDSS standard stars (Figure \ref{fig-sdssstar}). Thus,
no additional offsets were applied to the magnitude zero points.

{\bf RXJ1347.5--1145:} This cluster was observed with GMOS-S, see 
Table \ref{tab-photgmoss} for the photometric calibration.
The $g'$-band imaging was obtained in non-photometric conditions in dark time.
The photometry was calibrated by means of a single exposure obtained in photometric conditions in bright time.

{\bf MS0451.6--0305:} No standard stars were observed during the night of the $z'$-band observations.
Based on the $z'$-band magnitude zero points from UT 2001 Nov 22 and UT 2002 Feb 17,
and assuming the degradation of the zero point is similar to that of $i'$-band during the period,
we adopted a zero point of 26.686. J\o rgensen \& Chiboucas (2013) used 26.66.
This offset has no effect on our previous results, as the $z'$-band is not used in
the calibration to the $B$-band rest frame.

{\bf RXJ0216.5--1747:} The $z'$-band imaging was obtained on UT 2004 Jul 20, which is not
covered in J\o rgensen (2009). We derived the magnitude zero points for the night
the same way as done in J\o rgensen (2009) and find zero points of 
($zp_{\rm r},~ zp_{\rm i},~ zp_{\rm z}) = (28.223, 27.959, 26.819)$ for 
the $r'$-, $i'$- and $z'$-band, respectively.

{\bf RXJ1334.5+3030:} The $r'$- and $i'$-band observations were obtained on nights without
observations of standard stars. 
We derived the photometry from the co-added images scaled to the images obtained 
UT 2001 Dec 26, which is noted in the observing log as being photometric.
We then adopt the UT 2001 Dec 25 magnitude zero points. 
The comparison with the SDSS photometry of the field shows that $r'$- and $i'$-band
photometry is in good agreement with SDSS, while the  $z'$-band 
photometry shows a significant offset, cf.\ Table \ref{tab-photoffsets}.
The photometry was corrected for this offset.

{\bf RXJ0152.7--1357:} 
In J\o rgensen \& Chiboucas (2013) we confirmed that the $i'$-band photometry is 
in agreement with the HST/ACS photometry.
This field has no SDSS photometry. However,
comparison with SDSS stellar colors indicate $(i'-z')$ is too small with $\approx 0.1$ mag. 
We therefore offset the nominal $z'$-band zero point with --0.1, cf.\ Table \ref{tab-photoffsets}.
The effect of this offset on our previous results is minimal, affecting the $B$-band rest frame
magnitudes with $\approx$ 0.05 mag.

{\bf RXJ1226.9+3332:} The photometry was corrected with the offsets determined from the 
SDSS comparison, cf.\ Table \ref{tab-photoffsets}. 
The effect of this offset on our previous results is minimal, affecting the $B$-band rest frame
magnitudes with $\approx$ 0.05 mag.

{\bf RXJ1415.1+3612:} Offsets were applied based on both the direct comparison with the SDSS photometry,
and, for the $z'$-band, to optimize the match with the stellar colors, cf.\ Table \ref{tab-photoffsets}.

\section{Fully Calibrated Photometric Parameters \label{SEC-FINALPHOT} }

Table \ref{tab-finalphot} shows the content of the electronically available machine readable table
of the final calibrated photometric parameters.
For each cluster, objects classified as galaxies are listed first, followed by those classified 
as stars. The columns are as follows:

\begin{enumerate}
\item Cluster -- Cluster name
\item ID -- GCP ID number for the galaxy.

\item --4.
R.A.\ (J2000), Decl.\ (J2000) -- Right ascension and Declination calibrated to consistency
with USNO (Monet et al.\ 1998) with an rms scatter of $\approx 0.7$ arcsec.

\setcounter{enumi}{4}

\item --12. $g'_{\rm total}$, $r'_{\rm total}$,  $i'_{\rm total}$, $z'_{\rm total}$ -- Total magnitude in  
$g'$, $r'$, $i'$, and $z'$ determined as SExtractor {\tt mag\_auto} and associated uncertainties using
the corrections from Table \ref{tab-noise}. 

\setcounter{enumi}{12}

\item --20. $g'_{\rm aper}$, $r'_{\rm aper}$,  $i'_{\rm aper}$, $z'_{\rm aper}$  -- aperture magnitudes
derived using the aperture sizes defined in Equation \ref{eq-daper}, and associated uncertainties.

\setcounter{enumi}{20}

\item --28. $g'_{\rm 2.5}$, $r'_{\rm 2.5}$,  $i'_{\rm 2.5}$, $z'_{\rm 2.5}$  -- aperture magnitudes
derived using an aperture with diameter 2.5 arcsec, and associated uncertainties.

\setcounter{enumi}{28}

\item $P$ -- $P({\tt class\_star})$ product of {\tt class\_star} for the available passbands, cf.\ equation (\ref{eq-pstar}).

\item $M$ -- $M({\tt class\_star})$ median of {\tt class\_star} for the available passbands.

\end{enumerate}

For those objects classified as galaxies, the table also contains the following columns:

\begin{enumerate}
\setcounter{enumi}{30}

\item $\log r_{\rm iso}$ -- The logarithm (base 10) of isophotal circularized radius in arcseconds in the detection band,
cf.\ equation (\ref{eq-riso}).
See Table \ref{tab-photoverview} for the surface brightness limit at the isophote.

\item -- 35. $g'_{\rm iso}$,  $r'_{\rm iso}$,  $i'_{\rm iso}$,  $z'_{\rm iso}$ -- Isophotal magnitudes derived using
the surface brightness limits listed in Table \ref{tab-photoverview}.

\setcounter{enumi}{35}

\item $\epsilon$ -- Ellipticity in the detection band, as derived by SExtractor from semi-major and minor-axes, $\epsilon = 1-a/b$.

\item PA -- Position angle in the detection band in degrees measured North through East.

\end{enumerate}

\begin{figure}
\epsfxsize 8cm
\epsfbox{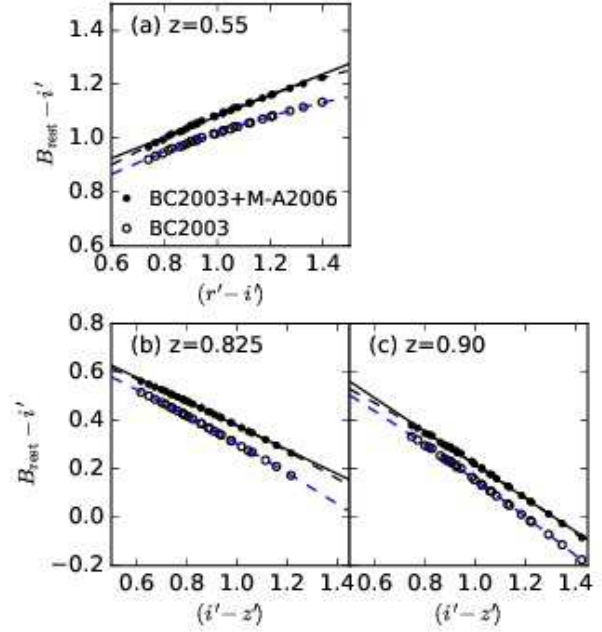}
\caption{Comparison of $B$-band rest frame calibrations. Solid points -- SSP model values using
SDSS filter functions as supplied by Bruzual \& Charlot (2003) and Johnson $U$, $B$, and $V$
filter functions from Ma\'{i}z Apell\'{a}niz (2006). Open points -- SSP model values using 
SDSS and Johnson $U$, $B$, and $V$ filter functions as supplied by Bruzual \& Charlot (2003), 
see text.
Solid black lines -- linear fits to the solid points, the calibrations adopted in the present paper. 
Dashed black lines -- second order fits to the black points. 
Dashed blue lines -- second order fits to the open points. These are the calibrations used in 
J\o rgensen et al.\ (2005) and J\o rgensen \& Chiboucas (2013), see text for discussion.
\label{fig-bc2003_models} }
\end{figure}

\begin{figure}
\epsfxsize 8.5cm
\epsfbox{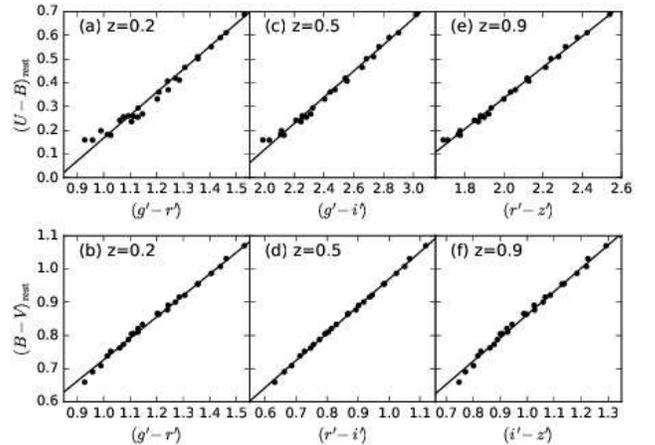}
\caption{Color rest frame calibrations.
Solid points -- SSP model values using SDSS filter functions as supplied by 
Bruzual \& Charlot (2003) and Johnson $U$, $B$, and $V$
filter functions from Ma\'{i}z Apell\'{a}niz (2006).
Solid lines -- linear fits showing the calibrations adopted in the present paper.
\label{fig-bc2003_colormodels} }
\end{figure}

All magnitudes and colors in the table are AB magnitudes.
Magnitude measurements with uncertainties larger than 1 mag are omitted from the table.
Galactic extinction for each field and filter are listed in Table \ref{tab-imdata} in Appendix \ref{SEC-OBSLOG}.
The data in Table \ref{tab-finalphot} have {\it not} been corrected for Galactic extinction.

Uncertainties are included for total magnitudes and colors. 
Uncertainties on isophotal magnitudes are similar to those on the total magnitudes.
The typical uncertainties on the logarithm of isophotal radii are 0.003 with the largest uncertainties 0.01-0.015.
For completeness, we list the isophotal radii even when they are smaller than the seeing of the image in the detection filter.
The uncertainties on the ellipticities are typically of similar size as the 
magnitude uncertainties in the detection band. The ellipticities have not been corrected for the effect of
the image quality, thus they are expected to be affected by systematic errors, especially
for galaxies smaller than about twice the seeing of the images.

\begin{deluxetable*}{lr rrrr r lll}
\tablecaption{Rest Frame Calibrations at the Cluster Redshifts\label{tab-clusrestcalib} }
\tabletypesize{\scriptsize}
\tablewidth{0pc}
\tablehead{
\colhead{Cluster} & \colhead{Redshift} & \multicolumn{4}{c}{$\lambda _{\rm eff}$ in cluster rest frame\tablenotemark{a}} &
\colhead{DM($z$)} & \colhead{$B_{\rm rest}$} & \colhead{$(U-B)$} & \colhead{$(B-V)$} \\
 & & $g'$ & $r'$  & $i'$  & $z'$ &  
}
\startdata
Reference       & 0.0000 & 475 & 630 & 780 & 925 & \\
Abell 1689      & 0.1865 & 400 & 531 & 657 & \nodata & 39.78 & $g'+0.336-0.440(g'-r')$ & $~~\, 1.005 (g'-r') - 0.801$ & $~~\, 0.673 (g'-r') + 0.079$ \\ % calib 9, 1, 6
                &        &     &     &     &         &       & $~~\, \pm 0.007  \pm 0.006$ & $\pm 0.028 ~~~~~~~~~~~ \pm 0.034$ & $\pm 0.007 ~~~~~~~~~~~  \pm 0.008$ \\
RXJ0056.2+2622  & 0.1922 & 398 & 528 & 654 & \nodata & 39.86 & $g'+0.343-0.461(g'-r')$ & $~~\, 0.989 (g'-r') - 0.800$ & $~~\, 0.662 (g'-r') + 0.079$ \\ % calib 9, 1, 6
                &        &     &     &     &         &       & $~~\, \pm 0.008  \pm 0.006$ & $\pm 0.027 ~~~~~~~~~~~ \pm 0.033$ & $\pm 0.007 ~~~~~~~~~~~  \pm 0.009$ \\
RXJ0142.0+2131  & 0.2794 & 371 & 492 & 610 & \nodata & 40.78 & $r'+0.434+0.297(g'-r')$ & $~~\, 0.852 (g'-r') - 0.845$ & $~~\, 1.460 (r'-i') + 0.127$ \\ % calib 10, 1, 7
                &        &     &     &     &         &       & $~~\, \pm 0.011  \pm 0.008$ & $\pm 0.015 ~~~~~~~~~~~ \pm 0.021$ & $\pm 0.020 ~~~~~~~~~~\,  \pm 0.011$ \\
RXJ0027.6+2616  & 0.3650 & 348 & 462 & 571 & \nodata & 41.45 & $r'+0.508+0.123(g'-r')$ & $~~\, 0.798 (g'-r') - 0.919$ & $~~\, 1.300 (r'-i') + 0.127$ \\ % calib 10, 1, 7
                &        &     &     &     &         &       & $~~\, \pm 0.007  \pm 0.005$ & $\pm 0.004 ~~~~~~~~~~~ \pm 0.006$ & $\pm 0.005 ~~~~~~~~~~\,  \pm 0.003$ \\
Abell 851       & 0.4050 & 338 & 448 & 555 & \nodata & 41.72 & $r'+0.574+0.054(r'-i')$ & $~~\, 0.867 (g'-r') - 1.035$ & $~~\, 1.163 (r'-i') + 0.139$ \\ % calib 11, 1, 7
                &        &     &     &     &         &       & $~~\, \pm 0.001  \pm 0.005$ & $\pm 0.005 ~~~~~~~~~~~ \pm 0.008$ & $\pm 0.004 ~~~~~~~~~~\,  \pm 0.002$ \\
RXJ1347.5--1145 & 0.4506 & 327 & 434 & 538 & \nodata & 41.99 & $r'+0.609-0.211(r'-i')$ & $~~\, 0.576 (g'-i') - 0.975$ & $~~\, 0.992 (r'-i') + 0.141$ \\ % calib 11, 2, 7
                &        &     &     &     &         &       & $~~\, \pm 0.001  \pm 0.001$ & $\pm 0.007 ~~~~~~~~~~\, \pm 0.017$ & $\pm 0.006 ~~~~~~~~~~\,  \pm 0.004$ \\
RXJ2146.0+0423  & 0.532  & 310 & 411 & 509 & \nodata & 42.42 & $r'+0.677-0.553(r'-i')$ & $~~\, 0.541 (g'-i') - 1.000$ & $~~\, 0.751 (r'-i') + 0.148$ \\ % calib 11, 2, 7
                &        &     &     &     &         &       & $~~\, \pm 0.005  \pm 0.005$ & $\pm 0.010 ~~~~~~~~~~\, \pm 0.026$ & $\pm 0.007 ~~~~~~~~~~\,  \pm 0.007$ \\
MS0451.6--0305  & 0.5398 & 308 & 409 & 507 & 601     & 42.46 & $i'+0.683+0.422(r'-i')$ & $~~\, 1.103 (r'-i') - 0.700$ & $~~\, 0.739 (r'-i') + 0.146$ \\ % calib 12, 3, 7
                &        &     &     &     &         &       & $~~ \pm 0.005  \pm 0.005$ & $\pm 0.033 ~~~~~~~~~~\, \pm 0.033$ & $\pm 0.008 ~~~~~~~~~~\,  \pm 0.008$ \\
RXJ0216.5--1747 & 0.578  & 301 & 399 & 494 & 586     & 42.64 & $i'+0.706+0.322(r'-i')$ & $~~\, 1.029 (r'-i') - 0.709$ & $~~\, 1.513 (i'-z') + 0.151$  \\ % calib 12, 3, 8
                &        &     &     &     &         &       & $~~ \pm 0.007  \pm 0.006$ & $\pm 0.024 ~~~~~~~~~~\, \pm 0.026$ & $\pm 0.019 ~~~~~~~~~~\,  \pm 0.009$ \\
RXJ1334.3+5030  & 0.620  & \nodata & 389 & 481 & 571 & 42.83 & $i'+0.710+0.237(r'-i')$ & $~~\, 0.985 (r'-i') - 0.741$ & $~~\, 1.500 (i'-z') + 0.122$ \\ % calib 12, 3, 8
                &        &     &     &     &         &       & $~~ \pm 0.007  \pm 0.006$ & $\pm 0.016 ~~~~~~~~~~\, \pm 0.018$ & $\pm 0.011 ~~~~~~~~~~\,  \pm 0.006$ \\
RXJ1716.6+6708  & 0.809  & \nodata & 348 & 431 & 511 & 43.53 & $i'+0.865-0.442(i'-z')$ & $~~\, 0.690 (r'-z') - 0.973$ & $~~\, 0.854 (i'-z') + 0.169$ \\ % calib 13, 4, 8
                &        &     &     &     &         &       & $~~ \pm 0.002  \pm 0.003$ & $\pm 0.004 ~~~~~~~~~~~ \pm 0.009$ & $\pm 0.005 ~~~~~~~~~~\,  \pm 0.004$ \\
MS1610.4+6616   & 0.8300 & \nodata & 344 & 426 & 505 & 43.60 & $i'+0.879-0.513(i'-z')$ & $~~\, 0.681 (r'-z') - 0.976$ & $~~\, 0.790 (i'-z') + 0.174$ \\ % calib 13, 4, 8
                &        &     &     &     &         &       & $~~ \pm 0.003  \pm 0.003$ & $\pm 0.004 ~~~~~~~~~~~ \pm 0.009$ & $\pm 0.005 ~~~~~~~~~~\,  \pm 0.005$ \\
RXJ0152.7--1357 & 0.8350 & \nodata & 343 & 425 & 504 & 43.62 & $i'+0.881-0.528(i'-z')$ & $~~\, 0.678 (r'-z') - 0.975$ & $~~\, 0.780 (i'-z') + 0.175$ \\ % calib 13, 4, 8
                &        &     &     &     &         &       & $~~ \pm 0.003  \pm 0.003$ & $\pm 0.005 ~~~~~~~~~~~ \pm 0.010$ & $\pm 0.006 ~~~~~~~~~~\,  \pm 0.005$ \\
RXJ1226.9+3332  & 0.8908 & \nodata & 333 & 413 & 489 & 43.79 & $i'+0.901-0.665(i'-z')$ & $~~\, 0.652 (r'-z') - 0.965$ & $~~\, 0.706 (i'-z') + 0.165$ \\ % calib 13, 4, 8
                &        &     &     &     &         &       & $~~ \pm 0.005  \pm 0.005$ & $\pm 0.006 ~~~~~~~~~~~ \pm 0.013$ & $\pm 0.010 ~~~~~~~~~~\,  \pm 0.010$ \\
RXJ1415.1+3612  & 1.0269 & \nodata & 311 & 385 & 456 & 44.17 & $z'+0.981-0.024(i'-z')$ & $~~\, 0.610 (r'-z') - 0.980$ & $~~\, 0.688 (i'-z') + 0.076$ \\ % calib 14, 4, 8
                &        &     &     &     &         &       & $~~\, \pm 0.002  \pm 0.002$ & $\pm 0.012 ~~~~~~~~~~~ \pm 0.026$ & $\pm 0.025 ~~~~~~~~~~\,  \pm 0.030$ \\
\enddata
\tablecomments{The second line for each cluster lists the uncertainties on the calibration coefficients.}
\tablenotetext{a}{Wavelengths noted only for the passbands that were obtained for each of the clusters. }
\end{deluxetable*}

Based on internal comparisons, we evaluate that the uncertainties on the position angles are 
$<3\degr$ for galaxies with $\epsilon \ge 0.3$ and total magnitude in the detection band
of 23 mag or brighter. Uncertainties are $<5\degr$ for galaxies with $\epsilon =0.1-0.3$ 
and total magnitude in the detection band of 22 mag or brighter. 
Position angles of fainter or less elliptical galaxies are subject to higher uncertainties.

\section{Calibration of the Photometry to the Rest Frames of the Galaxies \label{SEC-BREST} }

We calibrate the photometry, total magnitudes and colors, to the rest frames of the galaxies
using calibrations based on stellar population models from Bruzual \& Charlot (2003). 
We first described our method in J\o rgensen et al.\ (2005). 
Here we generalize the method to calibrate the total magnitudes to rest frame $B$ band for
all clusters and also establish the calibration of the colors to rest frame $(U-B)$ and $(B-V)$.
The rest frame $B$ magnitudes, $(U-B)$ and $(B-V)$ used here are Vega magnitudes.

We use single stellar population (SSP) models from Bruzual \& Charlot for a Chabrier (2003)
initial mass function, ages of 2--13 Gyr, metallicities of Z=0.004, 0.008, 0.02 (solar), 0.04,
and Padova 1994 evolutionary tracks.
In our previous calibrations (J\o rgensen et al.\ 2005, J\o rgensen \& Chiboucas 2013)
we used filter functions included in the software distributed by Bruzual \& Charlot.
The filter functions for the SDSS filters $g'$, $r'$, $i'$, and $z'$ are identical to those 
supplied by the SDSS. Thus, we maintain use of these filter functions.
However, the filter functions for $U$, $B$, and $V$ are from Buser \& Kurucz (1978).
Filter functions for these filters were shown by Ma\'{i}z Apell\'{a}niz (2006) to give inaccurate descriptions of data. 
Ma\'{i}z Apell\'{a}niz derived better filter functions, and also eliminated the 
internally inconsistent use of two filter functions for the $B$-filter.
We have here adopted these newer filter functions for $U$, $B$, and $V$. Below we comment
on the effect of this compared to our previously used calibrations.

The Bruzual \& Charlot SSP models were used to derive rest frame $B$, $(U-B)$, and $(B-V)$ (Vega magnitudes),
as well as observed AB magnitudes $g'$, $r'$, $i'$, and $z'$, and colors.
This was done in steps of 0.025 in redshift and for the redshift range spanning our observations.
For each of these redshifts, we established the calibration to rest frame $B$ as
\begin{equation}
B_{\rm rest} = m_{\rm obs} + \alpha _1 \cdot color_{\rm obs,1} + \beta _1 \\
\end{equation}
where $m_{\rm obs}$ is the magnitude in the observed band closest matching the rest frame $B$ at the redshift,
$color_{\rm obs}$ is the observed color best complementing $m_{\rm obs}$ to achieve coverage of the full rest frame $B$ band.
Inclusion of a second order color term, as we did in
J\o rgensen et al.\ (2005) and J\o rgensen \& Chiboucas (2013), does not significantly improve
the calibrations, when using the Ma\'{i}z Apell\'{a}niz (2006) $U$, $B$, and $V$ filter functions.
From $B_{\rm rest}$, the absolute $B$-band magnitude is derived as 
\begin{equation}
M_B = B_{\rm rest} - {\rm DM}(z)
\end{equation}
where DM($z$) is the distance modulus for a given redshift.
Similarly, we establish calibrations to rest frame  $(U-B)$ and $(B-V)$ at each of redshift as
\begin{equation}
(U-B) = \alpha _2 \cdot color_{\rm obs,2} + \beta _2 
\end{equation}
and
\begin{equation}
(B-V) = \alpha _3 \cdot color_{\rm obs,3} + \beta _3 
\end{equation}
where $color_{\rm obs,2}$ and $color_{\rm obs,3}$ are the observed colors from the passbands 
closest matching the passbands for rest frame $(U-B)$ and $(B-V)$, respectively.

The calibrations to rest frame $B$, $(U-B)$, and $(B-V)$, are applied to the data by interpolating 
the calibration coefficients to the exact redshift of each of the galaxies.
It is important to note that the validity of the calibrations do not rely on the models being successful
at modeling the ages and metallicities of the stellar populations in the observed galaxies. 
Rather the models only have to provide correct relative color information over the wavelength 
range spanned by the desired rest frame passbands and the observed passbands used in the calibration. 
As long as extrapolations from the observed passbands to the desired rest frame passbands 
are kept to minimum and the available models do span the observed colors, 
any short comings of models to reproduce the exact colors of galaxies for physically 
believable ages and metallicities are of less importance.
Additional information on how to calibrate photometry to a ``fixed-frame'' system, ie.\ rest frame $B$,
can be found in Blanton et al.\ (2003).

Figure \ref{fig-bc2003_models} shows our previous $B$-band calibration compared with the one established here
using filter functions from Ma\'{i}z Apell\'{a}niz (2006). The calibrations are shown at the model redshifts
closest to the redshifts of the three clusters analyzed in J\o rgensen \& Chiboucas (2013).
The difference between two calibrations is typically 0.05 mag in rest frame $B$ magnitudes, with 
the new calibration leading to fainter magnitudes. This change has no significant effect on our
previously published results. Future analysis of the GCP data will use the calibrations
established in the present paper.
Figure \ref{fig-bc2003_colormodels} shows the color calibrations for three typical redshifts
spanning the GCP cluster sample, demonstrating that linear calibrations are sufficient to
fit the model data and provide reliable calibrations.

In Table \ref{tab-clusrestcalib} we provide the calibrations 
matching the cluster redshifts, as well as the distance moduli for the clusters.
As guidance on how the optimal calibrations were chosen, we also list the effective 
wavelength of each of the observed bands in the cluster rest frames.
In most cases the observed colors used in the calibrations match the optimal redshift intervals
except for RXJ2146.0+0423 for which no $z'$ imaging was obtained. 
For the highest redshift clusters the calibrations to $(B-V)$ in all cases 
rely on the same photometry as the calibrations to $(U-B)$ and $B$. Thus, they rely
on extrapolation of the data using the SSP models.

\begin{figure*}
\epsfxsize 15.0cm
\begin{center}
\epsfbox{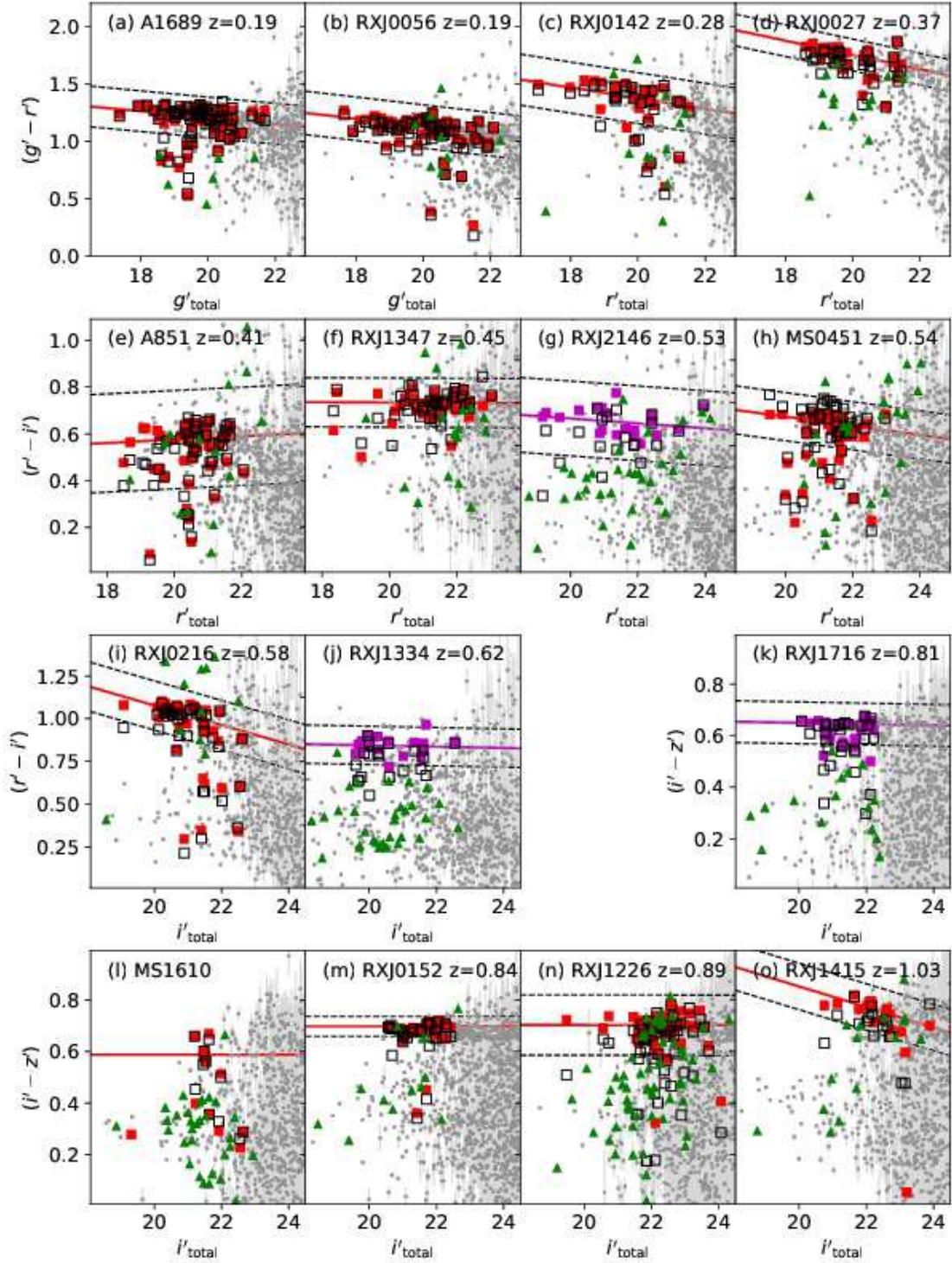}
\end{center}
\caption{
Color-magnitude diagrams, showing the aperture colors used for the rest frame $B$-band 
calibration versus the total magnitudes.
Small grey squares -- aperture colors for all galaxies in the field;
red squares -- aperture colors for confirmed members from our spectroscopy; 
magenta squares -- aperture colors for clusters without processed spectroscopy, the spectroscopic 
sample members selected for the red sequence fitting, see text;
green triangles -- aperture colors for either confirmed non-members from our spectroscopy
or galaxies omitted in the fitting of the red sequence, see text.
Red and magenta lines -- best fit red sequence to the red or magenta points, iteratively 
determined as described in the text.
Dashed black lines are offset from the best-fit color-magnitude relations with $\pm 3$ times the scatter.
Black open squares -- colors from {\tt mag\_auto} for confirmed members or for clusters without processed
spectroscopy the the spectroscopic sample members selected for the red sequence fitting.
\label{fig-CMp1} }
\end{figure*}

\begin{figure}
\epsfxsize 8cm
\epsfbox{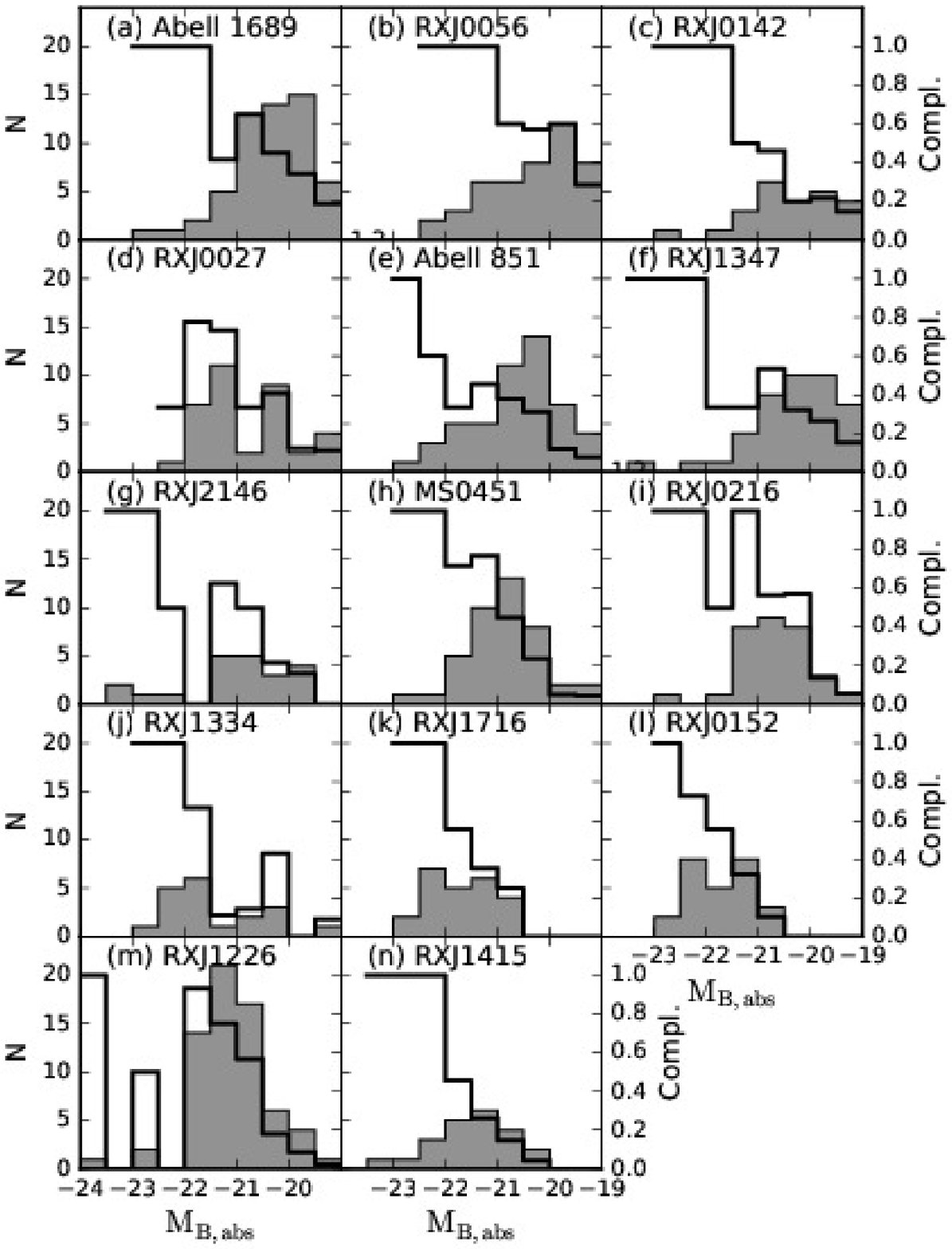}
\caption{Distribution of absolute $B$-band magnitudes.
Grey histograms -- the spectroscopic sample members on the red sequence of the the clusters. The y-axes on the left 
show the number of galaxies in each bin.
Black solid lines -- completeness in each magnitude bin on the red sequence of the clusters. The y-axes on the right
show the completeness fraction.
\label{fig-photometry_distributions} }
\end{figure}

\section{Spectroscopic Samples and Color-Magnitude Relations \label{SEC-CM} }

The spectroscopic samples for the GCP were selected based on magnitudes and colors, and when available
at the time of sample selection redshift information from the literature.
The aim was to include the maximum number of galaxies on the red sequence.
Our previous papers describe the sample selection for the clusters for which we have published 
the spectroscopic data, see Barr et al.\ (2005) for RXJ0142.0+2131,
J\o rgensen et al.\ (2005) for RXJ0152.7--1357, J\o rgensen \& Chiboucas (2013) for 
MS0451.6--0305 and RXJ1226.9+3332, and J\o rgensen et al.\ (2017) for Abell 1689, 
RXJ0056.2+2622, RXJ0027.6+2616, and RXJ1347.5--1145.
These papers also include grey scale images of the clusters with the spectroscopic samples labeled.
For the remainder of the clusters, we summarize the sample selection in Table \ref{tab-spsel}
in Appendix \ref{SEC-SAMPLE}, and in Appendix \ref{SEC-GREYSCALE} provide grey scale images with the spectroscopic samples marked, Figures 
\ref{fig-RXJ0142grey}-\ref{fig-RXJ1415grey}.
The grey scale image of RXJ0142.0+2131 is included in the present paper, as the X-ray data were not available at the 
time of publication of our previous paper on the cluster (Barr et al.\ 2005).

Figure  \ref{fig-CMp1} shows the color-magnitude relations for the clusters, using the colors
for the $B$-band rest frame calibration. 
For galaxies on the red sequence, the figure shows both aperture colors from aperture sizes defined in Equation (\ref{eq-daper}) and total colors.
We fit the color-magnitude relations, using aperture colors, for the members iteratively (red symbols on Figure \ref{fig-CMp1}), 
rejecting galaxies deviating more than three times the scatter relative to the relation. 
The rejection was iterated four times to reach a stable fit of the red sequence.
The best fits to the cluster members are shown as red solid lines and summarized in Table \ref{tab-cm}.
For clusters for which our spectroscopic data have not yet been processed, we instead
fit the relations to the spectroscopic sample, excluding those galaxies with blue colors in at least one of the 
available colors. 
The fits for these clusters were also determined iteratively with rejection.
The difference between using aperture colors and colors based on {\tt mag\_auto} 
is in median 0.03 on the zero points (total colors being bluer), with an rms scatter of 0.05. 
The slopes and scatter of the relations are not significantly different
for the two sets of colors. Thus, we proceed using only the aperture colors.  

\begin{deluxetable*}{l lrr}
\tablecaption{Color Magnitude Relations in the Observed Frame\label{tab-cm} }
\tabletypesize{\scriptsize}
\tablewidth{0pc}
\tablehead{
\colhead{Cluster} & \colhead{Relations} & \colhead{rms} & \colhead{N} 
}
\startdata
Abell 1689      &  $(g'-r') = (-0.027\pm 0.009) (g'-19) + (1.233\pm 0.010)$ & 0.059 & 61 \\
                &  $(r'-i') = (-0.015\pm 0.005) (g'-19) + (0.476\pm 0.006)$ & 0.032 & 64 \\
RXJ0056.2+2622  &  $(g'-r') = (-0.034\pm 0.008) (g'-19) + (1.161\pm 0.013)$ & 0.063 & 52 \\
                &  $(r'-i') = (-0.010\pm 0.004) (g'-19) + (0.412\pm 0.005)$ & 0.027 & 51 \\
RXJ0142.0+2131  &  $(g'-r') = (-0.046\pm 0.015) (r'-20) + (1.374\pm 0.017)$ & 0.074 & 24 \\
                &  $(r'-i') = (-0.016\pm 0.005) (r'-20) + (0.549\pm 0.006)$ & 0.027 & 25 \\
RXJ0027.6+2616  &  $(g'-r') = (-0.062\pm 0.011) (r'-20) + (1.750\pm 0.009)$ & 0.046 & 28 \\
                &  $(r'-i') = (-0.008\pm 0.005) (r'-20) + (0.650\pm 0.004)$ & 0.022 & 30 \\
Abell 851       &  $(g'-r') =  (0.001\pm 0.027) (r'-20) + (1.718\pm 0.027)$ & 0.137 & 41 \\
                &  $(r'-i') =  (0.007\pm 0.013) (r'-20) + (0.575\pm 0.013)$ & 0.070 & 44 \\
RXJ1347.5--1145 &  $(g'-r') = (-0.028\pm 0.025) (r'-20) + (1.540\pm 0.041)$ & 0.136 & 41 \\
                &  $(r'-i') = (-0.001\pm 0.007) (r'-20) + (0.734\pm 0.011)$ & 0.035 & 40 \\
RXJ2146.0+0423  &  $(g'-r') = (-0.063\pm 0.015) (r'-21) + (1.725\pm 0.020)$ & 0.076 & 20 \\
                &  $(r'-i') = (-0.014\pm 0.014) (r'-21) + (0.894\pm 0.019)$ & 0.073 & 20 \\
MS0451.6--0305  &  $(g'-r') = (-0.059\pm 0.031) (r'-21) + (1.686\pm 0.024)$ & 0.121 & 39 \\  
                &  $(r'-i') = (-0.027\pm 0.012) (r'-21) + (0.892\pm 0.009)$ & 0.047 & 36 \\
                &  $(i'-z') = (-0.001\pm 0.009) (r'-21) + (0.381\pm 0.007)$ & 0.036 & 42 \\
RXJ0216.5--1747 &  $(g'-r') = (-0.115\pm 0.032) (i'-21) + (1.813\pm 0.023)$ & 0.118 & 30 \\
                &  $(r'-i') = (-0.057\pm 0.013) (i'-21) + (1.021\pm 0.010)$ & 0.048 & 28 \\
                &  $(i'-z') = (-0.071\pm 0.011) (i'-21) + (0.386\pm 0.008)$ & 0.043 & 33 \\
RXJ1334.3+5030  &  $(r'-i') = (-0.005\pm 0.015) (i'-22) + (1.115\pm 0.027)$ & 0.050 & 18 \\
                &  $(i'-z') =  (0.020\pm 0.012) (i'-22) + (0.441\pm 0.022)$ & 0.033 & 18 \\
RXJ1716.6+6708  &  $(r'-i') = (-0.002\pm 0.021) (i'-22) + (1.271\pm 0.020)$ & 0.060 & 27 \\
                &  $(i'-z') = (-0.002\pm 0.010) (i'-22) + (0.645\pm 0.009)$ & 0.027 & 24 \\
MS1610.4+6616\tablenotemark{a}  & $(r'-i') = 1.329 \pm 0.019$ & 0.038 & 4 \\
                                & $(i'-z') = 0.588 \pm 0.022$ & 0.043 & 4 \\
RXJ0152.7--1357 &  $(r'-i') = (-0.027\pm 0.026) (i'-22) + (1.358\pm 0.022)$ & 0.075 & 27 \\
                &  $(i'-z') =  (0.009\pm 0.005) (i'-22) + (0.698\pm 0.004)$ & 0.013 & 25 \\
RXJ1226.9+3332  &  $(r'-i') = (-0.012\pm 0.016) (i'-22) + (1.141\pm 0.013)$ & 0.085 & 49 \\
                &  $(i'-z') =  (0.000\pm 0.007) (i'-22) + (0.702\pm 0.006)$ & 0.039 & 50 \\
RXJ1415.1+3612  &  $(r'-i') =  (0.000\pm 0.042) (i'-22) + (1.097\pm 0.028)$ & 0.086 & 14 \\  
                &  $(i'-z') = (-0.040\pm 0.011) (i'-22) + (0.773\pm 0.009)$ & 0.030 & 16 \\
\enddata
\tablenotetext{a}{Median colors of the four passive galaxies in the $z=0.83$ group.}
\end{deluxetable*}

\begin{figure}
\epsfxsize 5.5cm
\begin{center}
\epsfbox{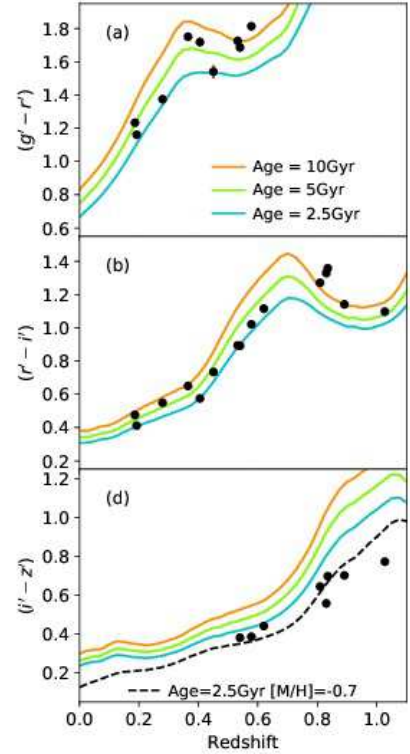}
\end{center}
\caption{
Observed colors as a function of redshift. The colors are the zero points from Table \ref{tab-cm} and correspond
to the colors at $M_{\rm B,abs} \approx -21$ mag. Models from Bruzual \& Charlot (2003) are overlaid: 
Orange, green and cyan lines -- [M/H]=0 and ages of 10 Gyr, 5 Gyr and 2.5 Gyr, respectively. In panel (c) we also 
show the model for [M/H]=--0.7 and age=2.5 Gyr (black dashed line), see text for discussion.
\label{fig-cm_redshift_obs} }
\end{figure}

We then evaluate the completeness of the spectroscopic samples along the red sequence, including galaxies within
$\pm 3$ times the scatter for the color-magnitude relations as marked on Figure \ref{fig-CMp1}. 
Figure \ref{fig-photometry_distributions} shows the distribution of the absolute $B$-band magnitudes, $M _{\rm B,abs}$,
of the spectroscopic samples, together with the completeness. In general the samples are at least 80\% complete
for galaxies brighter than $M _{\rm B,abs} \le -22$ mag, except when the spatial distribution of these
galaxies made it impossible to include all of them in the mask designs for the spectroscopic observations.
This was the case for RXJ0027.6+2616 and RXJ1226.9+3332.
For $-22 < M _{\rm B,abs} \le -19.5$ mag the samples for clusters at $z<0.5$ 
typically include 20-50\% of the red sequence galaxies. For higher redshift clusters, the samples are
limited at $M _{\rm B,abs} \approx -20$ mag, but reach the same completeness.

\begin{figure}
\epsfxsize 8.0cm
\epsfbox{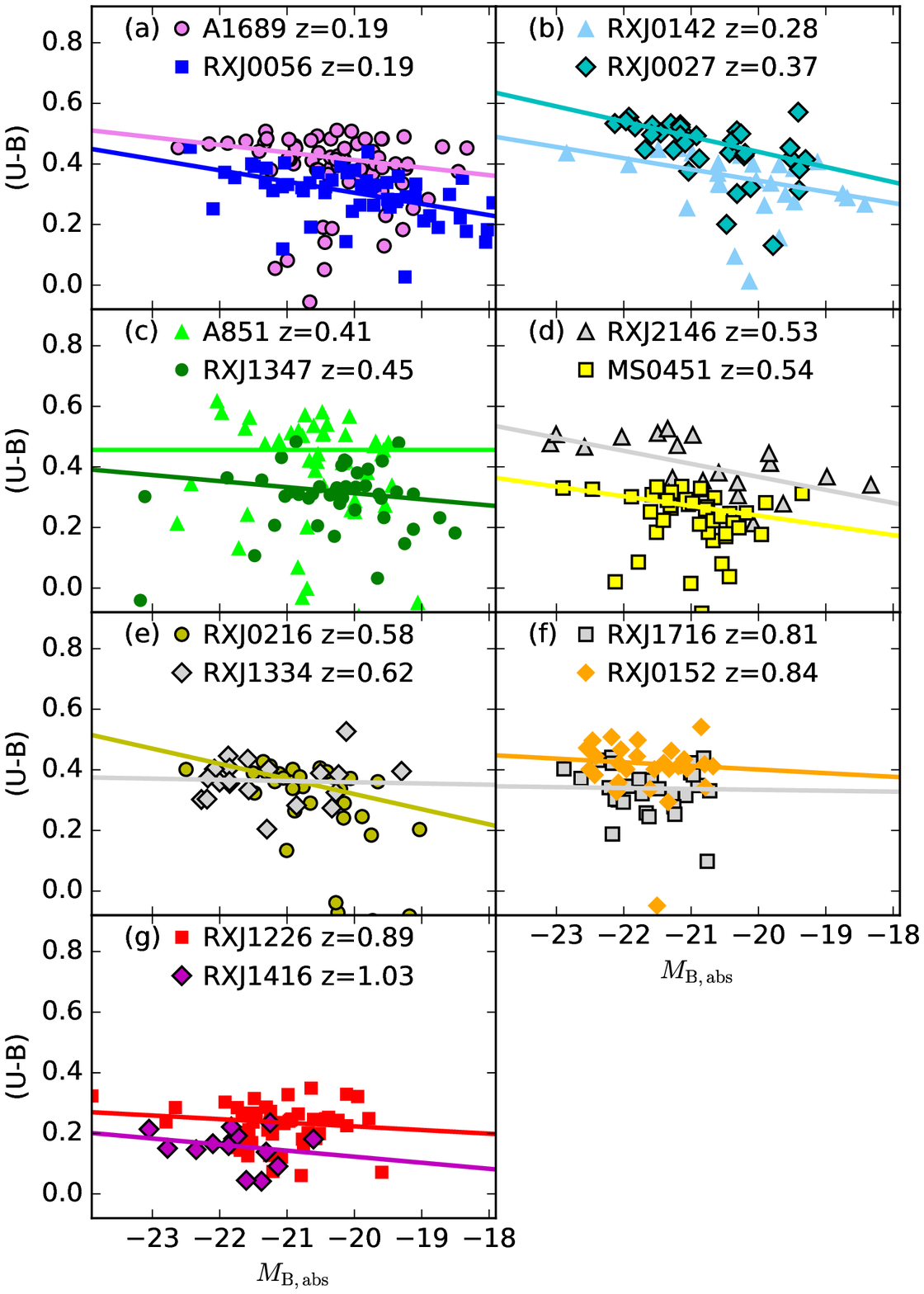}
\caption{
Color magnitude relations as rest frame $(U-B)$ versus the absolute total magnitude in the $B$-band, $M_{\rm B,abs}$.
Plum circles - Abell 1689;
blue squares -- RXJ0056.2+2622; cyan diamonds -- RXJ0027.6+2616; light green triangles -- Abell 851; 
dark green circles -- RXJ1347.5--1347; grey triangles -- RXJ2146.0+0432; yellow squares -- MS0451.6--0305; 
dark yellow circles -- RXJ0216.5--1747; grey diamonds -- RXJ1334.3+5030; grey squares -- RXJ1716.6+6708;
orange diamonds -- RXJ0152.7--1157; red squares -- RXJ1226.9+3332; magenta diamonds -- RXJ1415.1+3612.
Lines showing the best fits are color coded as the data points.
For clusters shown in grey, the figure shows data for the spectroscopic
sample members selected for the red sequence fitting, and the matching fits. 
For all other clusters all confirmed members are shown and included in the fits.
\label{fig-ub} }
\end{figure}

\begin{figure}
\epsfxsize 8.0cm
\epsfbox{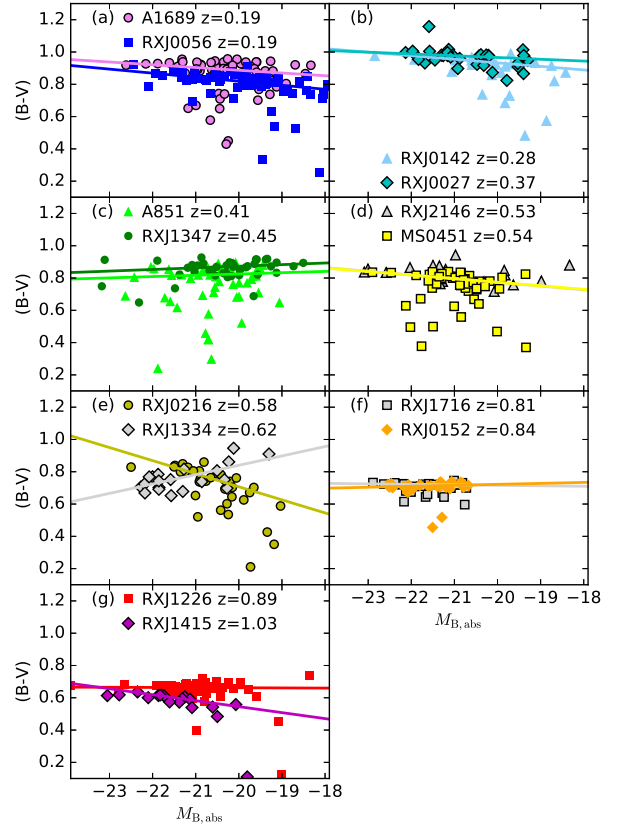}
\caption{
Color magnitude relations as rest frame $(B-V)$ versus the absolute total magnitude in the $B$-band,  $M_{\rm B,abs}$.
Symbols as in Figure \ref{fig-ub}.
\label{fig-bv} }
\end{figure}

\section{Red-Sequence Color-Magnitude Relations as a Function of Redshift \label{SEC-CMREDSHIFT} }

While the main purpose of this paper is to present the consistently calibrated X-ray measurements for the GCP clusters 
and the full photometric catalog from the optical imaging,
we take the opportunity to briefly discuss the changes in the color-magnitude 
relations as a function of redshift.

Figure \ref{fig-cm_redshift_obs} shows the observed mean colors of the red sequence as a function
of cluster redshift.
The colors are the zero points listed for the color magnitude relations in Table \ref{tab-cm} and
correspond to the colors at $M_{\rm B,abs}\approx -21$ mag.
Models from Bruzual \& Charlot (2003) are overlaid for ages of 2.5, 5, and 10 Gyr, and 
solar metallicity [M/H]=0.
The $(g'-r')$ and $(r'-i')$ colors follow the expected variation with redshift, consistent with
mean ages of 2.5-10 Gyr and solar metallicity.
The $(i'-z')$ colors are systematically bluer than predicted by the models. Figure \ref{fig-cm_redshift_obs}c
includes a low metallicity model with [M/H]=--0.7 and age=2.5 Gyr. 
Comparing with the models from Vazdekis et al.\ (2012)
and available from the MILES web site, we find that the MILES models for a Chabrier IMF and the BaSTI isochrones
is 0.05--0.10 bluer in $(i'-z')$ than the Bruzual \& Charlot models. 
Further, Mei et al.\ (2009) find the red sequence of RXJ0152.7--1357 to have $(r_{625}-z_{850})=1.93$ at $i_{775}=22.5$ in
AB magnitudes.
We find $(r'-z')=2.04$ at $i'=22.5$ for this cluster. While the two photometric systems are not completely identical,
we take the comparison as an indication that it is unlikely that our colors are significantly too blue. 
Our $z'$-band {\tt mag\_auto} for the galaxies may be too faint at the 0.12 mag level, cf.\ Section \ref{SEC-SDSS}. 
However, the $(i'-z')$ aperture colors for the stars are consistent with SDSS, see Figure \ref{fig-sdssstar}.
In conclusion, it is possible that the Bruzual \& Charlot (2003) models do not correctly model the $z'$-band
magnitudes. 

\begin{deluxetable*}{l rrrr rrrr}
\tablecaption{Rest Frame Color Magnitude Relations \label{tab-cmrest} }
\tabletypesize{\scriptsize}
\tablewidth{0pc}
\tablehead{
\colhead{Cluster} & \multicolumn{4}{c}{$(U-B)$ relation} & \multicolumn{4}{c}{$(B-V)$ relation}  \\
                  & \colhead{Slope} & \colhead{Zero point} & \colhead{rms} & \colhead{N} 
                  & \colhead{Slope} & \colhead{Zero point} & \colhead{rms} & \colhead{N} \\
}
\startdata
Abell 1689      &  $-0.025\pm 0.008$ &  $0.438\pm 0.009$ & 0.051 & 60 & $-0.017\pm 0.005$ & $0.908\pm 0.006$ & 0.034 & 60 \\
RXJ0056.2+2622  &  $-0.037\pm 0.007$ &  $0.342\pm 0.011$ & 0.055 & 51 & $-0.025\pm 0.005$ & $0.844\pm 0.007$ & 0.037 & 51 \\
RXJ0142.0+2131  &  $-0.037\pm 0.013$ &  $0.382\pm 0.018$ & 0.063 & 24 & $-0.022\pm 0.008$ & $0.955\pm 0.010$ & 0.037 & 25 \\
RXJ0027.6+2616  &  $-0.050\pm 0.008$ &  $0.490\pm 0.007$ & 0.035 & 28 & $-0.011\pm 0.007$ & $0.977\pm 0.006$ & 0.029 & 30 \\
Abell 851       &  $ 0.000\pm 0.024$ &  $0.456\pm 0.022$ & 0.119 & 41 &  $0.008\pm 0.014$ & $0.817\pm 0.013$ & 0.075 & 43 \\
RXJ1347.5--1145 &  $-0.020\pm 0.015$ &  $0.333\pm 0.018$ & 0.078 & 39 &  $0.010\pm 0.006$ & $0.863\pm 0.008$ & 0.037 & 40 \\
RXJ2146.0+0423  &  $-0.043\pm 0.014$ &  $0.410\pm 0.018$ & 0.071 & 20 & $-0.023\pm 0.009$ & $0.796\pm 0.011$ & 0.041 & 18 \\
MS0451.6--0305  &  $-0.032\pm 0.012$ &  $0.271\pm 0.009$ & 0.049 & 36 & $-0.022\pm 0.008$ & $0.796\pm 0.006$ & 0.033 & 36 \\
RXJ0216.5--1747 &  $-0.050\pm 0.014$ &  $0.370\pm 0.010$ & 0.046 & 28 & $-0.081\pm 0.011$ & $0.788\pm 0.009$ & 0.039 & 28 \\
RXJ1334.3+5030  &  $-0.004\pm 0.014$ &  $0.363\pm 0.014$ & 0.049 & 18 & $ 0.058\pm 0.016$ & $0.782\pm 0.015$ & 0.059 & 20 \\
RXJ1716.6+6708  &  $-0.003\pm 0.022$ &  $0.337\pm 0.020$ & 0.064 & 27 & $-0.003\pm 0.008$ & $0.719\pm 0.008$ & 0.023 & 24 \\
RXJ0152.7--1357 &  $-0.012\pm 0.019$ &  $0.413\pm 0.018$ & 0.056 & 27 & $ 0.006\pm 0.004$ & $0.716\pm 0.004$ & 0.012 & 26 \\
RXJ1226.9+3332  &  $-0.012\pm 0.013$ &  $0.235\pm 0.010$ & 0.067 & 50 & $-0.001\pm 0.005$ & $0.663\pm 0.004$ & 0.027 & 50 \\
RXJ1415.1+3612  &  $-0.020\pm 0.026$ &  $0.143\pm 0.028$ & 0.056 & 14 & $-0.037\pm 0.009$ & $0.582\pm 0.007$ & 0.025 & 17 \\
\enddata
\end{deluxetable*}

\begin{figure*}
\epsfxsize 15cm
\begin{center}
\epsfbox{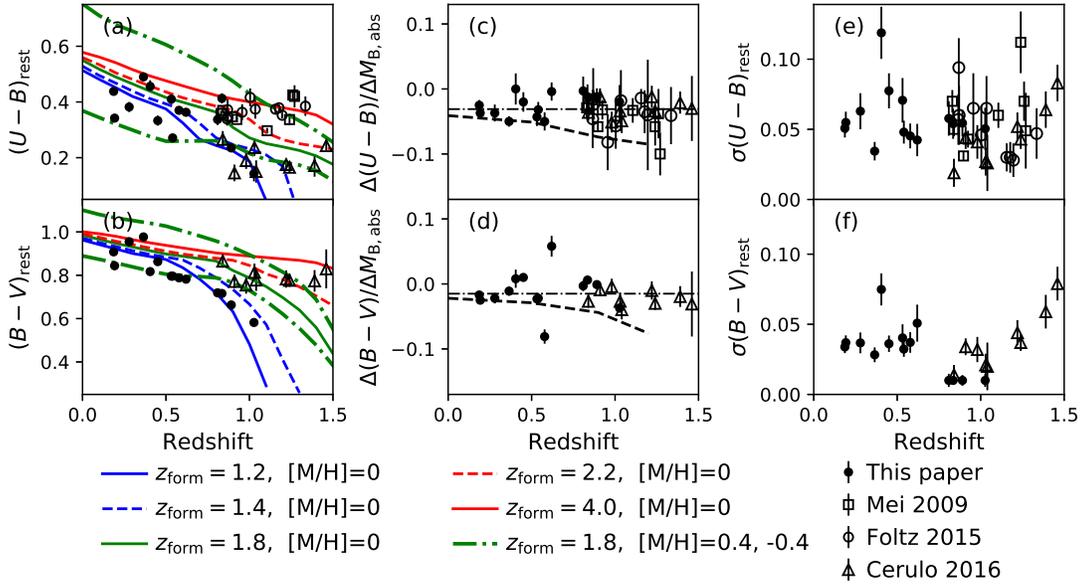}
\end{center}
\caption{
(a)-(b) Rest frame colors $(U-B)$ and $(B-V)$ of the red sequence as a function of redshift. Passive evolution models
based on Bruzual \& Charlot (2003) are overlaid:
Blue solid, blue dashed, green solid, red dashed, and red solid lines -- [M/H]=0 and formation redshifts of
$z_{\rm form} = 1.2, 1.4, 1.8, 2.2$ and 4.0, respectively. Dot-dashed green lines show models for 
$z_{\rm form} = 1.8$ with [M/H]=--0.4 or 0.4, with lower metallicity leading to bluer colors.
(c)-(d) Slopes of the rest frame color-magnitude relations.
Dashed lines -- predicted slopes as a function of redshift, assuming passive evolution and 
relations between age, metallicity and velocity dispersions as found by Thomas et al.\ (2005) at low redshift, see text.
Dot-dashed lines -- predicted slopes adopting only the metallicty-velocity dispersion relation from Thomas et al.\ (2005)
and no age variation with velocity dispersion.
(e)-(f) Internal scatter of the rest frame color-magnitude relations. In panel (f) the four clusters shown at 0.01 
formally has color-magnitude relations with no internal scatter.
On all panels: Solid circles -- our data; 
open squares -- Mei et al.\ (2009); open circles -- Foltz et al.\ (2015); open triangles -- Cerulo et al.\ (2016). 
\label{fig-cm_redshift_rest} }
\end{figure*}

To further investigate the changes in the color-magnitude relations with redshift, we 
establish the relations based on the absolute $B$-band magnitudes, $(U-B)$ and $(B-V)$ colors
in the rest frames of the clusters. Figures \ref{fig-ub}--\ref{fig-bv} and Table \ref{tab-cmrest} 
summarize the slopes, zero points and scatter of the color-magnitude relations.
The fits are based on confirmed members for those clusters with fully processed spectroscopy.
For clusters without processed spectroscopy, the fits are based on the same selection
of spectroscopic samples as used for establishing the color-magnitude relations in the observed frames, cf.\ Section \ref{SEC-CM}.
The unusually positive slope for RXJ1334.3+5030 $(B-V)$-magnitude relation may be a result of inclusion of faint non-members. 
However, the zero point at $M_{\rm B,abs} = -21$ appears to be affected less than 0.05 mag, so we make no attempt here
to exclude additional galaxies from the fit.
In Figure \ref{fig-cm_redshift_rest} we show the colors at $M_{\rm B,abs} = -21$,
the slopes, and internal scatter of the relations, as a function of cluster redshift. 
The internal scatter was derived by subtracting off in quadrature the median uncertainty
at the observed magnitude corresponding to $M_{\rm B,abs} = -21$. We do not include any
contribution from the calibration to the rest frame, as random errors from the observations
dominate over random errors from the rest frame calibration. 
The figure also includes data from Cerulo et al.\ (2016), Foltz et al.\ (2015), and Mei et al.\ (2009) covering redshifts from 0.8 to 1.5.
The literature data have been calibrated to also show colors at $M_{\rm B,abs} = -21$ (Vega magnitudes)
and slopes of the relations relative to the absolute $B$-band magnitude.

In Figure \ref{fig-cm_redshift_rest}ab passive evolution models based on Bruzual \& Charlot (2003) are shown for 
formation redshifts of $z_{\rm form}=1.4-4.0$ and solar metallicity. For $z_{\rm form}=1.8$ we also show a low and 
high metallicity model to illustrate how the assumed metallicity affects the predicted colors.
Our results are generally in agreement with passive evolution and a formation redshift of $1.5-2.0$,
consistent with our results based on the absorption line indices (J\o rgensen et al.\ 2017).
At redshifts $z=0.8-1.0$ where our coverage overlaps with Cerulo et al.\ (2016) our $(U-B)$ results
are in agreement, while our $(B-V)$ colors are $\approx 0.15$ bluer than found by Cerulo et al.
We caution that our wavelength coverage for the highest redshift clusters does not overlap with the $V$ band
and therefore the $(B-V)$ colors rely on extra\-polation based on the Bruzual \& Charlot models.
The $(U-B)$ colors from Mei et al.\ and Foltz et al.\ are $\approx 0.1-0.15$ redder than our results and those from
Cerulo et al.
In general, the zero points of the color magnitude relations provide a much less stringent constraint on the ages 
of intermediate redshift cluster galaxies, than the absorption line indices (e.g., J\o rgensen \& Chiboucas 2013;
J\o rgensen et al.\ 2017). However, the results still serve a role as a consistency check of the results.

We find no change in the slope of color-magnitude relations as a function of redshift, see Figure \ref{fig-cm_redshift_rest}cd. 
This is in agreement with results from Cerulo et al.\ (2016), Foltz et al.\ (2015), and Mei et al.\ (2009).
Thomas et al.\ (2005) established age-velocity dispersion and metallicity-velocity dispersion relations
at $z \approx 0$.  
We use those relations, and the Faber-Jackson relation (1976) (luminosity-velocity dispersion relation) established
for the joint sample of Abell 1689 and RXJ0056.2+2622 members, to derive the expected slopes of the color-magnitude relations, 
under the assumption of passive evolution. The predictions are shown as the dashed lines on Figure \ref{fig-cm_redshift_rest}cd.
Assuming no age variation with velocity dispersion and adopting the metallicity-velocity dispersion relation from
Thomas et al.\ gives predicted slopes of the color-magnitude relations indicated by the dot-dashed lines on 
Figure \ref{fig-cm_redshift_rest}cd.
The joint data from the GCP (this paper), Cerulo et al.\ (2016), Foltz et al.\ (2015), and Mei et al.\ (2009) are inconsistent with
the low redshift age-velocity dispersion relation seen simply as a consequence of passive evolution.
This limits the allowable age change along the color magnitude relation from the brightest
cluster galaxies to galaxies four magnitudes fainter to $<0.05$ dex.
Alternatively, a steep slope of the low redshift age-velocity dispersion relation must be maintained by adding younger galaxies 
to the red sequence, possibly primarily at low masses, cf.\ McDermid et al.\ (2015).

We also find no change in the internal scatter of the  color-magnitude relations as a function of redshift, see Figure \ref{fig-cm_redshift_rest}ef.
One might expect that the addition of younger galaxies to the red sequence at later epochs would lead to a higher scatter
at lower redshifts. However, the samples used for establishing the color-magnitude relations are incomplete at low 
luminosities and also biased against galaxies far from the red sequence, as they are simply our spectroscopic samples aimed at galaxies on the red sequence. 
In addition, the transition from blue star forming galaxy to passive red galaxy may be too fast to result in 
significantly higher scatter. Only Abell 851 in the GCP sample contains a significant number of post-star burst bulge-dominated
galaxies (Hibon et al.\ 2018). The color-magnitude relations for this cluster does have significantly higher scatter than found for the other GCP clusters.
The results from Cerulo et al.\ (2016), Foltz et al.\ (2015), and Mei et al.\ (2009) are based on larger photometric samples.
However, these authors also use sigma-clipping when fitting the color-magnitude relations, most likely affecting the
estimates of the scatter.

\section{Summary \label{SEC-SUMMARY} }

In this paper we have given an overview of the science goals for the  Gemini/HST Galaxy Cluster Project (GCP),
summarized the cluster selection, and assembled consistently calibrated X-ray measurements for the clusters.
We present the photometric catalog based on the ground-based imaging of the GCP clusters
in $g'$, $r'$, $i'$ and $z'$. 
The photometry has been calibrated to consistency with the SDSS photometric system.
The sample selection for the spectroscopic observations are summarized, and provided for those 
clusters not included in prior publications.

We established the calibration of the photometry to rest frame magnitudes and colors
and provide calibration coefficients at the relevant cluster redshifts.

Finally we have derived the color-magnitude relations for all the clusters and briefly 
discussed the redshift dependence the red sequence mean color and scatter, and compare our results to results from the 
literature for higher redshift clusters, and to stellar population models.
The absence of change in the slopes of the rest frame color magnitude relations with redshifts
limits the allowable age differences along the color magnitude to $<0.05$ dex from the 
brightest cluster galaxies to those four magnitudes fainter.
The data add evidence to the need for younger, low mass, galaxies to be added to the red sequence between $z\approx 1$ and the 
present in order to obtain a relatively steep age-velocity dispersion relation at low redshift.

\acknowledgements
The Gemini TACs and the former Gemini Director Matt Mountain are thanked for generous 
time allocations to carry out the Gemini/HST Galaxy Cluster Project.
The anonymous referee is thanked for comments and suggestions that helped improve this paper.
The following collaborators are thanked for their contributions to proposals, 
obtaining the observations and/or data processing during the early phases of the project: 
Jordi Barr, Marcel Bergmann, Maela Collobert, David Crampton, Roger Davies, and Bryan Miller.
We thank Harald Ebeling for confirming ahead of publication that RXJ1415.1+3612 was a 
$z\approx 1$ cluster suitable for inclusion in the GCP.
Pierluigi Cerulo is thanked for comments on the rest frame calibration and the
redshift evolution of the color-magnitude relations.

Based on observations obtained at the Gemini Observatory, which is operated by the 
Association of Universities for Research in Astronomy, Inc., under a cooperative agreement 
with the NSF on behalf of the Gemini partnership: the National Science Foundation 
(United States), the National Research Council (Canada), CONICYT (Chile), 
Minist\'{e}rio da Ci\^{e}ncia, Tecnologia e Inova\c{c}\~{a}o 
(Brazil) and Ministerio de Ciencia, Tecnolog\'{i}a e Innovaci\'{o}n Productiva (Argentina).
The data were obtained while also the Science and Technology Facilities Council (United Kingdom), 
and the Australian Research Council (Australia) contributed to the Gemini Observatory.

The data presented in this paper originate from the following Gemini programs:
GN-2001B-Q-10, GN-2001B-DD-3, GN-2002A-Q-34, GN-2002B-Q-29, GN-2002B-DD-4, 
GN-2003A-DD-4, GN-2003B-Q-21, GN-2003B-DD-3, GS-2003B-Q-26, 
GN-2004A-Q-45, and GS-2005A-Q-27, 
System Verification program GN-2001B-SV-51,
engineering programs GN-2002B-SV-90 and GN-2003A-SV-80.
The data were processed using the Gemini IRAF package.

Photometry from SDSS is used for comparison purposes. Funding for the SDSS and SDSS-II
has been provided by the Alfred P. Sloan Foundation, the Participating Institutions,
the National Science Foundation, the U.S. Department of Energy, the National
Aeronautics and Space Administration, the Japanese Monbukagakusho, the Max
Planck Society, and the Higher Education Funding Council for England.
The SDSS Web Site is http://www.sdss.org/.

Observations have been used that were obtained with {\it XMM-Newton}, an ESA science mission
funded by ESA member states and NASA, the {\it Chandra X-ray Observatory}, and the {\it ROSAT} Mission.
The observations were obtained from these missions data archives.

% get the next table to start in a sensible place
\clearpage

\appendix

\section{Log of available observations \label{SEC-OBSLOG} }

Table \ref{tab-imdata} gives detailed information on the available observations, exposure times, image quality,
and sky brightness. The table also lists the adopted Galactic extinction for each field and filter.

\startlongtable
\begin{deluxetable*}{llllrccc}
\tablecaption{GMOS-N and GMOS-S Imaging Data \label{tab-imdata} }
\tabletypesize{\scriptsize}
\tablewidth{0pc}
\tablehead{
\colhead{Cluster} & \colhead{Program ID\tablenotemark{a}} & \colhead{Dates} & \colhead{Filter} & \colhead{Exposure time} & \colhead{FWHM\tablenotemark{b}} & \colhead{Sky brightness} & \colhead{$A_n$\tablenotemark{c}} \\
\colhead{} & \colhead{} & \colhead{(UT)} & \colhead{} & \colhead{(sec)} & \colhead{(arcsec)} & \colhead{(mag arcsec$^{-2}$)} & \colhead{(mag)} }
\startdata
Abell 1689 F1\tablenotemark{d}  & GN-2003B-DD-3 & 2003 Dec 24 & $g'$ & 4 $\times$ 180   & 0.54 & 21.55 & 0.089 \\  % East ; updated 2015nov13
               & GN-2001B-Q-10 & 2001 Dec 24 & $r'$     & 5 $\times$ 300   & 1.05 & 20.32 & 0.062 \\  % unbinned
               & GN-2001B-Q-10 & 2001 Dec 25 & $i'$     & 3 $\times$ 300   & 1.04 & 19.49 & 0.046 \\  % unbinned
Abell 1689 F2\tablenotemark{d}  & GN-2003B-DD-3 & 2003 Dec 24 & $g'$     & 4 $\times$ 180   & 0.57 & 21.54 & 0.089 \\  % West
               & GN-2001B-Q-10 & 2001 Dec 24 & $r'$     & 6 $\times$ 300   & 1.05 & 19.80 & 0.062 \\  % unbinned
               & GN-2001B-Q-10 & 2001 Dec 25 & $i'$     & 3 $\times$ 300   & 0.81 & 19.51 & 0.046 \\  % unbinned
RXJ0056.2+2622 F1\tablenotemark{e} & GN-2003B-Q-21 & 2003 Jul 2 & $g'$  & 4 $\times$ 120   & 0.79 & 21.94 & 0.192 \\ % South; updated 2014may28
               & GN-2003B-Q-21 & 2003 Jul 2 & $r'$     & 4 $\times$ 120   & 0.70 & 21.24 & 0.133 \\
               & GN-2003B-Q-21 & 2003 Jul 2 & $i'$     & 4 $\times$ 120   & 0.62 & 20.34 & 0.099 \\  % sky updated 2015dec10
RXJ0056.2+2622 F2\tablenotemark{e} & GN-2003B-Q-21 & 2003 Jul 30 & $g'$  & 4 $\times$ 120   & 1.04 & 22.06 & 0.192 \\ % North; updated 2014may28
               & GN-2003B-Q-21 & 2003 Jul 30 & $r'$     & 4 $\times$ 120   & 0.85 & 21.20 & 0.133 \\
               & GN-2003B-Q-21 & 2003 Jul 30 & $i'$     & 5 $\times$ 120   & 1.05 & 19.98 & 0.099 \\  % 2016jan07 seeing estimated, has some extended halo that confuses SExtractor
RXJ0142.0+2131 & GN-2001B-SV-51 & 2001 Oct 22 & $g'$  & 6 $\times$ 600   & 0.67 & 21.58 & 0.225 \\  % updated 2015dec04 
               & GN-2001B-SV-51 & 2001 Oct 22 & $r'$  & 8 $\times$ 300   & 0.52 & 20.65 & 0.156 \\
               & GN-2001B-SV-51 & 2001 Oct 22 & $i'$  & 8 $\times$ 300   & 0.53 & 19.90 & 0.116 \\
RXJ0027.6+2616 & GN-2003B-Q-21 & 2003 Jul 1, 2003 Jul 2 & $g'$     & 4 $\times$ 360   & 0.60 & 21.97 & 0.134 \\  % updated 2014may16 
               & GN-2003B-Q-21 & 2003 Jul 1, 2003 Jul 2 & $r'$     & 4 $\times$ 300   & 0.47 & 21.22 & 0.099 \\
               & GN-2002B-SV-90,3B-Q-21 & 2002 Sep 30, 2003 Jul 1 & $i'$  & 32$\times$ 120   & 0.65 & 20.19 & 0.069 \\
Abell 851      & GN-2001B-Q-10  & 2001 Nov 21 & $g'$  & 6 $\times$ 600   & 0.73 & 21.65 & 0.055 \\  % updated 2015dec04 
               & GN-2001B-Q-10  & 2001 Nov 17 & $r'$  & 6 $\times$ 300   & 0.69 & 20.97 & 0.038 \\
               & GN-2001B-Q-10  & 2001 Nov 17 & $i'$  & 7 $\times$ 300   & 0.71 & 19.98 & 0.028 \\
RXJ1347.5--1145& GS-2005A-Q-27 & 2005 Apr 13 & $g'$     & 4 $\times$ 450   & 1.16 & 22.15 & 0.204 \\  % updated 2014jun12
               & GS-2005A-Q-27\tablenotemark{f} & 2005 Jan 11 & $g'$  & 1 $\times$ 450 & 0.99 & 19.58 & 0.204 \\ 
               & GS-2005A-Q-27 & 2005 Jan 11 & $r'$     & 4 $\times$ 300   & 0.72 & 20.93 & 0.141 \\
               & GS-2005A-Q-27 & 2005 Jan 11 & $i'$     & 4 $\times$ 300   & 0.73 & 20.38 & 0.105 \\
RXJ2146.0+0423 & GN-2003B-Q-21  & 2003 Jul 1 & $g'$  & 6 $\times$ 600   & 0.55 & 22.17 & 0.197 \\  % updated 2015dec04 
               & GN-2003B-Q-21  & 2003 Jul 1 & $r'$  & 6 $\times$ 300   & 0.49 & 21.18 & 0.136 \\
               & GN-2003B-Q-21  & 2003 Jul 1 & $i'$  & 7 $\times$ 300   & 0.46 & 20.25 & 0.101 \\
MS0451.6--0305 & GN-2003B-Q-21 & 2003 Dec 24 & $g'$     &  6 $\times$ 600 & 0.80 & 22.28 & 0.127 \\
               & GN-2002B-Q-29 & 2002 Sep 12-16 & $r'$     & 15 $\times$ 600 & 0.57 & 20.82 & 0.099 \\
               & GN-2001B-DD-3,2B-Q-29 & 2001 Dec 26, 2002 Sept 15 & $i'$     & 6 $\times$ 600, 2 $\times$ 300 & 0.71 & 18.38 & 0.078 \\ %grey
               & GN-2001B-DD-3 & 2001 Dec 26 & $z'$     & 19 $\times$ 600  & 0.72 & 18.54 & 0.064 \\ %grey
RXJ0216.5--1747 & GN-2003B-Q-21 & 2003 Aug 1-5 & $g'$    & 11 $\times$ 300 & 0.67 & 21.68 & 0.119 \\ % updated 2015dec18 checked
               & GN-2003B-Q-21 & 2003 Aug 27-28 & $r'$\tablenotemark{g}   & 5 $\times$ 300 & 0.62 & 20.78 & 0.083 \\
               & GN-2003B-Q-21 & 2003 Jul 31, 2003 Aug 1 & $i'$     &  6 $\times$ 300 & 0.52 & 19.64 & 0.061 \\ % Updated this 2015dec18 
               & GN-2004A-Q-45 & 2004 Jul 20 & $z'$     & 10 $\times$ 300 & 0.58 & 18.80 & 0.046 \\
RXJ1334.3+5030 & GN-2001B-DD-3,2A-Q-34 & 2001 Dec 26, 2002 Feb 11 & $r'$     & 11 $\times$ 600  & 1.06 & 21.12 & 0.022 \\ %updated 2015dec22
               & GN-2001B-DD-3,2A-Q-34 & 2001 Dec 26, 2002 Feb 11 & $i'$     & 4 $\times$ 480, 6 $\times$ 300 & 0.87 & 20.62 & 0.016 \\
               & GN-2003B-DD-3         & 2003 Dec 23              & $z'$     & 8 $\times$ 300  & 0.54 & 19.46 & 0.012 \\
RXJ1716.6+6708 & GN-2003A-DD-4 & 2003 May 3 & $r'$     & 6 $\times$ 600  & 0.72 & 20.92 & 0.080 \\ %updated 2015dec04
               & GN-2003A-DD-4 & 2003 May 3, 2003 May 8 & $i'$     & 11 $\times$ 420 & 0.72 & 19.58 & 0.059 \\
               & GN-2003A-DD-4 & 2003 May 3-8 & $z'$     & 19 $\times$ 210 & 0.73 & 18.54 & 0.044 \\
MS1610.4+6616  & GN-2003A-DD-4 & 2003 Apr 25 & $r'$     & 6 $\times$ 450  & 0.73 & 21.00 & 0.068 \\ %updated 2015dec04
               & GN-2003A-DD-4 & 2003 Apr 25 & $i'$     & 6 $\times$ 450  & 0.69 & 19.80 & 0.050 \\
               & GN-2003A-DD-4 & 2003 Apr 26 & $z'$     & 12 $\times$ 210 & 0.76 & 18.81 & 0.038 \\  % updated 2016feb01 - z-sky
RXJ0152.7--1357 & GN-2002B-Q-29 & 2002 Sep 14-15 & $r'$     & 12 $\times$ 600           & 0.68 & 20.59 & 0.033\\
               & GN-2002B-Q-29,SV-90 & 2002 Jul 17-19, 2002 Sep 14-25 & $i'$& 7 $\times$ 450, 100 $\times$ 120 & 0.56 & 19.31 & 0.024 \\ %grey
               & GN-2002B-Q-29 & 2002 Jul 19, 2002 Sep 14-17 & $z'$   & 25 $\times$ 450 & 0.59 & 19.03 & 0.018 \\ %grey
RXJ1226.9+3332 & GN-2003A-DD-4 & 2003 Apr 26 & $r'$     & 9 $\times$ 600                     & 0.75 & 21.16 & 0.044 \\
               & GN-2003A-DD-4,SV-80& 2003 Jan 31-Feb 1, 2003 Mar 13 & $i'$     & 7 $\times$ 300, 3 $\times$ 360 & 0.78 & 20.53 & 0.033 \\
               & GN-2003A-DD-4,SV-80& 2003 Mar 13, 2003 May 6 & $z'$     & 29 $\times$ 120   & 0.68 & 18.80 & 0.024 \\ %grey
RXJ1415.1+3612 & GN-2003A-DD-4      & 2003 Apr 27, 2003 May 5 & $r'$     & 11 $\times$ 600 & 0.63 & 20.82 & 0.023 \\
               & GN-2003A-DD-4,SV-80& 2003 Apr 27, 2003 May 5 & $i'$     & 94 $\times$ 120 & 0.71 & 19.51 & 0.017 \\ %grey
               & GN-2003A-DD-4      & 2003 Apr 27, 2003 May 5 & $z'$     & 13 $\times$ 210 & 0.62 & 18.78 & 0.013 \\ %grey
\enddata
\tablenotetext{a}{Observations with program IDs starting with GN and GS were obtained with GMOS-N and GMOS-S, respectively.}
\tablenotetext{b}{Image quality measured as the average FWHM of 7-10 stars in the field from the final stacked images.}
\tablenotetext{c}{Galactic extinction at cluster center, Schlafly \& Finkbeiner (2011) as available through the NASA/IPAC Extragalatic Database.}
\tablenotetext{d}{F1 pointing Eastern field (RA,DEC)$_{\rm J2000}$ = (13\,11\,37.0, --1\,20\,29),
F2 pointing Western field  (RA,DEC)$_{\rm J2000}$ = (13\,11\,23.5, --1\,20\,29) }
\tablenotetext{e}{F1 pointing Southern field (RA,DEC)$_{\rm J2000}$ = (0\,55\,59.0, 26\,20\,30), 
F2 pointing Northern field  (RA,DEC)$_{\rm J2000}$ = (0\,56\,00.5, 26\,25\,10) }
\tablenotetext{f}{Observation obtained in twilight, used for photometric calibration, only.}
\tablenotetext{g}{Observations in $r'$-band (12 $\times$ 240sec) obtained under GS-2003B-Q-26 
were not used as the images have significantly worse image quality than those from GN-2003B-Q-21.}
\end{deluxetable*}

\section{Spectroscopic Sample Selection \label{SEC-SAMPLE} }

The sample selection for the GCP spectroscopic observations is based on the color-magnitude diagrams
and any redshift information available in the literature at the time of sample selection.
Table \ref{tab-spsel} summarizes the sample selection for those clusters not included in our previous
papers.
Objects in classes 1 and 2 have highest priority, and span the brighter 2-2.5 magnitude of the red sequence.
Available redshifts were used to give higher priority to members and if possible excluded non-members from this selection. 
We used redshifts from Gioia et al.\ (1999) for RXJ1716.6+6708, and from Dressler et al.\ (1999) for Abell 851.
Objects in class 3 are typically fainter galaxies on the red sequence. However, for clusters with some prior redshift information, brighter red sequence
galaxies without prior redshift determinations are included in class 3. Objects in class 4 are only added to fill available space in the mask design.
They are typically fainter and/or bluer than the objects in classes 1--3.
For the field of MS1610.4+6616 the sample selection was aimed at the published cluster redshift of 0.65 
(Luppino \& Gioia 1995).
Object class 1 were assigned to targets with colors that could match this redshift and with a limit of $i'=21.8$ mag.
However, because the field does not contain a rich cluster, there is no well-defined red sequence at the expected colors. 
Redder and bluer, or fainter objects were assigned object class 2, and given lower priority in the mask design.
The redder objects turned out to be part of the poor group that we identified at $z=0.83$.
The sample in RXJ1334.3+5030 was selected without a blue limit on the colors and using only
two intervals of total magnitude to ensure coverage in luminosities. Thus, the sample contains a much
larger fraction of blue galaxies than the samples in the other fields.

\begin{deluxetable*}{lcl}
\tablecaption{Selection Criteria for Spectroscopic Samples \label{tab-spsel} }
\tablewidth{0pc}
\tablehead{
\colhead{Cluster} & \colhead{Obj.Class} & \colhead{Selection criteria} }
\startdata
Abell 851\tablenotemark{a} & 1 & Confirmed member based on redshift $\wedge ~ 18.4 \le r' \le 20.9 ~ \wedge ~ 0.3 \le (r'-i') \le 0.8 $ \\
 & 2 & Confirmed member based on redshift $\wedge ~ 20.9 < r' \le 21.7 ~ \wedge ~ 0.3 \le (r'-i') \le 0.8 $ \\
 & 3 & No redshift $\wedge ~ 18.4 \le r' \le 21.7 ~ \wedge ~ 0.3 \le (r'-i') \le 0.8 $ \\
 & 4 & $\left[ 18.4 \le r' \le 21.7 ~ \wedge ~ \left( (r'-i')< 0.3 \vee (r'-i')>0.8 \right) \right]  ~\vee ~21.7 < r' \le 22.7  $ \\
RXJ2146.0+0423 & 1 & $i' \le 20.55 ~\wedge ~(g'-r') \ge 1.0 ~\wedge ~(r'-i') \ge 0.55 ~\wedge ~(g'-r') \ge 0.4 + 1.3(r'-i')$ \\
& 2 & $20.55 < i' \le 21.65 ~\wedge ~(g'-r') \ge 1.0 ~\wedge ~(r'-i') \ge 0.55 ~\wedge ~(g'-r') \ge 0.4 + 1.3(r'-i')$ \\
& 3 & $21.65 < i' \le 22.55 ~\wedge ~(g'-r') \ge 1.0 ~\wedge ~(r'-i') \ge 0.55 ~\wedge ~(g'-r') \ge 0.4 + 1.3(r'-i')$ \\
& 4 & $\left[ 23. < i' \le 22.55 ~\wedge  (g'-r') \ge 1.0 ~\wedge ~(r'-i') \ge 0.55 ~\wedge ~(g'-r') \ge 0.4 + 1.3(r'-i')\right] ~\vee$ \\
&   & $\left[ i' \le 23 ~\wedge ~\left( (g'-r') < 1.0 \vee (r'-i') < 0.55 \vee (g'-r') < 0.4 + 1.3(r'-i')\right) \right]$ \\
RXJ0216.5--1747\tablenotemark{b} & 1 & $i' \le 20.5 ~\wedge ~(g'-r') \ge 1.2 ~\wedge ~(r'-i') \ge 0.7$ \\
& 2 & $20.5 < i' \le 21.6 ~\wedge ~(g'-r') \ge 1.2 ~\wedge ~(r'-i') \ge 0.7$ \\
& 3 & $21.6 < i' \le 22.6 ~\wedge ~(g'-r') \ge 1.2 ~\wedge ~(r'-i') \ge 0.7$ \\
& 4 & $i' \le 22.6 ~\wedge ~ \left[ (g'-r') < 1.2 ~\vee ~(r'-i') < 0.7 \right]$ \\
RXJ1334.3+5030\tablenotemark{a} & 1 & $i' \le 20.7 ~\wedge ~(r'-i') \le 1.35$ \\
& 2 & $20.7 < i' \le 22.4 ~\wedge ~(r'-i') \le 1.35$ \\
RXJ1716.6+6708 & 1 &  Confirmed member based on redshift $ \wedge ~ i' \le 21.6 ~\wedge ~0.9 \le (r'-i') \le 1.5 ~\wedge ~0.45 \le (i'-z') \le 0.75 $ \\
& 2 & $\left[ {\rm Confirmed~member~based~on~redshift}~ \wedge ~ 21.6 < i' \le 22.5 ~\wedge ~0.9 \le (r'-i') \le 1.5 ~\wedge ~0.45 \le (i'-z') \le 0.75~ \right] ~ \vee $ \\
&   & $\left[ {\rm No~redshift} ~ \wedge ~ i' \le 21.6 ~ \wedge ~ 0.9 \le (r'-i') \le 1.5 ~\wedge ~0.45 \le (i'-z') \le 0.75 \right]$  \\
& 3 & No redshift $  \wedge ~ 21.6 < i' \le 22.5 ~\wedge ~0.9 \le (r'-i') \le 1.5 ~\wedge ~0.45 \le (i'-z') \le 0.75 $ \\
& 4 & $ i' \le 22.5 ~\wedge ~ \left[ (r'-i') < 0.9 ~\vee ~(i'-z') < 0.45 \right]$ \\
MS1610.4+6616 & 1 & $i' \le 21.8 ~\wedge ~0.8 \le (r'-i') \le 1.2 ~\wedge ~0.3 \le (i'-z') \le 0.45$ \\
& 2 & $i' < 23 ~ \wedge ~  !\left[ ~ 0.8 \le (r'-i') \le 1.2 ~\wedge ~0.3 \le (i'-z') \le 0.45~ \right] $ \\
RXJ1415.1+3612 & 1 & $i' \le 22 ~\wedge ~0.6 \le (i'-z') \le 0.8 ~\wedge ~(r'-i') \le 1.3$ \\
& 2 & $22 < i' \le 22.5 ~\wedge ~0.6 \le (i'-z') \le 0.8 ~\wedge ~(r'-i') \le 1.3$ \\
& 3 & $22.5 < i' \le 23.5 ~\wedge ~0.6 \le (i'-z') \le 0.8 ~\wedge ~(r'-i') \le 1.3$ \\
& 4 & $i' \le 23.5 ~\wedge ~(i'-z') < 0.6$ \\
\enddata
\tablenotetext{a}{At the time of sample selection only imaging in $r'$ and $i'$ was available.}
\tablenotetext{b}{At the time of sample selection only imaging in $g'$, $r'$ and $i'$ was available.}
\end{deluxetable*}

\section{Cluster Properties and Grey Scale Images \label{SEC-GREYSCALE} }

Here we summarize the properties of the clusters, with reference to the original discovery papers,
results regarding cluster structure (evidence for sub-clusters or merging) and
the star formation history.

\begin{figure*}
\epsfxsize 17.0cm
\epsfbox{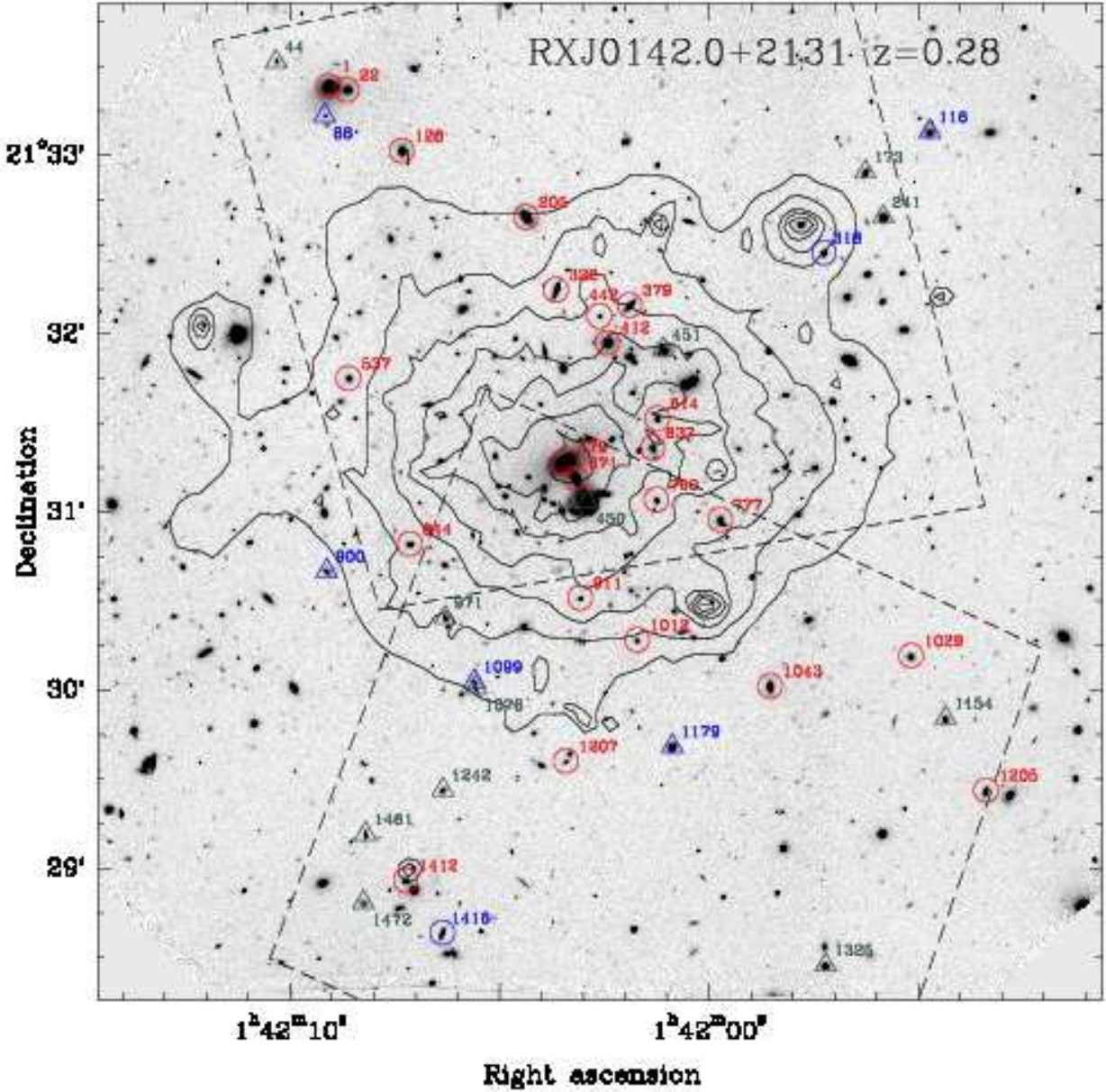}
\caption{
GMOS-N $r'$-band image of RXJ0142.0+2132 with the spectroscopic samples marked.
Contours of the {\it Chandra} X-ray data are overlaid.
Red circles -- confirmed bulge-dominated members with EW[\ion{O}{2}]$\le 5${\AA}. 
Blue circles -- confirmed bulge-dominated members with EW[\ion{O}{2}] $> 5${\AA}.
Blue triangles -- confirmed disk-dominated members 
Dark green triangles -- confirmed non-members. 
The approximate location of the {\it HST}/ACS fields observed in F775W are marked with dashed lines,
Chiboucas et al.\ (2009).
The X-ray image is from the {\it Chandra} ACIS camera
\dataset[ADS/Sa.CXO\#obs/10440]{Chandra ObsId 10440}.
The X-ray image was smoothed; any structure seen is significant at 
the 3$\sigma$ level or higher. The spacing between the contours is logarithmic with a factor of 
1.5 between each contour.
\label{fig-RXJ0142grey} }
\end{figure*}

\begin{figure*}
\epsfxsize 17.0cm
\epsfbox{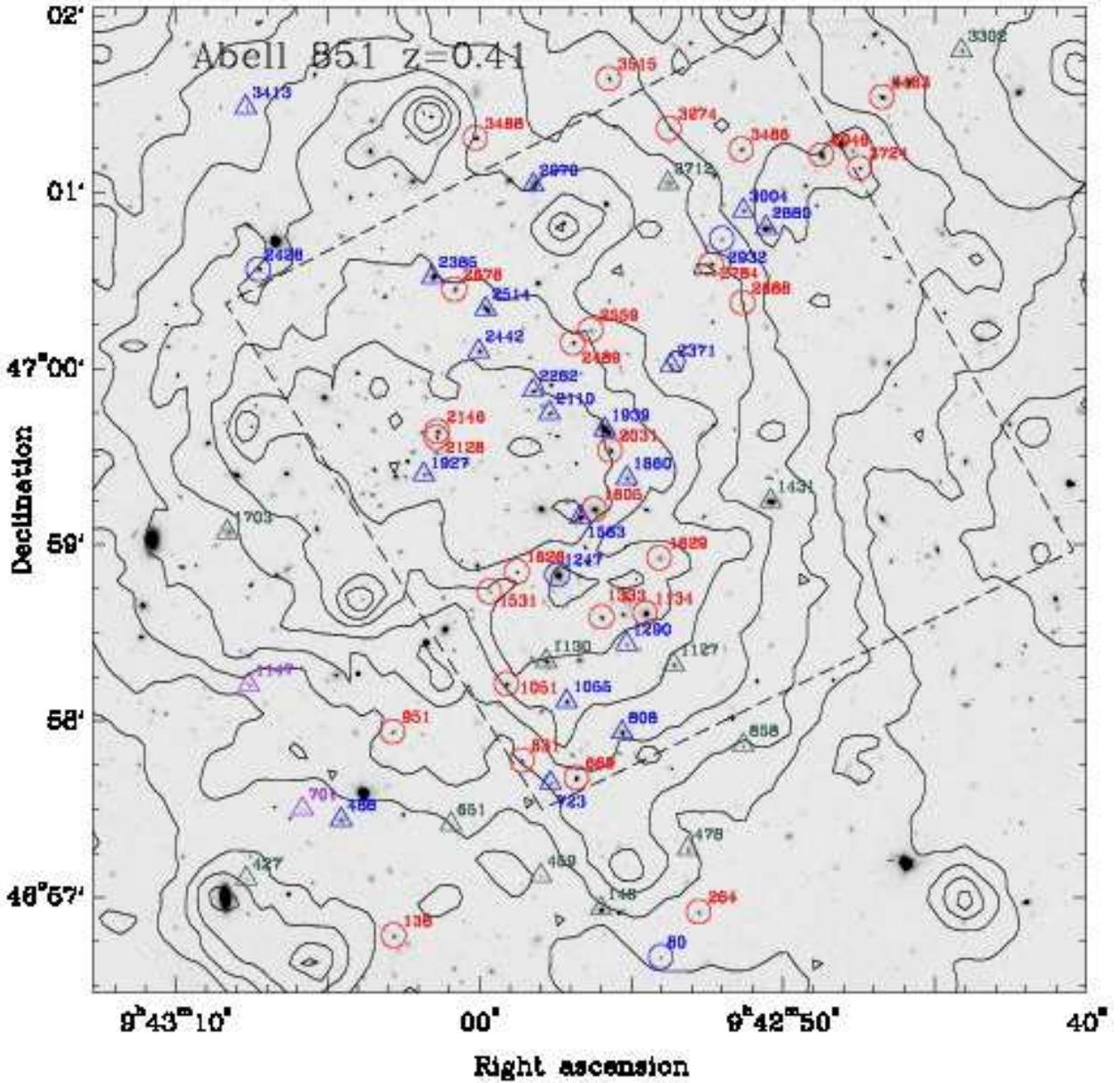}
\caption{
GMOS-N $r'$-band image of Abell 851 with the spectroscopic samples marked.
Contours of the {\it XMM-Newton} X-ray data are overlaid.
Red circles -- confirmed bulge-dominated members with EW[\ion{O}{2}]$\le 5${\AA}. 
Blue circles -- confirmed bulge-dominated members with EW[\ion{O}{2}] $> 5${\AA}.
Blue triangles -- confirmed disk-dominated members 
Dark green triangles -- confirmed non-members. 
Purple triangles -- targets for which the spectra do not allow redshift determination. 
The approximate location of the {\it HST}/ACS field observed in F814W is marked with dashed lines. 
Most of the GMOS-N field is also covered by {\it HST}/WFPC2 observations (Hibon et al.\ in prep.).
The X-ray image is the sum of the images from the two {\it XMM-Newton} EPIC-MOS cameras. 
The X-ray image was smoothed; any structure seen is significant at 
the 3$\sigma$ level or higher. The spacing between the contours is logarithmic with a factor of 
1.5 between each contour.
\label{fig-A0851grey} }
\end{figure*}

\begin{figure*}
\epsfxsize 17.0cm
\epsfbox{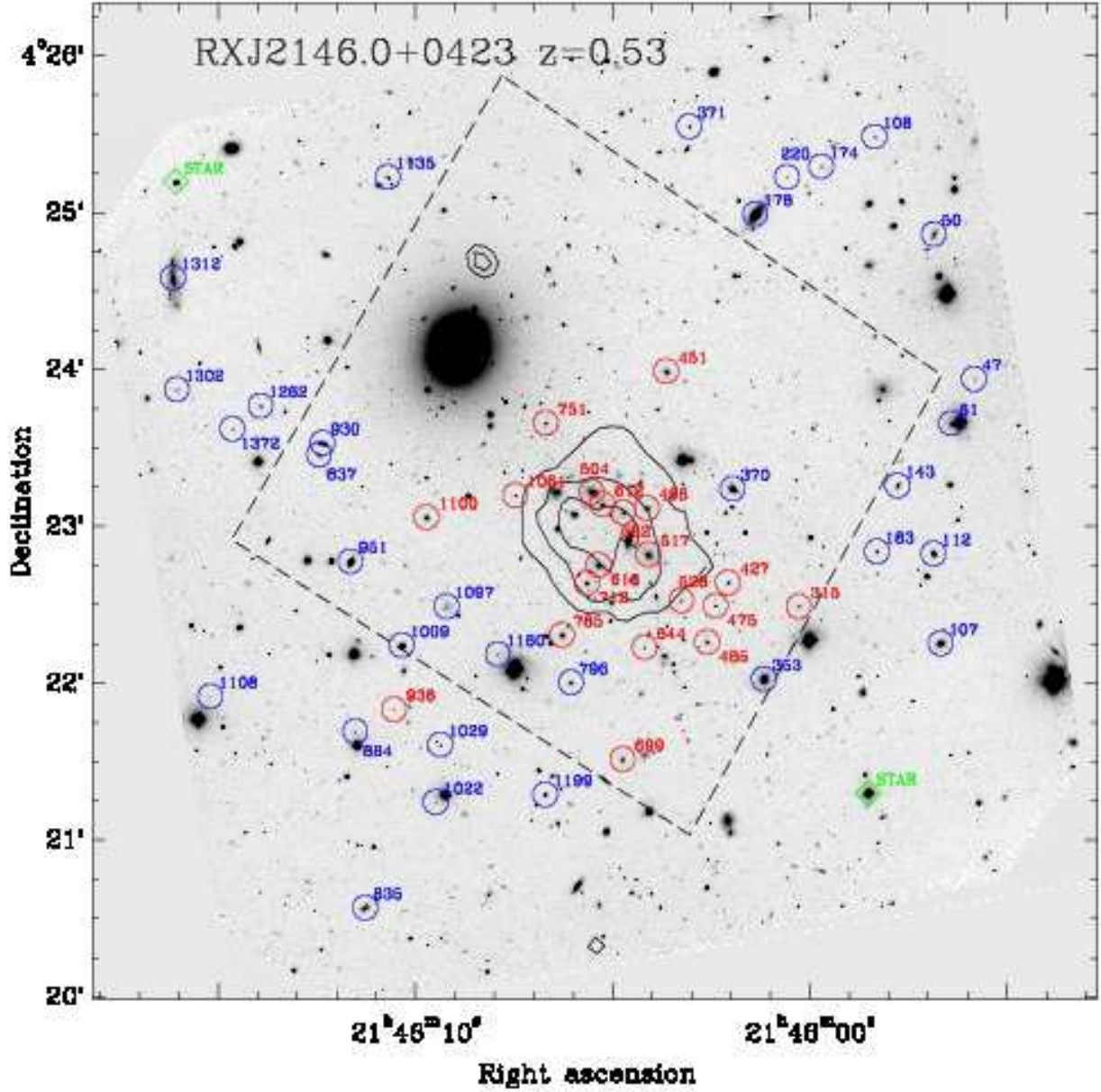}
\caption{
GMOS-N $r'$-band image of RXJ2146.0+0423 with the spectroscopic samples marked.
Contours of the {\it XMM-Newton} X-ray data are overlaid.
Red circles -- galaxies used to fit the red sequence ($(r'-i') \ge 1.0$ and $(i'-z') \ge 0.5$).
Blue circles -- blue galaxies in the spectroscopic sample.
Green diamonds -- blue stars included in the mask to facilitate correction for telluric absorption lines. 
The approximate location of the {\it HST}/ACS field observed in F814W is marked with dashed lines. 
The X-ray image is the sum of the images from the two {\it XMM-Newton} EPIC-MOS cameras. 
The X-ray image was smoothed; any structure seen is significant at 
the 3$\sigma$ level or higher. The spacing between the contours is logarithmic with a factor of 
1.5 between each contour.
\label{fig-RXJ2146grey} }
\end{figure*}

\begin{figure*}
\epsfxsize 17.0cm
\epsfbox{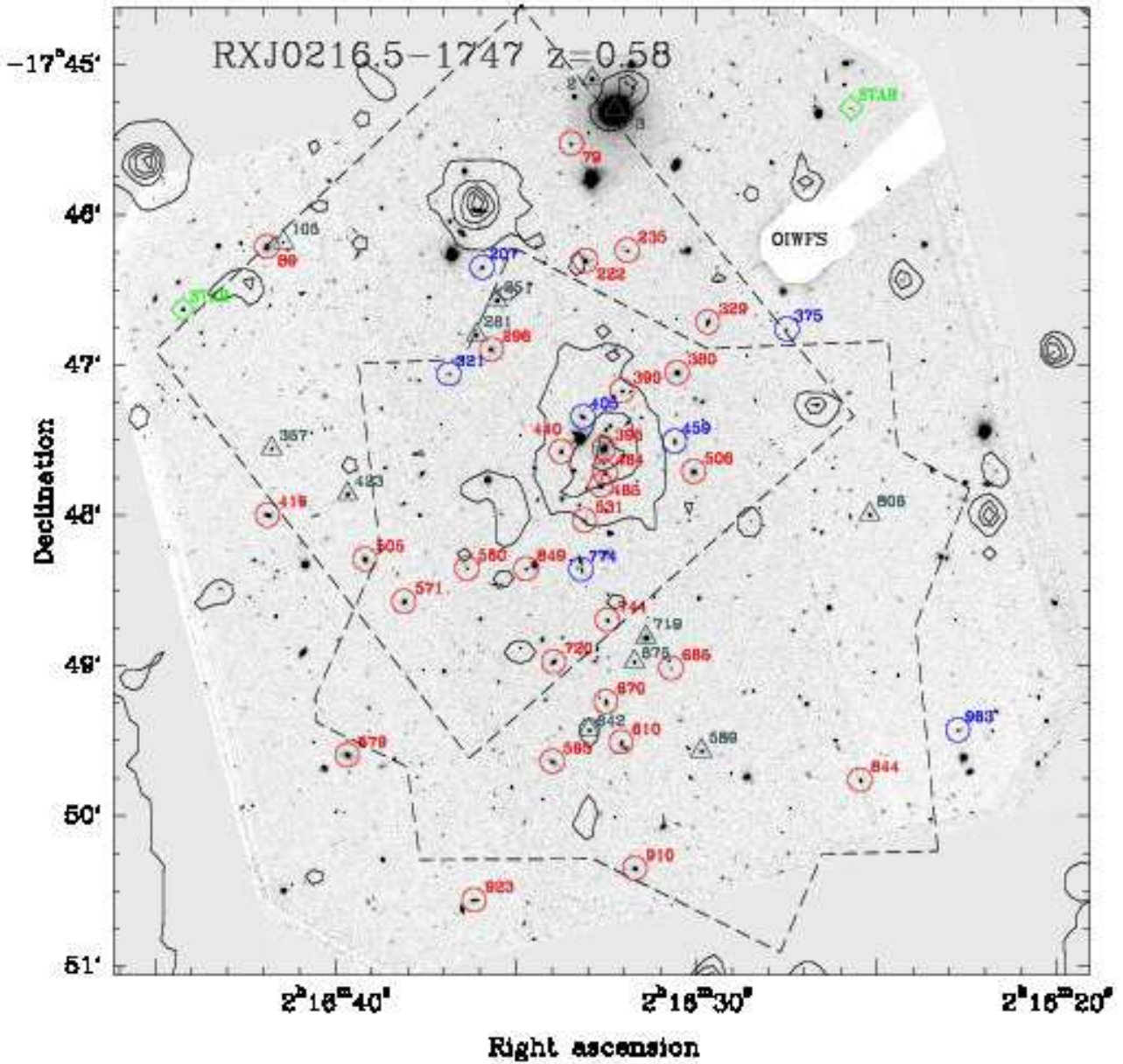}
\caption{
GMOS-S $i'$-band image of RXJ0216.5--1747 with the spectroscopic samples marked.
Contours of the {\it Chandra} X-ray data are overlaid.
Red circles -- confirmed members on the red sequence, $(r'-i')\ge 0.7$.
Blue circles -- confirmed blue members with $(r'-i') < 0.7$.
Dark green triangles -- confirmed non-members. 
Green diamonds -- blue stars included in the mask to facilitate correction for telluric absorption lines. 
The approximate location of the {\it HST}/ACS fields observed in F775W are marked with dashed lines.
The southern {\it HST}/ACS field was observed at two different roll-angles of {\it HST}.
The vignetting of the GMOS-S OIWFS is marked.
The X-ray image is from the {\it Chandra} ACIS camera, and is the sum of 
\dataset [ADS/Sa.CXO\#obs/05760] {Chandra ObsId 5760} and
\dataset [ADS/Sa.CXO\#obs/06393] {Chandra ObsId 6393}.
The X-ray image was smoothed; any structure seen is significant at 
the 3$\sigma$ level or higher. The spacing between the contours is logarithmic with a factor of 
1.5 between each contour.
\label{fig-RXJ0216grey} }
\end{figure*}

\begin{figure*}
\epsfxsize 17.0cm
\epsfbox{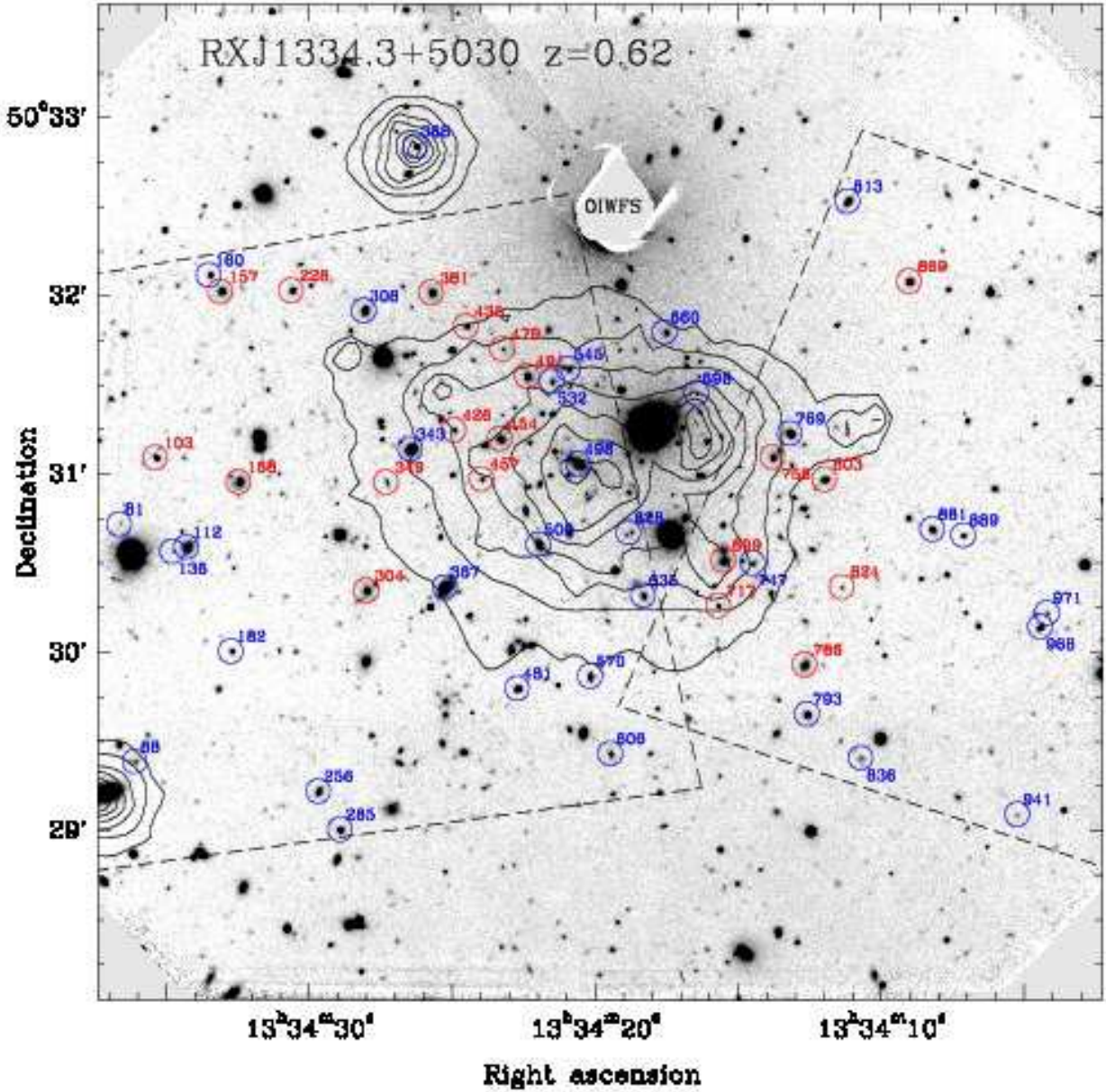}
\caption{
GMOS-N $i'$-band image of RXJ1334.3+5030 with the spectroscopic samples marked.
Contours of the {\it XMM-Newton} X-ray data are overlaid.
Red circles -- galaxies used to fit the red sequence ($(r'-i') \ge 0.9$ and $(i'-z') \ge 0.3$).
Blue circles -- blue galaxies in the spectroscopic sample.
The vignetting of the GMOS-N OIWFS is marked.
The approximate location of the {\it HST}/ACS fields observed in F775W is marked with dashed lines. 
The location of the {\it HST}/ACS fields were chosen to avoid the three bright foreground stars 
(one of which is vignetted by the OIWFS on this image) and optimize the inclusion of the red galaxies
in the spectroscopic sample.
The X-ray image is the sum of the images from the two {\it XMM-Newton} EPIC-MOS cameras. 
The X-ray image was smoothed; any structure seen is significant at 
the 3$\sigma$ level or higher. The spacing between the contours is logarithmic with a factor of 
1.5 between each contour.
\label{fig-RXJ1334grey} }
\end{figure*}

\begin{figure*}
\epsfxsize 17.0cm
\epsfbox{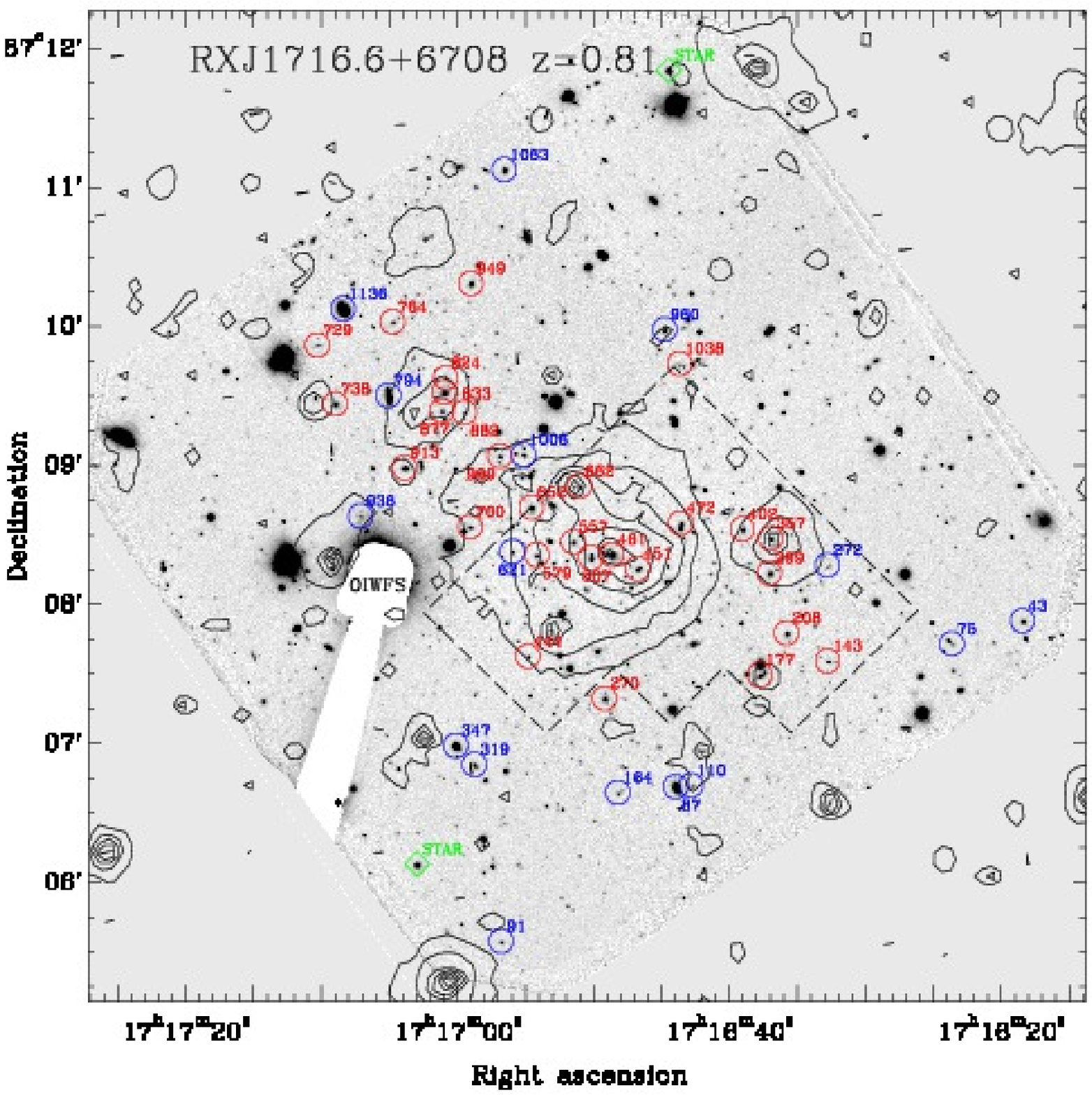}
\caption{
GMOS-N $i'$-band image of RXJ1716.6+6708 with the spectroscopic samples marked.
Contours of the {\it Chandra} X-ray data are overlaid.
Red circles -- galaxies used to fit the red sequence ($(r'-i') \ge 1.0$ and $(i'-z') \ge 0.5$).
Blue circles -- blue galaxies in the spectroscopic sample.
Green diamonds -- blue stars included in the mask to facilitate correction for telluric absorption lines. 
The vignetting of the GMOS-N OIWFS is marked.
The approximate location of the {\it HST}/WFPC2 field observed in F814W is marked with dashed lines. 
The X-ray image is from the {\it Chandra} ACIS camera
 \dataset [ADS/Sa.CXO\#obs/00548] {Chandra ObsId 548}.
The X-ray image was smoothed; any structure seen is significant at 
the 3$\sigma$ level or higher. The spacing between the contours is logarithmic with a factor of 
1.5 between each contour.
\label{fig-RXJ1716grey} }
\end{figure*}

\begin{figure*}
\epsfxsize 17.0cm
\epsfbox{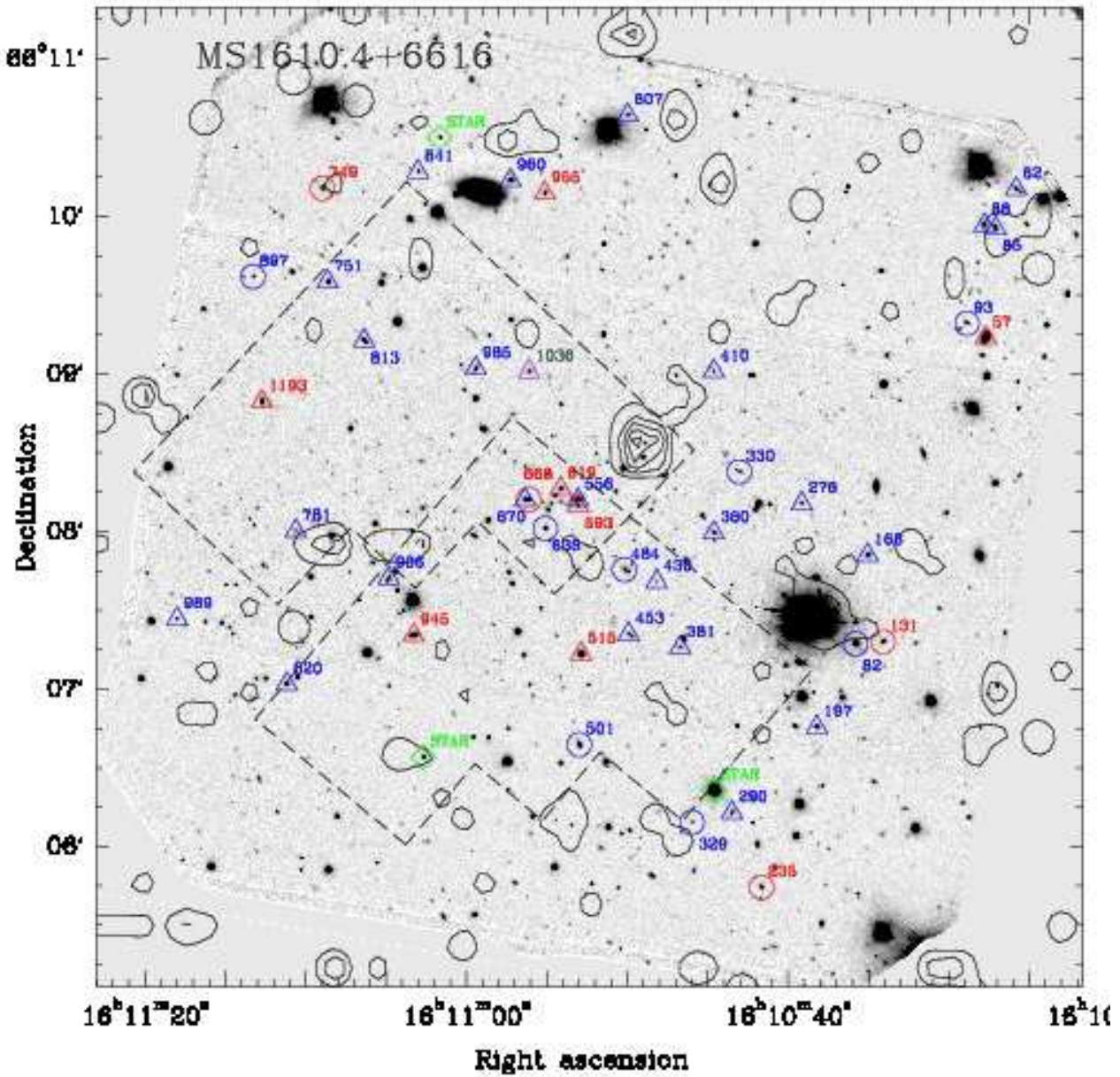}
\caption{
GMOS-N $i'$-band image of MS1610.4+6616 with the spectroscopic samples marked.
Contours of the {\it ROSAT} X-ray data are overlaid.
Red circles -- passive galaxies in the $z=0.83$ group. 
Blue circles -- emission line galaxies in the  $z=0.83$ group.
Red triangles -- passive galaxies at other redshifts.
Blue triangles -- emission line galaxies at other redshifts.
Purple triangles -- targets for which the spectra do not allow redshift determination. 
Green diamonds -- blue stars included in the mask to facilitate correction for telluric absorption lines. 
The approximate location of the {\it HST}/WFPC2 fields observed in F702W are marked with dashed lines. 
The X-ray image is from the {\it ROSAT} HRI camera. 
The X-ray image was smoothed; any structure seen is significant at 
the 3$\sigma$ level or higher. The spacing between the contours is logarithmic with a factor of 
1.5 between each contour.
The field contains no extended X-ray sources.
There are two X-ray bright point sources at (RA,DEC) = (16:10:49, 66:08:32) and
(16:10:38, 66:07:26). The latter correponds to the position of a bright foreground star.
\label{fig-MS1610grey} }
\end{figure*}

\begin{figure*}
\epsfxsize 17.0cm
\epsfbox{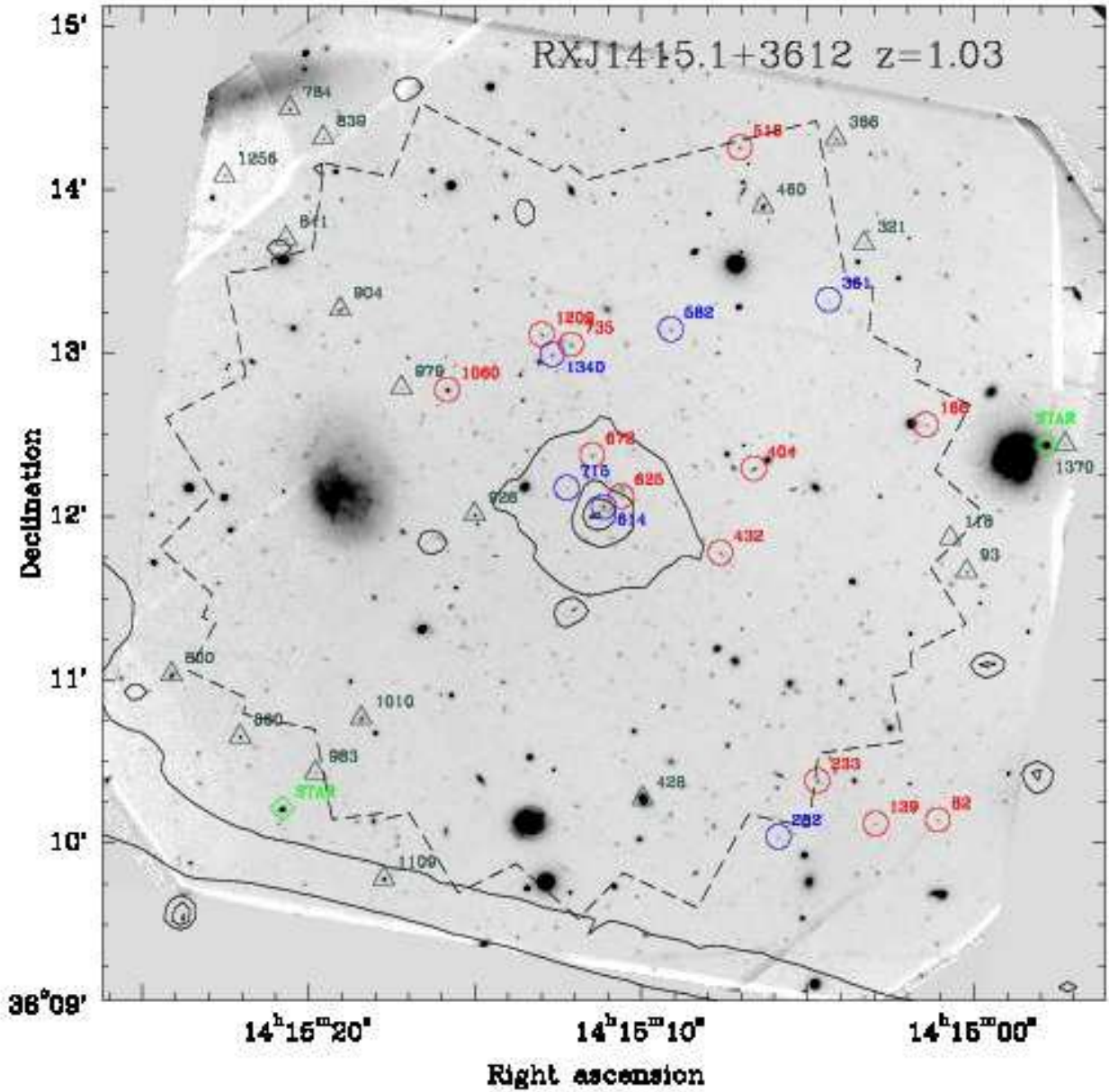}
\caption{
GMOS-N $i'$-band image of RXJ1415.1+3612 with the spectroscopic samples marked.
Contours of the {\it Chandra} X-ray data are overlaid.
Red circles -- confirmed passive member galaxies.
Blue circles -- confirmed members with significant emission.
Dark green triangles - non-members.
Green diamonds -- blue stars included in the mask to facilitate correction for telluric absorption lines. 
The approximate location of the {\it HST}/ACS field observed in F850LP is marked with dashed lines. 
The field was observed in several visits with different roll-angles of {\it HST}.
The X-ray image is from the {\it Chandra} ACIS camera and is the sum of 
\dataset [ADS/Sa.CXO\#obs/04163] {Chandra ObsId 4163},
\dataset [ADS/Sa.CXO\#obs/12255] {Chandra ObsId 12255},
\dataset [ADS/Sa.CXO\#obs/12256] {Chandra ObsId 12256},
\dataset [ADS/Sa.CXO\#obs/13118] {Chandra ObsId 13118}, and
\dataset [ADS/Sa.CXO\#obs/13119] {Chandra ObsId 13119}
The X-ray image was smoothed; any structure seen is significant at 
the 3$\sigma$ level or higher. The spacing between the contours is logarithmic with a factor of 
1.5 between each contour.
\label{fig-RXJ1415grey} }
\end{figure*}

In the descriptions we make use of Figures \ref{fig-RXJ0142grey}--\ref{fig-RXJ1415grey} showing the 
grey scale images of the clusters for which such information was not included in our previous papers. 
The images include overlaid contours of X-ray data from either {\it XMM-Newton} or {\it Chandra}, 
or in the case of the field MS1610.4+6616 X-ray data from {\it ROSAT}.
We also show the greyscale image for RXJ0142.0+2131 since at the time of the original publication 
(Barr et al.\ 2005), the {\it Chandra} X-ray data were not available.
Grey scale images of the remaining GCP clusters are available in
J\o rgensen et al.\ (2005, RXJ0152.7--1357), 
J\o rgensen \& Chiboucas (2013, MS0451.6--0305, RXJ1226.9+3332),
and J\o rgensen et al.\ (2017, Abell 1689, RXJ0056.2+2622, RXJ0027.6+2626, RXJ1347.5--1145).

For the clusters RXJ0142.0+2131, Abell 851, RXJ0216.5--1747, and RXJ1415.1+3612 our spectroscopic
samples are marked with information about cluster membership.
For MS1610.4+6616 we show the spectroscopic sample with the 12 members of the poor group indicated.
The processing of the spectroscopic data for RXJ2146.0+0423, RXJ1334.3+5060 and RXJ1716.6+6708
is pending. Thus, for these clusters we show the spectroscopic sample divided in galaxies on the 
red sequence and those bluer than the red sequence. The labeling matches our selection
for the fits to the red sequence, see Section \ref{SEC-CM}.

{\bf Abell 1689 / RXJ1311.4--0120:}
The cluster is included in the Abell catalog of northern clusters (Abell et al.\ 1989).
The cluster has been observed with {\it XMM-Newton} and {\it Chandra},
see J\o rgensen et al.\ (2017) for the X-ray data overlaid on our imaging data.
The cluster velocity dispersion is very high, $\approx 2100\, {\rm km\,s^{-1}}$ (J\o rgensen et al.; Czoske 2004).
Analysis of the cluster kinematic data and lensing data (Lemze et al.\ 2009; Umetsu et al.\ 2015)
shows that the central structure is complex and that the X-ray mass estimate is likely too low.
The cluster may be a merger, see discussion in Andersson \& Madejski (2004).
The cluster is included in our spectroscopic analysis of GCP data (J\o rgensen et al.\ 2017)
where we find that the stellar populations of the bulge-dominated galaxies are consistent
with the median metallicities and abundance ratios for the GCP clusters.

{\bf RXJ0056.2+2622 / Abell 115:}
This cluster is also included in the Abell catalog of northern clusters (Abell et al.\ 1989).
This is a binary cluster, see grey scale image in our analysis paper J\o rgensen et al.\ (2017).
Barrena et al.\ (2007) fund that the two sub-clusters are in the plane of the sky as based
on the kinematic structure of the cluster.
The Northern sub-cluster brightest galaxy is the powerful radio galaxy 3C\,28 and hosts an 
active galactic nucleus (AGN), see e.g.\ Hardcastle et al.\ (2009).
The galaxy is ID 1054 in our spectroscopic sample.
The cluster has been observed with {\it XMM-Newton} and {\it Chandra}.
The cluster X-ray emission is quite diffuse showing the presence of the two sub-clusters, see
the overlay of {\it XMM-Newton} data on our optical imaging in J\o rgensen et al.\ (2017).
We find that the stellar populations of the bulge-dominated galaxies in the cluster are consistent
with the median metallicities and abundance ratios for the GCP clusters (J\o rgensen et al.).

{\bf RXJ0142.0+2131:} 
The cluster is included in Northern {\it ROSAT} All-Sky Galaxy Cluster Survey (NORAS, B\"{o}hringer et al.\ 2000) 
and the extended {\it ROSAT} Brightest Cluster Sample (eBCS, Ebeling et al.\ 2000a).
A bright foreground galaxy is superimposed on the cluster near the center. 
This lead B\"{o}hringer et al.\ to mistakenly list the cluster redshift as 0.0696, the 
redshift of the foreground galaxy, while the correct cluster redshift is $z=0.28$.
At the time of the publication of our spectroscopic study of the cluster, Barr et al.\ (2005),
{\it XMM-Newton} and {\it Chandra} data were not available. 
{\it Chandra} data has since been obtained. Figure \ref{fig-RXJ0142grey} shows the X-ray 
data overlaid on our optical imaging. 
The morphological appearance is that of a relaxed cluster, with X-ray point sources 
associated with optical counterparts. 
No detailed analysis of the X-ray data seems to be available in the literature.

{\bf RXJ0027.6+2616:}
This cluster was discovered in {\it ROSAT} observations and 
first listed in NORAS by B\"{o}hringer et al.\ (2000) and also 
included in the eBCS (Ebeling et al.\ 2000a).
Later observations with {\it Chandra} show little substructure, except for possibly
some X-ray emission associated with members of the foreground group at $z=0.34$,
which we identified from our optical spectroscopy J\o rgensen et al.\ (2017).
The full spectroscopic analysis is included in J\o rgensen et al.

{\bf Abell 851:} 
The cluster is included in the Abell catalog of northern clusters (Abell et al.\ 1989).
The cluster is very massive and contains substantial sub-structure. 
Based on {\it XMM-Newton} observations, De Filippis et al.\ (2003) identified two main 
sub-clusters with internal structure, 
see also Figure \ref{fig-A0851grey} of the X-ray data overlaid on our optical imaging.
The cluster contains a large fraction of disk galaxies as well as post-starburst galaxies
(Andreon et al.\ 1997; Oemler et al.\ 2009), and as such is quite atypical for a massive cluster at 
this redshift. Andreon et al.\ find the spiral fraction to be close to 50 percent and 
hypothesize that the reason may be that the relatively low density inter-cluster gas failed to 
stop the star formation in the cluster members.
Oemler et al.\ focus on star-burst and post-starburst galaxies in the cluster, and in
particular find that the youngest star-burst galaxies reside in the center of the cluster.
Our spectral analysis of the GCP data will be presented in Hibon et al.\ (in prep.).

{\bf RXJ1347.5--1147:} 
This cluster was discovered as the most X-ray luminous of the {\it ROSAT} clusters (Schindler et al.\ 1997).
It has been studied extensively in both the optical and X-ray.
Weak and strong lensing studies have been conducted in an attempt to better estimate
the cluster mass and understand the dynamical structure of the cluster, e.g., Brada\v{c} et al.\ (2008).
As discussed in J\o rgensen et al.\ (2017), there is evidence that
the cluster contains an infalling sub-structure to the south east of the cluster center
(Ettori et al.\ 2004; Kreisch et al.\ 2016). We also found that the 
velocity dispersion of the cluster is in agreement with expectations from the X-ray
luminosity, once corrected for the diffuse emission from the sub-structure.
The cluster is included in the analysis in J\o rgensen et al.

{\bf RXJ2146.0+0423:}
The cluster was first mentioned by Gunn, Hoessel \& Oke (1986) in their photographic survey
for intermediate redshift clusters.
The cluster appeared in the 160 square degree {\it ROSAT} survey (Vikhlinin et al.\ 1998), 
with the cluster redshift listed by Mullis et al.\ (2003).
The cluster was also included in the Wide Angle {\it ROSAT} Pointed Survey (WARPS; Perlman et al.\ 2002). 
Figure \ref{fig-RXJ2146grey} shows our optical image of the cluster overlaid with the X-ray data from {\it XMM-Newton}.
The cluster is one of the lowest mass clusters included in the GCP and appears relatively compact
with the majority of the red galaxies in our spectroscopic sample within one arcminute of the cluster center.
Our spectroscopic analysis of the cluster will be presented in a future paper.

{\bf MS0451.6--0305:} 
This cluster was the most X-ray luminous cluster included in the Einstein Extended Medium Sensitive Survey
(EMSS, Gioia \& Luppino 1994).
Based on {\it ROSAT} data, it was estimated to be among the most X-ray luminous clusters above redshift 0.5 
(Ebeling et al.\ 2007).
The CNOC survey found a very large cluster velocity
dispersion, $1330\pm 100 {\rm km\,s^{-1}}$, confirming the high mass of the cluster (Ellingson et al.\ 1998; Borgani et al.\ 1999).
This is in agreement with our result for the velocity dispersion, $1450_{-159}^{+100}\,{\rm km\,s^{-1}}$
(J\o rgensen \& Chiboucas 2013).
Strong lensing modeling by Zitrin et al.\ (2011) shows that the central mass distribution may be elliptical,
while weak lensing studies show that the brightest cluster galaxy is slightly offset from the peak of the X-ray emission
(Comerford et al.\ 2010; Hoekstra et al.\ 2012; Soucail et al.\ 2015).
Thus, the cluster is most likely not relaxed.
Moran et al.\ (2007a, 2007b) used wide-field {\it HST}/ACS data and optical spectroscopy to study
the morphological evolution and star formation history. These authors concluded that the star formation history
is truncated, and that star formation stopped at an epoch corresponding to a formation redshift of $\approx 2$.
The cluster is included in our analysis in J\o rgensen \& Chiboucas (2013)
and J\o rgensen et al.\ (2017). Based on the absorption line strengths we find that the bulge-dominated
galaxies in this cluster on average have $\approx 0.1$ dex lower metallicity than found for other GCP clusters.

{\bf RXJ0216.5--1747:}
The cluster was included in WARPS (Perlman et al.\ 2002).  Together with \\
RXJ2146.0+0423, RXJ1334.3+5030 and RXJ1716.6+6708, this cluster
is among the four lowest mass clusters in the GCP.
Figure \ref{fig-RXJ0216grey} shows our optical image of the cluster overlaid with the X-ray data from {\it Chandra}.
The cluster appears less compact than RXJ2146.0+0423 and RXJ1716.6+6708.
Our spectroscopic analysis of the cluster will be presented in a future paper.

{\bf RXJ1334.3+5030:}
The cluster was included in the Bright Bright Serendipitous High-Redshift Archival Cluster (SHARC) survey (Romer et al.\ 2000).
The cluster is characterized as non-relaxed by Parekh et al.\ (2015) based on the
morphology of the X-ray emission.
Figure \ref{fig-RXJ1334grey} shows our optical image of the cluster overlaid with the X-ray data from {\it XMM-Newton}.
Our spectroscopic analysis of the cluster will be presented in a future paper.

{\bf RXJ1716.6+6708:}
This cluster was discovered as part of the {\it ROSAT} North Ecliptic Pole Survey (Henry et al.\ 1997).
At the time of discovery the cluster was among only four known $z>0.75$ X-ray clusters.
Gioia et al.\ (1999) find from 37 member galaxies a cluster velocity dispersion of $\approx 1500 {\rm km\,s^{-1}}$,
which is significantly higher than expected given the X-ray luminosity of the cluster.
Figure \ref{fig-RXJ1716grey} shows our optical image of the cluster overlaid with the X-ray data from {\it Chandra}.
The X-ray morphology shows some sub-structure possibly associated with two concentrations of red galaxies about
1.5 arcminute from the otherwise compact core of the cluster.
The cluster was included, together with the two higher redshift GCP clusters RXJ0152.7--1357 and RXJ1226.9+3332,
in the investigation of the $K$-band luminosity function for $z=0.6-1.3$
clusters by De Propris et al.\ (2007), who found the luminosity functions to be consistent with
massive galaxies being fully in place by $z\approx 1.3$, but also that the 
epoch of major star formation was as recent as $z=1.5-2$.
Our spectroscopic analysis of the cluster will be presented in a future paper.

{\bf MS1610.4+6616:} 
The EMSS (Gioia \& Luppino 1994) included MS1610.4+6616 as a cluster at redshift larger than 0.5,
while Luppino \& Gioia (1995) gives a redshift of 0.65. 
However, subsequent observations have shown that this is not a rich cluster.  
Donahue et al.\ (1999) state that the X-ray source is a point source and possibly originates from an AGN.
Our spectroscopic observations confirm that the field does not contain a rich cluster.
Our sample contains 27 galaxies with redshifts of $z=0.60-0.86$, 12 of which are clustered at $z=0.83$.
However, there is no clear clustering of these galaxies and no obvious extended X-ray emission
associated with them, see Figure \ref{fig-MS1610grey}.
For completeness, we include the photometry obtained of this field in the present paper.

{\bf RXJ0152.7--1357:} 
The cluster was originally discovered from {\it ROSAT} data and is included in 
three different {\it ROSAT} surveys: the {\it ROSAT} Deep Cluster Survey, 
the Bright SHARC survey (Nichol et al. 1999), and WARPS (Ebeling et al. 2000b). 
The {\it XMM-Newton} and {\it Chandra} X-ray observations show that the cluster consists
of two sub-clusters, probably in the process of merging (Maughan et al.\ 2003).
See also J\o rgensen et al.\ (2005) for the {\it XMM-Newton} X-ray data overlaid on our optical imaging.
Nantais et al.\ (2013) studied the morphologies of the member galaxies and put forward the hypothesis
that infalling galaxies are transformed directly from peculiar systems into bulge-dominated galaxies.
The cluster is included in our analysis in J\o rgensen et al.\ (2005, 2006, 2007), J\o rgensen \& Chiboucas (2013)
and J\o rgensen et al.\ (2017). The steep slope of Fundamental Plane for the cluster relative to that of our low redshift reference sample
supports that low mass bulge-dominated galaxies contain younger stellar populations than those of the higher mass galaxies.
This is also supported by the analysis of spectroscopic and photometric data presented by Demarco et al.\ (2010).
In addition, we find that the average of the abundance ratios $[\alpha /{\rm Fe}]$ derived from our spectra are $\approx 0.2$ dex
higher than that of the other GCP clusters.

{\bf RXJ1226.9+3332:}
This cluster was discovered in WARPS (Ebeling et al.\ 2001). 
The X-ray structure is due to AGNs. However, Maughan et al.\ (2007) analyzed the temperature map of the X-ray gas
and concluded that it showed evidence of a recent merger event, which is associated with the overdensity
of the galaxies south-west of the cluster center, see the grey scale image of the cluster in
our analysis paper J\o rgensen \& Chiboucas (2013). 
Recent analysis based on the Sanyaev-Zel'dovich effect of the cluster supports this view (Adam et al.\ 2015).
The cluster is included in our analysis in J\o rgensen et al.\ (2006, 2007), J\o rgensen \& Chiboucas (2013)
and J\o rgensen et al.\ (2017). This Fundamental Plane for this cluster is consistent with our
results for RXJ0152.7--1357, and indicates presence of younger stellar populations in the lower mass galaxies than those
in higher mass galaxies.

{\bf RXJ1415.1+3612:}
The highest redshift cluster in our $z=0.2-1.0$ GCP sample 
was first listed in WARPS (Perlman et al.\ 2002), with the note added in proof giving the 
spectroscopic confirmation of $z=1.013$ for the brightest cluster galaxy.
Huang et al.\ (2009) studied strong lensing created by the cluster.
These authors find the cluster redshift to be $z=1.026$ and 
cluster velocity dispersion of $\sigma _{\rm cl}=807 \pm 185 {\rm km\,s^{-1}}$, 
in agreement with our measurements, cf.\ Table \ref{tab-redshifts}.
Figure \ref{fig-RXJ1415grey} shows our optical image of the cluster overlaid with the X-ray data from {\it Chandra}.
Our spectroscopic analysis of the cluster will be presented in a future paper.

\end{document}